\documentclass{lmcs}
\pdfoutput=1

\usepackage{lastpage}
\lmcsdoi{18}{1}{11}
\lmcsheading{}{\pageref{LastPage}}{}{}%
{Apr.~07,~2021}{Jan.~17,~2022}{}
\usepackage[utf8]{inputenc}


\usepackage{tikz}
\usetikzlibrary{decorations,arrows,shapes,automata,calc}

\usepackage{amssymb}
\usepackage{color}

\newcommand{\IN}{\mathbb{N}}
\newcommand{\IZ}{\mathbb{Z}}

\newcommand{\UCol}{\mathsf{UCol}}
\newcommand{\DCol}{\mathsf{DCol}}
\newcommand{\emptyPth}{\ensuremath{\lambda}}
\newcommand{\pth}{\ensuremath{\rho}}
\newcommand{\play}{\ensuremath{\pi}}
\newcommand{\lang}{\ensuremath{K}}
\newcommand{\clr}{\ensuremath{c}}
\newcommand{\colorsSubset}{\ensuremath{B}}
\newcommand{\colors}{\ensuremath{C}}
\newcommand{\state}{\ensuremath{s}}
\newcommand{\states}{\ensuremath{S}}

\newcommand{\covStates}{\ensuremath{\states_{\mathsf{cov}}}}
\newcommand{\edge}{\ensuremath{e}}
\newcommand{\edges}{\ensuremath{E}}
\newcommand{\edgeIn}{\ensuremath{\mathsf{in}}}
\newcommand{\edgeOut}{\ensuremath{\mathsf{out}}}
\newcommand{\arena}{\ensuremath{\mathcal{A}}}
\newcommand{\col}{\ensuremath{\mathsf{col}}}
\newcommand{\colhat}{\ensuremath{\mathsf{\widehat{col}}}}
\newcommand{\arenafull}{\ensuremath{\arena = (\states_1, \states_2, \edges)}}

\newcommand{\pref}{\ensuremath{\sqsubseteq}} 
\newcommand{\strictpref}{\ensuremath{\sqsubset}}
\newcommand{\inverse}[1]{{#1}^{-1}}
\newcommand{\invpref}{\inverse{\pref}}

\newcommand{\game}{\ensuremath{\mathcal{G}}}
\newcommand{\gamefull}{\ensuremath{\game = (\arena, \pref)}}
\newcommand{\strat}{\ensuremath{\sigma}}
\newcommand{\strats}{\ensuremath{\Sigma}}
\newcommand{\stratsFM}{\ensuremath{\Sigma^{\mathsf{FM}}}}
\newcommand{\stratsML}{\ensuremath{\Sigma^{\mathsf{ML}}}}

\newcommand*{\player}[1]{\ensuremath{\mathcal{P}_{#1}}}
\newcommand*{\Pone}{\ensuremath{\mathcal{P}_1}}
\newcommand*{\Ptwo}{\ensuremath{\mathcal{P}_2}}

\newcommand{\Rec}[1]{\ensuremath{\mathcal{R}(#1)}}
\newcommand{\Reg}[1]{\ensuremath{\textsf{Reg}(#1)}}
\newcommand{\Prefixes}[1]{\ensuremath{\textsf{Prefs}(#1)}}
\newcommand{\PrefClosure}[1]{[#1]}

\newcommand{\Plays}{\ensuremath{\mathsf{Plays}}}
\newcommand{\Paths}{\ensuremath{\mathsf{Hists}}}

\newcommand{\famelt}{\ensuremath{L}}

\newcommand{\word}{\ensuremath{w}}
\newcommand{\Words}{\ensuremath{W}}
\newcommand{\memstate}{\ensuremath{m}}
\newcommand{\memstates}{\ensuremath{M}}
\newcommand{\memupd}{\alpha_{\mathsf{upd}}}
\newcommand{\memnxt}{\alpha_{\mathsf{nxt}}}
\newcommand{\memupdhat}{\widehat{\memupd}}
\newcommand{\meminit}{\memstate_{\mathsf{init}}}
\newcommand{\mealy}{\ensuremath{\Gamma}}
\newcommand{\memskel}{\ensuremath{\mathcal{M}}}
\newcommand{\memskelCyc}{\ensuremath{\mathcal{M}^\mathsf{c}}}
\newcommand{\memskelPref}{\ensuremath{\mathcal{M}^\mathsf{p}}}
\newcommand{\memskelfull}{\ensuremath{\memskel = (\memstates, \meminit, \memupd)}}

\newcommand{\prodAS}[2]{\ensuremath{#1 \ltimes #2}}
\newcommand{\memProduct}{\otimes}

\newcommand{\atmtnArena}[1]{\ensuremath{\mathsf{Arena}(#1)}}
\newcommand{\atmtn}{\ensuremath{\mathcal{N}}}
\newcommand{\atmtnIndex}[1]{\ensuremath{\atmtn^{#1}}}
\newcommand{\atmtnStates}{\ensuremath{Q}}
\newcommand{\atmtnState}{\ensuremath{q}}
\newcommand{\atmtnTrans}{\ensuremath{\delta}}
\newcommand{\atmtnTransHat}{\ensuremath{\widehat{\delta}}}
\newcommand{\atmtnInitState}{\ensuremath{\atmtnState_{\mathsf{init}}}}
\newcommand{\atmtnInitStates}{\ensuremath{\atmtnStates_{\mathsf{init}}}}
\newcommand{\atmtnFinalStates}{\ensuremath{Q_\mathsf{fin}}}
\newcommand{\atmtnFinalState}{\ensuremath{\atmtnState_\mathsf{fin}}}
\newcommand{\atmtnfull}{\ensuremath{\atmtn = (\atmtnStates, \colorsSubset, \atmtnTrans,
\atmtnInitStates, \atmtnFinalStates)}}
\newcommand{\atmtnfullIndex}[1]{\ensuremath{\atmtnIndex{#1} =
(\atmtnStates^{#1}, \colorsSubset^{#1}, \atmtnTrans^{#1}, \atmtnInitState^{#1},
\atmtnFinalStates^{#1})}}

\newcommand*{\forest}{\ensuremath{\mathcal{F}}}
\newcommand*{\stratBis}{\ensuremath{\tau}}
\newcommand*{\pathFun}{\ensuremath{\mathcal{H}}}
\newcommand*{\stratFun}{\ensuremath{f}}
\newcommand{\memskelIndexFull}[1]{\ensuremath{\memskel^{#1} = (\memstates^{#1}, \meminit^{#1}, \memupd^{#1})}}
\newcommand*{\wc}{\ensuremath{W}} 
\newcommand*{\comp}[1]{\overline{#1}}
\newcommand*{\reachT}{\ensuremath{T_1}}
\newcommand*{\reachTT}{\ensuremath{T_2}}
\newcommand*{\memstatePref}{\ensuremath{\memstate^\mathsf{p}}}
\newcommand*{\memstateCyc}{\ensuremath{\memstate^\mathsf{c}}}
\newcommand*{\meminitPref}{\ensuremath{\meminit^\mathsf{p}}}
\newcommand*{\meminitCyc}{\ensuremath{\meminit^\mathsf{c}}}
\newcommand*{\memupdPref}{\ensuremath{\memupd^\mathsf{p}}}
\newcommand*{\memupdCyc}{\ensuremath{\memupd^\mathsf{c}}}

\newcommand*{\memupdhatCyc}{\ensuremath{\widehat{\memupd^\mathsf{c}}}}
\newcommand*{\memstatesPref}{\ensuremath{\memstates^\mathsf{p}}}
\newcommand*{\memstatesCyc}{\ensuremath{\memstates^\mathsf{c}}}
\newcommand*{\memskelfullPref}{\ensuremath{\memskelPref = (\memstatesPref, \meminitPref, \memupdPref)}}
\newcommand*{\memskelfullCyc}{\ensuremath{\memskelCyc = (\memstatesCyc, \meminitCyc, \memupdCyc)}}
\newcommand*{\memskelTriv}{\memskel_{\mathsf{triv}}}
\newcommand*{\arenaClass}{\ensuremath{\mathfrak{A}}}

\makeatletter
\tikzset{circle split part fill/.style  args={#1,#2}{%
 alias=tmp@name, 
  postaction={%
    insert path={
     \pgfextra{%
     \pgfpointdiff{\pgfpointanchor{\pgf@node@name}{center}}%
                  {\pgfpointanchor{\pgf@node@name}{east}}%
     \pgfmathsetmacro\insiderad{\pgf@x}
      \fill[#1] (\pgf@node@name.base) ([xshift=-\pgflinewidth]\pgf@node@name.east) arc
                          (0:180:\insiderad-\pgflinewidth)--cycle;
      \fill[#2] (\pgf@node@name.base) ([xshift=\pgflinewidth]\pgf@node@name.west)  arc
                           (180:360:\insiderad-\pgflinewidth)--cycle;            
         }}}}}
 \makeatother

\colorlet{drouge}{red}
\colorlet{frouge}{red!20!white}
\colorlet{dbleu}{blue}
\colorlet{fbleu}{blue!40!white}
\colorlet{dviolet}{blue!50!red}
\colorlet{fviolet}{blue!50!red!40!white}
\colorlet{dvert}{green!80!black}
\colorlet{fvert}{green!80!black!20!white}
\colorlet{djaune}{yellow!80!black}
\colorlet{fjaune}{yellow!80!black!20!white}
\colorlet{dgris}{white!60!black}
\colorlet{fgris}{white!90!black}
\colorlet{dgrisf}{white!30!black}
\colorlet{fgrisf}{white!70!black}
\colorlet{dorange}{red!50!yellow}
\colorlet{forange}{red!50!yellow!30!white}

\tikzstyle{ptrond}=[draw,circle,minimum height=2mm]
\tikzstyle{ptcarre}=[draw,minimum width=3mm,minimum height=3mm]
\tikzstyle{moyrond}=[draw,circle,minimum height=5mm]
\tikzstyle{moycarre}=[draw,minimum width=4mm,minimum height=4mm]
\tikzstyle{rond}=[draw,circle,minimum height=7mm]
\tikzstyle{oval}=[draw,ellipse,minimum height=7mm]
\tikzstyle{carre}=[draw,minimum width=6mm,minimum height=6mm]
\tikzstyle{rouge}=[draw=drouge,fill=frouge]
\tikzstyle{vert}=[draw=dvert,fill=fvert]
\tikzstyle{jaune}=[draw=djaune,fill=fjaune]
\tikzstyle{bleu}=[draw=dbleu,fill=fbleu]
\tikzstyle{violet}=[draw=dviolet,fill=fviolet]
\tikzstyle{orange}=[draw=dorange,fill=forange]
\tikzstyle{gris}=[draw=dgris,fill=fgris]
\tikzstyle{grisf}=[draw=dgrisf,fill=fgrisf]
\tikzstyle{rvert}=[style=rond,style=vert]
\tikzstyle{rrouge}=[style=rond,style=rouge]
\tikzstyle{roundrect}=[draw,rounded rectangle, minimum width=6mm,minimum height=6mm]
\tikzstyle{splitrond}=[draw,circle split,minimum height=7mm,circle split part fill={blue!50,red!50}]
\tikzstyle{splitrondbv}=[draw,circle split,minimum height=7mm,circle split part fill={fbleu,fvert}]
\tikzstyle{splitrondrg}=[draw,circle split,minimum height=7mm,circle split part fill={frouge,fgris}]
\tikzstyle{splitrondbo}=[draw,circle split,minimum height=7mm,circle split part fill={fbleu,forange}]

\makeatletter
\def\listof#1{\expandafter\@listof#1+\@end}
\def\@listof#1+#2\@end{\def\@tempa{#1}\ifx\@tempa\@empty\else
    \langle #1\rangle \fi
  \def\@tempa{#2}\ifx\@tempa\@empty\else,\@listof#2\@end\fi}
\def\stackof#1{\begin{array}{@{}>{\scriptstyle}c@{}}\expandafter\@stackof#1+\@end}
\def\@stackof#1+#2\@end{\def\@tempa{#1}\ifx\@tempa\@empty\else
    \langle #1\rangle \fi
  \def\@tempa{#2}\ifx\@tempa\@empty\end{array}\else\\[-1mm]\@stackof#2\@end\fi}
\makeatother

\begin{document}

\title[Playing Optimally with Arena-Independent Finite Memory]{Games Where You Can Play Optimally\texorpdfstring{\\}{} with Arena-Independent Finite Memory\texorpdfstring{\rsuper*}{}}
\titlecomment{{\lsuper*}Research supported by F.R.S.-FNRS under Grant n$^\circ$ F.4520.18 (ManySynth), F.R.S.-FNRS mobility funding for scientific missions (Y.~Oualhadj in UMONS, 2018), and ENS Paris-Saclay visiting professorship (M.~Randour, 2019). Mickael Randour is an F.R.S.-FNRS Research Associate and Pierre Vandenhove is an F.R.S.-FNRS Research Fellow.}

\author[P.~Bouyer]{Patricia Bouyer\rsuper{a}}

\author[S.~Le~Roux]{St\'ephane {Le Roux}\rsuper{a}}

\author[Y.~Oualhadj]{Youssouf Oualhadj\rsuper{b}}

\author[M.~Randour]{Mickael Randour\rsuper{c}}

\author[P.~Vandenhove]{Pierre Vandenhove\rsuper{a,c}}
\address{Université Paris-Saclay, CNRS, ENS Paris-Saclay, Laboratoire Méthodes Formelles, 91190, Gif-sur-Yvette, France}
\address{Univ Paris Est Creteil, LACL, F-94010 Creteil, France}
\address{F.R.S.-FNRS \& UMONS~-- Universit\'e de Mons, Mons, Belgium}
\email{mickael.randour@umons.ac.be}

\begin{abstract}
For decades, two-player (antagonistic) games on graphs have been a framework of choice for many important problems in theoretical computer science. A notorious one is controller synthesis, which can be rephrased through the game-theoretic metaphor as the quest for a winning strategy of the system in a game against its antagonistic environment. Depending on the specification, \textit{optimal} strategies might be simple or quite complex, for example having to use (possibly infinite) memory. Hence, research strives to understand which settings allow for simple strategies.

In 2005, Gimbert and Zielonka provided a complete characterization of preference relations (a formal framework to model specifications and game objectives) that admit \textit{memoryless} optimal strategies for both players. In the last fifteen years however, practical applications have driven the community toward games with complex or multiple objectives, where memory~--- finite or infinite~--- is almost always required. Despite much effort, the exact frontiers of the class of preference relations that admit \textit{finite-memory} optimal strategies still elude us.

In this work, we establish a complete characterization of preference relations that admit optimal strategies using \textit{arena-independent} finite memory, generalizing the work of Gimbert and Zielonka to the finite-memory case. We also prove an equivalent to their celebrated corollary of great practical interest: if both players have optimal (arena-independent-)finite-memory strategies in all one-player games, then it is also the case in all two-player games. Finally, we pinpoint the boundaries of our results with regard to the literature: our work completely covers the case of arena-independent memory (e.g., multiple parity objectives, lower- and upper-bounded energy objectives), and paves the way to the arena-dependent case (e.g., multiple lower-bounded energy objectives).
\end{abstract}

\maketitle

\section{Introduction}
\label{sec:intro}
\paragraph{Controller synthesis through the game-theoretic metaphor} Two-player games on (finite) graphs are studied extensively, in particular for their application to controller synthesis for reactive systems (see, e.g.,~\cite{DBLP:conf/dagstuhl/2001automata,rECCS,DBLP:conf/lata/BrenguierCHPRRS16,DBLP:reference/mc/BloemCJ18}). The seminal model is \textit{antagonistic} (i.e., \textit{zero-sum} if one chooses a quantitative view): player~1 ($\Pone$) is seen as the system to control, player~2 ($\Ptwo$) as its antagonistic environment, and the game models their interaction. Each vertex of the game graph (called \textit{arena}) models a \textit{state} of the system and belongs to one of the players. Players take turns moving a pebble from state to state along the edges, each player choosing the destination whenever the pebble is on one of his states. These choices are made according to the \textit{strategy} of the player, which, in general, might use memory (bounded or not) of the past moves to prescribe the next action.

The resulting infinite sequence of states, called \textit{play}, represents the execution of the system. The objective of $\Pone$ is to enforce a given \textit{specification}, often encoded as a \textit{winning condition} (i.e., a set of winning plays) or as a \textit{payoff function} to maximize (i.e., a quantitative performance to optimize). This paradigm focuses on the \textit{worst-case} performance of the system, hence $\Ptwo$'s goal is to prevent $\Pone$ from achieving his objective.

The goal of \textit{synthesis} is thus to decide if $\Pone$ has a \textit{winning strategy}, i.e., one ensuring a given winning condition or guaranteeing a given payoff threshold, against all possible strategies of $\Ptwo$, and to build such a strategy efficiently if it exists.

Winning strategies are essentially \textit{formal blueprints} for controllers to implement in practical applications. Therefore, the complexity of these strategies is of tremendous importance: the simpler the strategy, the easier and cheaper it will be to build the corresponding controller and maintain it. This explains why a lot of research effort is constantly put into identifying the exact complexity (in terms of memory and/or randomness) of strategies needed to play \textit{optimally} (i.e., to the best of the player's ability) for each specific class of games and objectives (e.g.,~\cite{GZ05,DBLP:journals/acta/ChatterjeeRR14,DBLP:journals/iandc/Chatterjee0RR15,DBLP:journals/iandc/VelnerC0HRR15,DBLP:conf/icalp/FijalkowHKS15,DBLP:journals/corr/BruyereHR16,DBLP:journals/iandc/AminofR17,DBLP:conf/fossacs/BouyerHMR017,DBLP:conf/fsttcs/0001PR18,DBLP:journals/acta/BouyerMRLL18,DBLP:conf/concur/BruyereHRR19}). Alongside the \textit{practical interest} of this question lies the \textit{theoretical puzzle}: understanding the underlying mechanisms and implicit properties of games that lead to ``simple'' strategies being sufficient. Given the numerous connections between two-player games and various branches of mathematics and computer science, this fundamental question has interest in its own right.

\paragraph{Preference relations} As hinted above, there are two prominent ways to formalize a game objective in the literature. The first one, dubbed \textit{quantitative} and inspired by games in economics, is to use \textit{payoff functions} mapping plays to numerical values, and to see $\Pone$ as a maximizer player. This is for example the case of mean-payoff games~\cite{EM79}. The second one, called \textit{qualitative}, is to define a \textit{set of winning plays}~--- called \textit{winning condition}~--- induced by some property, as in, e.g., parity games~\cite{DBLP:conf/focs/EmersonJ88,DBLP:journals/tcs/Zielonka98}. The two formalisms are strongly linked: the classical decision problem for quantitative games is to fix a payoff threshold and ask if $\Pone$ has a strategy to guarantee it, essentially transforming the problem into a qualitative one (where the winning plays are all those achieving a payoff at least equal to the threshold). To define payoff functions or winning conditions, one often uses weights, priorities, colors, etc, on states or edges of the arena.

In this work, we walk in the footsteps of Gimbert and Zielonka~\cite{GZ05}: we associate a \textit{color} to each edge of our arenas, and we adopt the abstract formalism of \textit{preference relations} over infinite sequences of colors (induced by plays). This general formalism permits to encode virtually all classical game objectives, both qualitative and quantitative, and lets us reason in a well-founded framework under minimal assumptions. See Example~\ref{ex:preferenceRelations} for illustrations of classical objectives encoded as preference relations.

\paragraph{Memoryless optimal strategies} Remarkably, several canonical classes of games that have been around for decades and proved their usefulness over and over~--- e.g., mean-payoff~\cite{EM79}, parity~\cite{DBLP:conf/focs/EmersonJ88,DBLP:journals/tcs/Zielonka98}, or energy games~\cite{DBLP:conf/emsoft/ChakrabartiAHS03}~--- share a desirable property: they all admit \textit{memoryless optimal strategies for both players}. That is, for every strategy $\strat_{i}$ of $\player{i}$, there is a strategy $\strat^\mathsf{ML}_{i}$ which is \textit{at least as good} (i.e., wins whenever $\strat_i$ wins or ensures at least the same payoff) and that uses no memory at all. Such a memoryless strategy always picks the same edge when in the same state, regardless of what happened earlier in the game.

Memoryless strategies are the simplest kind of strategies one can use in a turn-based game on a graph. Therefore, it is quite interesting that they suffice for objectives as rich as the ones we just discussed. Following this observation, a lot of effort has been put into understanding which games admit memoryless optimal strategies, and in identifying the exact frontiers of \textit{memoryless determinacy}. Let us mention, non-exhaustively, works by Gimbert and Zielonka~\cite{DBLP:conf/mfcs/GimbertZ04,GZ05} (culminating in a complete characterization), Aminof and Rubin~\cite{DBLP:journals/iandc/AminofR17} (through the prism of first-cycle games), and Kopczy\'nski~\cite{DBLP:conf/icalp/Kopczynski06} and Bianco et al.~\cite{DBLP:journals/amai/BiancoFMM11} (half-positional determinacy). All these advances were built by identifying the common underlying mechanisms in ad hoc proofs for specific classes of games, and generalizing them to wide classes (e.g., the first-cycle games of Aminof and Rubin are inspired by the seminal paper of Ehrenfeucht and Mycielski on mean-payoff games~\cite{EM79}).

\paragraph{Gimbert and Zielonka's approach} Arguably, the most important result in this direction is the \textit{complete characterization} of preference relations admitting memoryless optimal strategies, established in~\cite{GZ05}, fifteen years ago. By complete characterization, we mean sufficient \textit{and} necessary conditions on the preference relations.

This result can be stated as follows: a preference relation admits memoryless optimal strategies for both players on all arenas if and only if the relation (used by $\Pone$) and its inverse (used by $\Ptwo$) are \textit{monotone} and \textit{selective}. These concepts will be defined formally in Section~\ref{subsec:concepts}, but let us give an intuition here. Roughly, a preference relation is monotone if it is stable under \textit{prefix addition}: that is, given two sequences of colors such that one is \textit{strictly} preferred to the other, it is impossible to reverse this order of preference by adding the same prefix to both sequences. Selectivity is similarly defined with regard to \textit{cycle mixing}: if a preference relation is selective, then, starting from two sequences of colors, it is impossible to create a third one by mixing the first two in such a way that the third one is strictly preferred to the first two. Observe that these elegant notions coincide with the natural intuition that memoryless strategies suffice if there is no interest in behaving differently in a state depending on what happened earlier.

In addition to this complete characterization, Gimbert and Zielonka proved another great result, of high interest in practice~\cite[Corollary 7]{GZ05}: as a by-product of their approach, they obtain that if memoryless strategies suffice in all one-player games of $\Pone$ and all one-player games of $\Ptwo$, they also suffice in all two-player games. Such a \textit{lifting corollary} provides a neat and easy way to prove that a preference relation admits memoryless optimal strategies \textit{without proving monotony and selectivity at all}: proving it in the two one-player subcases, which is generally much easier as it boils down to graph reasoning, and then lifting the result to the general two-player case through the corollary.

\paragraph{The rise of memory} Over the last decade, the increasing need to model complex specifications has shifted research toward games where multiple (quantitative and qualitative) objectives co-exist and interact, requiring the analysis of \textit{interplay} and \textit{trade-offs} between several objectives. Hence, a lot of effort is put into studying games where objectives are actually conjunctions of objectives, or even richer Boolean combinations. See for example~\cite{DBLP:conf/fossacs/ChatterjeeHP07} for combinations of parity,~\cite{DBLP:journals/tcs/ChatterjeeD12,DBLP:journals/acta/ChatterjeeRR14,DBLP:conf/icalp/JurdzinskiLS15} for combinations of energy and parity,~\cite{DBLP:journals/iandc/VelnerC0HRR15} for combinations of mean-payoff,~\cite{DBLP:journals/acta/BouyerMRLL18,DBLP:conf/fossacs/BouyerHMR017} for combinations of energy and average-energy,~\cite{DBLP:conf/concur/BruyereHRR19} for combinations of energy and mean-payoff,~\cite{DBLP:journals/iandc/Chatterjee0RR15} for combinations of total-payoff, or~\cite{DBLP:journals/iandc/Chatterjee0RR15,DBLP:journals/corr/BruyereHR16,DBLP:conf/concur/BrihayeDOR19} for combinations of window objectives.

When considering such rich objectives, \textit{memoryless strategies usually do not suffice}, and one has to use an amount of memory that can quickly become an obstacle to implementation (e.g., exponential memory) or that can prevent it completely (infinite memory). Establishing precise memory bounds for such general combinations of objectives is tricky and sometimes counterintuitive. For example, while energy games and mean-payoff games are inter-reducible in the single-objective setting, exponential-memory strategies are both sufficient and necessary for conjunctions of energy objectives~\cite{DBLP:journals/acta/ChatterjeeRR14,DBLP:conf/icalp/JurdzinskiLS15} while \textit{infinite-memory} strategies are required for conjunctions of mean-payoff ones~\cite{DBLP:journals/iandc/VelnerC0HRR15}.

A natural question arises: \textit{which preference relations do admit finite-memory optimal strategies?} Surprisingly, whether an equivalent to Gimbert and Zielonka's characterization could be obtained in the finite-memory case or not has remained an open question up to now. It is worth noticing that such an equivalent could be of tremendous help in practice, especially if a \textit{lifting corollary} also holds: see for example~\cite{DBLP:journals/acta/BouyerMRLL18,DBLP:conf/fossacs/BouyerHMR017,DBLP:conf/concur/BruyereHRR19}, where proving that finite-memory strategies suffice in one-player games was fairly easy, in contrast to the high complexity of the two-player case~--- a lifting corollary could grant the two-player case for free!

Having said that, one has to hope that the following corollary can be established: ``if \textit{finite-memory} strategies suffice in all one-player games of $\Pone$ and all one-player games of $\Ptwo$, they also suffice in all two-player games.'' Unfortunately, this hope is but a delusion.

\paragraph{Lifting corollary: a counterexample} Consider games where the colors are integers, and the objective of $\Pone$ is to create a play such that (a) the running sum of weights grows up to infinity (e.g., consider its $\liminf$ to define it properly), \textit{or} (b) this running sum of weights takes value zero infinitely often. As this defines a qualitative objective, the corresponding preference relation induces only two equivalence classes: \textit{winning} and \textit{losing} plays. The inverse relation, used by $\Ptwo$, is trivial to obtain. It is fairly easy to prove that $\Pone$ always has finite-memory optimal strategies in his one-player games (i.e., games where $\Ptwo$ has no choice), and so does $\Ptwo$ in his one-player games. See Section~\ref{sec:counterexample} for formal details.

Now, consider the very simple two-player game depicted in Figure~\ref{fig:counterexample}. First, observe that $\Pone$ (circle) has an \textit{infinite-memory} strategy to win: $\Pone$ should keep track of the running sum of weights (which is unbounded, hence the need for infinite memory) and loop in $s_1$ up to the point where this sum hits zero, when $\Pone$ should then go to $s_2$. This strategy ensures victory because either $\Ptwo$ always goes back to $s_1$, in which case (b) is satisfied; or $\Ptwo$ eventually loops forever on $s_2$, in which case (a) is satisfied. It remains to argue that $\Pone$ has no \textit{finite-memory} winning strategy in this game. This can be done using a standard argument: whatever the amount of memory used by $\Pone$, $\Ptwo$ may loop in $s_2$ long enough as to exceed the bound up to which $\Pone$ can track the sum accurately; thus dooming $\Pone$ to fail to reset the sum to zero in $s_1$ infinitely often.

\begin{figure}[tbh]
\begin{center}
\begin{tikzpicture}[every node/.style={font=\small,inner sep=1pt}]
\draw (0,0) node[carre] (s1) {$s_2$};
\draw (2,0) node[rond] (s2) {$s_1$};
\draw (-1.2,0) node[] (s1lab) {$1$};
\draw (1,-0.6) node[] (s1lab) {$1$};
\draw (3.3,0) node[] (s2lab) {$-1$};
\draw (1,0.6) node[] (s2lab) {$-1$};
\draw (s1) edge[-latex',out=30,in=150] (s2);
\draw (s2) edge[-latex',out=210,in=330](s1);
\draw (s1) edge[-latex',out=150,in=210,looseness=4,distance=1cm] (s1);
\draw (s2) edge[-latex',out=330,in=30,looseness=4,distance=1cm] (s2);
\end{tikzpicture}
\caption{$\Pone$ (circle) needs infinite memory to win in this game (by always resetting the sum of weights to zero by looping long enough in $s_1$ before going back to $s_2$), whereas both players have finite-memory optimal strategies in all one-player games using the same preference relations.}
\label{fig:counterexample}
\end{center}
\end{figure}

This modest example proves that \textit{Gimbert and Zielonka's approach cannot work in full generality in the finite-memory case}, and for good reasons. Informally, in this case, the corollary breaks down because of (the absence of some sort of) monotony. In the case of \textit{memoryless} strategies, as in~\cite{GZ05}, $\Pone$ is already doomed in one-player games in the absence of monotony: two prefixes to distinguish~--- in order to play optimally~--- can be hardcoded as different paths leading to the same state in a game arena, as if they were chosen by $\Ptwo$ in a two-player game. In the case of \textit{finite-memory} strategies, however, the situation is different. In one-player games, the number of such paths that can be hardcoded in an arena is always bounded, hence finite memory might suffice to react, i.e., to keep track of which prefix is the current one and how to behave accordingly. However, in two-player games, $\Ptwo$ might create an \textit{infinite} number of prefixes to distinguish (using a cycle), thus requiring $\Pone$ to use infinite memory to be able to do so. This is exactly what happens in the example above: in any one-player game, the largest sum that $\Pone$ has to track is bounded, whereas $\Ptwo$ can make this sum as large as he wants in two-player games.

\paragraph{Our approach} In a nutshell, we generalize Gimbert and Zielonka's results~--- characterization \textit{and} lifting corollary~--- to the case of \textit{arena-independent} finite memory. That is, we encompass \textit{all} situations where the memory needed by the two players is \textit{solely dependent on the preference relation} (e.g., colors, dimensions of weight vectors), and \textit{not} on the game arena (i.e., number of edges/states). Let us take some classical examples to illustrate this notion.
\begin{itemize}
\item All memoryless-determined relations~--- studied in~\cite{GZ05}~--- use arena-independent memory: the memory required, \textit{none}, is the same for all arenas.
\item Combinations of parity objectives use arena-independent memory~\cite{DBLP:conf/fossacs/ChatterjeeHP07}: the memory only depends on the number of objectives and the number of priorities~--- both parameters of the preference relation, not on the size of the arena.
\item Lower- and upper-bounded energy objectives also use arena-independent memory (see, e.g.,~\cite{DBLP:conf/formats/BouyerFLMS08,DBLP:journals/acta/BouyerMRLL18,DBLP:conf/fossacs/BouyerHMR017}): the memory only depends on the bounds and the weights~--- parameters of the preference relation, not on the size of the arena.
\item On the contrary, combinations of lower-bounded energy objectives (with no upper bound) require \textit{arena-dependent} memory~\cite{DBLP:journals/acta/ChatterjeeRR14,DBLP:conf/icalp/JurdzinskiLS15}: the memory depends on the size of the arena in addition to the weights used in it. Such a setting falls outside the scope of our results.
\end{itemize}
This informal concept of arena-independent memory is transparent in our work: in all our results,  we use \textit{memory skeletons}~--- essentially Mealy machines without a next-action function (Section~\ref{sec:prelims})~--- that suffice for \textit{all} arenas, and that are at the basis of the strategies we build. A quick look at our main concepts (Section~\ref{subsec:concepts}) and results (Section~\ref{subsec:results}) suffices to grasp the formalism behind this intuition.

This restriction to arena-independent memory is natural given the counterexample to a general approach presented above. It is also important to note that it is not as restrictive as it may seem, as hinted by the examples above: we are \textit{not restricted to constant memory} but to memory only depending on the \textit{parameters of the preference relation} (or equivalently, objective), and not of the arena. This framework thus already encompasses many objectives from the literature~--- e.g.,~\cite{EM79,DBLP:conf/focs/EmersonJ88,DBLP:journals/tcs/Zielonka98,DBLP:conf/emsoft/ChakrabartiAHS03,DBLP:journals/acta/BouyerMRLL18,DBLP:journals/corr/abs-1010-2420,DBLP:conf/fossacs/ChatterjeeHP07,DBLP:journals/corr/BruyereHR16,DBLP:journals/iandc/Chatterjee0RR15,DBLP:conf/formats/BouyerFLMS08,DBLP:journals/acta/BouyerMRLL18,DBLP:conf/fossacs/BouyerHMR017}, as well as possible extensions. We discuss this topic in more details in Section~\ref{sec:discussion}, where we provide a precise description of the frontiers of our results within the current research landscape.

Let us also highlight that the \textit{arena-independent} case, which we solve here, is an exact equivalent to Gimbert and Zielonka's results in the finite-memory case: the memoryless case is de facto arena-independent. Therefore, this paper strictly generalizes~\cite{GZ05} by allowing to study any arena-independent memory skeleton instead of the unique trivial one corresponding to memoryless strategies.

\paragraph{Outline of our contributions} Informally, our characterization can be stated as follows: given a preference relation and a memory skeleton $\memskel$, both players have optimal finite-memory strategies based on skeleton $\memskel$ in all games if and only if the relation and its inverse are $\memskel$-monotone and $\memskel$-selective.

These last two concepts are keys to our approach. Intuitively, they correspond to Gimbert and Zielonka's monotony and selectivity, \textit{modulo a memory skeleton}. Recall that monotony and selectivity are related to stability of the preference relation with regard to prefix addition and cycle mixing, respectively. Our more general concepts of \textit{$\memskel$-monotony} and \textit{$\memskel$-selectivity} serve the same purpose, but they only compare sequences of colors that are deemed equivalent by the memory skeleton. For the sake of illustration, take selectivity: it implies that one has no interest in mixing different cycles of the game arena. For its generalization, the memory skeleton is taken into account: $\memskel$-selectivity implies that one has no interest in mixing cycles of the game arena \textit{that are read as cycles on the same memory state in the skeleton} $\memskel$.

Let us give a quick breakdown of our approach. In Section~\ref{sec:prelims}, we introduce all basic notions, including the memory skeletons, and we establish several technical results. We also discuss \textit{optimal strategies} and \textit{Nash equilibria}, their relationship, and their roles in our approach.

Section~\ref{sec:charac} is dedicated to our characterization, and consists of three parts. In Section~\ref{subsec:concepts}, we introduce the concepts of $\memskel$-monotony and $\memskel$-selectivity, cornerstones of our work. We also present two essential tools to establish the characterization: \textit{prefix-covers} and \textit{cyclic-covers} of arenas. Section~\ref{subsec:results} states formally our characterization (Theorem~\ref{thm:equivalence}), as well as the corresponding lifting corollary (Corollary~\ref{cor:lifting}), from one-player to two-player games. We close this overview with an example of application, in Section~\ref{subsec:example}.

The proof of the characterization (Theorem~\ref{thm:equivalence}) is split in two. In Section~\ref{sec:FMToConcepts}, we establish the implication from (the sufficiency of) finite memory based on $\memskel$ to $\memskel$-monotony (Theorem~\ref{thm:FMtoMonotony}) and $\memskel$-selectivity (Theorem~\ref{thm:FMtoSelectivity}) of the preference relation. The main idea here is to build game arenas based on automata recognizing the languages involved in the two concepts, and to use the existence of finite-memory optimal strategies in these arenas to prove that $\memskel$-monotony and $\memskel$-selectivity hold.

In Section~\ref{sec:conceptsToFM}, we prove the converse implication. We proceed in two steps, first establishing the existence of \textit{memoryless} optimal strategies in ``covered'' arenas (Lemma~\ref{lem:inductionStep} and Theorem~\ref{thm:conceptsToML}), and then building on it to obtain the existence of \textit{finite-memory} optimal strategies in general arenas (Corollary~\ref{cor:UFM}). The main technical tools we use are Nash equilibria and the aforementioned notions of prefix-covers and cyclic-covers.

We close the paper with a discussion of our characterization, presented in Section~\ref{sec:discussion}: we highlight some limitations and interesting features, compare its scope with the current research landscape, and sketch directions for future work.

\paragraph{Technical overview} Naturally, our technical approach is inspired by the one of Gimbert and Zielonka for the memoryless case~\cite{GZ05}, which can actually be rediscovered through our results using a trivial memory skeleton. Two of the most important challenges we had to overcome were:
\begin{enumerate}
\item establishing natural concepts of \textit{monotony and selectivity modulo memory} that are exactly as powerful as required to maintain a complete characterization (i.e., sufficient \textit{and} necessary conditions) in the finite-memory case;
\item circumventing the seemingly unavoidable \textit{coupling between the memory skeleton and the arena} in the inductive argument needed to prove the implication from  $\memskel$-monotony and $\memskel$-selectivity to finite-memory optimal strategies~--- which we were able to do using our notions of prefix-covers and cyclic-covers.
\end{enumerate}
All along our paper, we highlight the similarities and discrepancies between our work and Gimbert and Zielonka's~\cite{GZ05}. Whenever possible, we also go further, using weaker hypotheses and proving stronger results, along with addressing core problems left untouched in~\cite{GZ05}~--- while they do have an important impact on the approach (e.g., the role of the zero-sum hypothesis). In that respect, we hope to shed a new light on the seminal results of~\cite{GZ05} while generalizing them.

\paragraph{Critical analysis} Before jumping to the technical part of this work, let us take a step back and assess the place of our work in its larger line of research. The natural endgame is characterizing all preference relations admitting finite-memory optimal strategies, including those using \textit{arena-dependent} memory, and pinpointing the frontiers of application of the lifting corollary~--- that is, under which conditions is finite-memory determinacy preserved when going from one-player to two-player games?

The road is long from Gimbert and Zielonka's characterization in the memoryless case~\cite{GZ05} to such a general result, and this work is but a first step. We have already established that Gimbert and Zielonka's approach cannot be fully transposed for finite memory. Our focus on \textit{arena-independent} memory is a way to study the frontiers of this approach while providing an extension of practical interest. While it may seem limited at first, note that our framework already encompasses arguably rich classes of games such as, e.g., generalized parity games and fully-bounded energy games.

Let us stress that our result~--- relating a memory skeleton $\memskel$ and preference relations for which this skeleton suffices~--- cannot be obtained by simply considering product arenas and invoking Gimbert and Zielonka's result on memoryless determinacy~\cite{GZ05}. While, of course, memoryless strategies on product arenas correspond to memoryfull strategies on original arenas (as we will formally establish in Lemma~\ref{lem:UFMiffMLonProduct}), invoking~\cite{GZ05} requires to be able to quantify on \textit{all} arenas, not only \textit{product} arenas. Filling this gap is exactly the goal of this paper, and it is made possible through the new concepts we sketched above.

From a practical point of view, our equivalence result has limitations as it inherently uses the memory skeleton $\memskel$. At this point, our approach neither helps in finding an appropriate skeleton, nor in determining the minimal one; two highly interesting questions from a practical standpoint. Nonetheless, to advance toward answering these questions and to be able to find good skeletons automatically, one first has to understand their theoretical characteristics, which we do here as a necessary stepping stone. Focusing on applications, let us note that the equivalence result is often not the most suited tool: this is instead where the lifting corollary shines. As noted before, reasoning on one-player games (i.e., graphs) is generally much easier than in two-player games (e.g., \cite{DBLP:journals/acta/BouyerMRLL18,DBLP:conf/fossacs/BouyerHMR017,DBLP:conf/concur/BruyereHRR19}). Hence, a reasonably easy way to tackle practical cases is to find skeletons sufficient for $\Pone$ and $\Ptwo$ in their respective one-player games and to use our constructive result to build a skeleton that suffices for both in two-player games: interestingly, the product of the two one-player-game skeletons is sufficient for both players in all two-player games. Hence the memory blowup is mild.

Finally, we believe it should be possible to generalize our approach to some extent to the \textit{arena-dependent} case, through some \textit{function} associating memory skeletons to arenas (e.g., skeletons encoding bounded counters, with bounds growing with the size of the arena, as for multiple lower-bounded energy objectives). Again, the previous example proves that this would not hold in full generality, but our hope is to establish conditions on this function (which is induced by the preference relation) under which the approach would hold. We leave this question open for now: this paper paves the way to this more general setting.

\paragraph{Related work} We already discussed the most important related papers, notably~\cite{GZ05}. Let us highlight here some works where similar approaches have been considered to establish ``meta-theorems'' applying to general classes of games. First and foremost is the determinacy theorem by Martin that guarantees determinacy (without considering the complexity of strategies) for Borel winning conditions~\cite{martin_AM75}.

Following the same motivation as our work~--- the need to characterize (combinations of) objectives admitting finite-memory optimal strategies, Le Roux et al.~\cite{DBLP:conf/fsttcs/0001PR18} take another road: whereas our work permits to lift results from one-player games to two-player games, they provide a lifting from the single-objective case to the multi-objective one.

Our work focuses on \textit{deterministic turn-based} two-player games. Our results were recently extended to \textit{stochastic} games
~\cite{DBLP:conf/concur/BouyerORV21} (both the characterization in terms of generalizations of the monotony and the selectivity concepts, and the lifting corollary).
Sufficient conditions for memoryless determinacy were also previously provided for stochastic models (e.g.,~\cite{DBLP:conf/stacs/Gimbert07,gimbert2014twoplayer}).
Some sufficient criteria, orthogonal to our approach, were studied for \textit{concurrent} games in~\cite{DBLP:conf/mfcs/Roux18}. A recent preprint~\cite{DBLP:journals/corr/abs-2110-01276} revisits our work in the context of \textit{infinite} arenas, providing a game-theoretic
characterization of $\omega$-regular objectives.

Finally, we recently discovered unpublished content in Kopczy\'nski's PhD thesis~\cite{KopThesis}. Kopczy\'nski distinguishes \textit{chromatic} memory (which corresponds to our definition of memory skeleton), and the more powerful \textit{chaotic} memory, where transitions of the memory can depend on the actual edges of the arenas, rather than simply on the colors of the edges. Chaotic memory is thus intrinsically \textit{arena-dependent}. Our notion of an arena being both prefix- and cyclic-covered by a memory skeleton $\memskel$ is equivalent to a notion in~\cite[Definition~8.12]{KopThesis}, which defines that an arena \textit{adheres to chromatic memory $\memskel$} if it is possible to assign a state of $\memskel$ to every state of the arena such that moving along the edges of the arena updates these memory states in a consistent way. Our definitions of \textit{prefix-} and \textit{cyclic-cover} can be seen as two distinct sides of this idea of \textit{adherence}, which when added up, are actually equivalent to it.

\paragraph{Comparison with conference version} Our paper presents in full details the contributions published in a preceding conference version~\cite{fmCONCUR}. All sections have been supplemented with extra explanations, remarks and examples. All proofs are now directly provided in the main text. In practice, the previous ``Technical sketch'' section has been replaced by two sections (Sections~\ref{sec:FMToConcepts} and~\ref{sec:conceptsToFM}), each detailing the proof of an implication of our main equivalence (Theorem~\ref{thm:equivalence}). These sections contain extra intermediate lemmas which, albeit more technical, have interest on their own.
A new section (Section~\ref{sec:counterexample}) was also added to formally prove statements about the counterexample (Figure~\ref{fig:counterexample}) sketched in this introduction.

\paragraph{Acknowledgments} We extend our warmest thanks to Mathieu Sassolas, for inspiring discussions that were essential in starting this work.

\section{Preliminaries}
\label{sec:prelims}

\paragraph{Automata and languages of colors}
Let $\colors$ be an arbitrary set of \textit{colors}.

We recall classical notions on automata on finite words. A \emph{non-deterministic finite-state automaton (NFA)} is a tuple $\atmtnfull$, where $\atmtnStates$ is a finite
set of \textit{states}, $\colorsSubset \subseteq \colors$ is a finite \textit{alphabet} of colors, $\atmtnTrans\subseteq \atmtnStates \times \colorsSubset \times
\atmtnStates$ is a set of \textit{transitions}, $\atmtnInitStates \subseteq \atmtnStates$ is a set of \textit{initial} states,
and $\atmtnFinalStates\subseteq\atmtnStates$ is a set of \textit{final} states.
Given a state $\atmtnState \in \atmtnStates$ and a word $\word \in \colorsSubset^*$, we denote by $\atmtnTransHat(\atmtnState, \word)$ the set of states that can be reached from $q$ after reading $w$. Without loss of generality, we assume all NFA to be \textit{coaccessible}, i.e., for all $\atmtnState \in \atmtnStates$,
there exists $\word \in \colorsSubset^\ast$, such that $\atmtnTransHat(\atmtnState, \word) \cap \atmtnFinalStates \neq \emptyset$. Recall that NFA precisely recognize \textit{regular languages}.

For any finite subset $\colorsSubset \subseteq \colors$, we denote by $\Reg{\colorsSubset}$ the set of all regular languages over $\colorsSubset$.
 Let
\(\Rec{\colors} = \bigcup_{\colorsSubset\, \subseteq\, \colors,\, \vert \colorsSubset\vert < \infty} \Reg{\colorsSubset},
\)  that is, all the regular languages built over $\colors$.

Let $\lang\subseteq \colors^*$ be a language of \textit{finite} words. We denote by $\Prefixes{\lang}$
the set of all \textit{prefixes} of the words in $\lang$.
	We define the set of \textit{infinite} words
	\(
	\PrefClosure{\lang} = \{\word = c_1 c_2\ldots{}\in\colors^\omega\mid \forall\, n \geq 1,\, c_1 \ldots{} c_n \in \Prefixes{\lang} \},
	\)
	which contains all infinite words for which every finite prefix is a prefix of a word in $\lang$. Intuitively, if $K$ is regular, $\PrefClosure{\lang}$  is the language of infinite words that correspond to infinite paths that can always branch and reach a final state, on an automaton for $\lang$: we will formalize this in Lemma~\ref{lem:arenaLanguage}. Given a finite word $\word \in \colors^*$ and a language $\lang\subseteq \colors^*$, we write $\word\lang$ for their concatenation, i.e., the language $\word\lang = \{\word' = \word \word'' \mid w'' \in \lang\} \subseteq C^*$.

The following observation, already noted in~\cite{GZ05}, will come in handy too.
\begin{lem}
\label{lem:languageSplit}
	Let $\lang_1,\lang_2\subseteq C^*$. Then $\PrefClosure{\lang_1\cup \lang_2} = \PrefClosure{\lang_1}\cup
	\PrefClosure{\lang_2}$.
\end{lem}
\begin{proof}
	Let $\word\in\PrefClosure{\lang_1\cup \lang_2}$. Every finite prefix of $\word$ is in
	$\Prefixes{\lang_1\cup \lang_2}$. Assume w.l.o.g.~that infinitely many prefixes of $\word$
	are in $\Prefixes{\lang_1}$. This implies that \emph{all} prefixes of $\word$ are in
	$\Prefixes{\lang_1}$ (intuitively, because there is a continuity in the prefix relation). Hence, $\word\in\PrefClosure{\lang_1}\cup
	\PrefClosure{\lang_2}$.

	Now, let $\word\in\PrefClosure{\lang_1}\cup\PrefClosure{\lang_2}$. If $\word\in\PrefClosure{\lang_1}$
	(resp.~$\PrefClosure{\lang_2}$), every finite prefix of $\word$ is in $\Prefixes{\lang_1}$
	(resp.~$\Prefixes{\lang_2}$), so in particular it is in $\Prefixes{\lang_1\cup \lang_2}$.
	Hence, $\word\in\PrefClosure{\lang_1\cup \lang_2}$.
\end{proof}

\paragraph{Arenas} We consider two players: player 1 ($\Pone$) and player 2 ($\Ptwo$). An \emph{arena} is a tuple $\arenafull$ such that $\states =
\states_1\uplus\states_2$
(disjoint union) is a finite set of \emph{states} partitioned into states of $\Pone$ ($\states_1$) and $\Ptwo$ ($\states_2$), and $\edges\subseteq \states\times
\colors\times \states$
is a finite set of \emph{edges}.
Let $\col\colon \edges\to\colors$ be the projection of edges to \textit{colors} and $\colhat$ its natural extension to \textit{sequences} of edges. For an edge $\edge \in \edges$, we use $\edgeIn(\edge)$ and $\edgeOut(\edge)$ to denote its starting state and arrival state respectively, i.e., $\edge = (\edgeIn(\edge), \col(\edge), \edgeOut(\edge))$. We assume all arenas to be non-blocking, i.e., for all $\state \in \states$, there exists $\edge \in \edges$ such that $\edgeIn(\edge) = \state$. For $i\in\{1, 2\}$, we call an arena $\arenafull$ a \emph{$\player{i}$'s one-player arena} if for all $\state\in\states_{3-i}$, $\lvert\{\edge\in\edges\mid \edgeIn(\edge) = \state\}\rvert = 1$~--- that is, $\player{3-i}$ has no choice.

Let $\Paths(\arena, \state)$ denote the set of \textit{histories} in $\arena$ from initial state $\state \in \states$, i.e., finite sequences of edges $\pth = \edge_1 \ldots{} \edge_n \in \edges^+$ such that $\edgeIn(\edge_1) = \state$ and for all $i$, $1 \leq i < n$, $\edgeOut(\edge_i) = \edgeIn(e_{i+1})$.
Let $\Plays(\arena, \state)$ denote the set of \textit{plays} in $\arena$ from initial state $\state \in \states$, i.e., infinite sequences of edges $\play = \edge_1 \edge_2 \ldots{} \in \edges^\omega$ such that $\edgeIn(\edge_1) = \state$ and for all $i \geq 1$, $\edgeOut(\edge_i) = \edgeIn(e_{i+1})$.
We write $\Paths(\arena, S')$ and $\Plays(\arena, S')$ for the unions over subsets of initial states $S' \subseteq \states$, and write $\Paths(\arena)$ and $\Plays(\arena)$ for the unions over all states of $\arena$.

Let $\pth = \edge_1 \ldots \edge_n \in \Paths(\arena)$ (resp.~$\play = \edge_1\edge_2 \ldots \in \Plays(\arena)$): we extend the operator $\edgeIn$ to histories (resp.~plays) by identifying $\edgeIn(\pth)$ (resp.~$\edgeIn(\play)$) to $\edgeIn(\edge_1)$. We proceed similarly for $\edgeOut$ and histories: $\edgeOut(\pth) = \edgeOut(\edge_n)$. For the sake of convenience, we consider that any set $\Paths(\arena, \state)$ contains the \textit{empty history} $\emptyPth_\state$ such that $\edgeIn(\emptyPth_\state) = \edgeOut(\emptyPth_\state) = \state$. We write $\Paths_i(\arena, \state)$ and $\Paths_i(\arena)$ for the subsets of histories $\pth$ such that $\edgeOut(\pth) \in \states_i$, $i \in \{1, 2\}$, i.e., histories whose last state belongs to $\player{i}$.

For any set of histories $H \subseteq \Paths(\arena)$, we write $\colhat(H)$ for its projection to colors, i.e., $\colhat(H) = \{\colhat(\pth) \mid \pth \in H\}$. We do the same for sets of plays.

\paragraph{Memory skeletons}
	A \emph{memory skeleton} is a tuple $\memskelfull$ where $\memstates$ is a
	\textit{finite} set of \textit{states}, $\meminit \in \memstates$ is a fixed \textit{initial} state and $\memupd\colon \memstates\times\colors\to\memstates$
	is an \textit{update function}. We write $\memupdhat$ for the natural extension of $\memupd$ to sequences of colors in $\colors^\ast$.
Note that memory skeletons are deterministic and might have an \textit{infinite} number of transitions, in contrast to NFA\@. We define the \textit{trivial memory skeleton} with only one state as $\memskel_{\mathsf{triv}} = (\memstates = \{\meminit\}, \meminit, \memupd\colon \{\meminit\} \times \colors \to \{\meminit\})$: it permits to formalize memoryless strategies~\cite{GZ05} in our framework.

	Let $\memskelfull$ be a memory skeleton. For $\memstate, \memstate' \in \memstates$, we define the language
	\(
	\famelt_{\memstate, \memstate'} = \left\lbrace \word \in \colors^\ast \mid \memupdhat(\memstate, \word) = \memstate'\right\rbrace
	\)
	that contains all words that can be read from $\memstate$ to $\memstate'$ in $\memskel$.

Let $\memskel^1 = (\memstates^1, \meminit^1, \memupd^1)$ and $\memskel^2 = (\memstates^2, \meminit^2, \memupd^2)$ be two memory skeletons. We define their \textit{product} $\memskel^1 \memProduct \memskel^2$ as the memory skeleton $\memskelfull$ obtained as follows: $\memstates = \memstates^1 \times \memstates^2$, $\meminit = (\meminit^1, \meminit^2)$, and, for all $\memstate^1 \in \memstates^1$, $\memstate^2 \in \memstates^2$, $c \in \colors$, $\memupd((\memstate^1, \memstate^2), c) = (\memupd^1(\memstate^1, c), \memupd^2(\memstate^2, c))$. That is, the memories are updated in parallel when a color is read.

\paragraph{Product arenas} Let $\arenafull$ be an arena and $\memskelfull$ be a memory skeleton. We
	define their \textit{product} $\prodAS{\arena}{\memskel}$ as the arena $(\states_1', \states_2', \edges')$ where $\states_1' = \states_1\times\memstates$, $\states_2' = \states_2\times\memstates$, and $\edges' \subseteq \states' \times \colors \times \states'$, with $\states' = \states_1' \uplus \states_2'$, is such that
	$((\state_1,\memstate_1), \clr,
	(\state_2,\memstate_2))\in\edges'$ if and only if
	$(\state_1,\clr, \state_2)\in\edges$ and
	$\memupd(\memstate_1,\clr) = \memstate_2$. That is, the memory is updated according to the colors of the edges in $\edges$. Note that even though $\memskel$ might contain an infinite number of transitions since $\colors$ might be infinite, $\prodAS{\arena}{\memskel}$ is always finite, as $\edges$ is finite in $\arena$. Since we assume arena $\arena$ is non-blocking, it is also the case of arena $\prodAS{\arena}{\memskel}$.

\paragraph{Arena induced by an NFA} Let $\atmtnfull$ be an NFA\@. We say that a state $\atmtnState \in \atmtnStates$ is \textit{essential} if there exists an infinite path in $\atmtn$ starting in $\atmtnState$. Let $\atmtnStates_{\mathsf{ess}} = \{\atmtnState \in \atmtnStates \mid \atmtnState \text{ is essential}\}$. We define the corresponding \textit{one-player} arena $\atmtnArena{\atmtn} = (\states_1 = \atmtnStates_\mathsf{ess}, \states_2 = \emptyset, \edges \subseteq \atmtnStates_\mathsf{ess} \times \colorsSubset \times \atmtnStates_\mathsf{ess})$, where $\edge = (\atmtnState, \clr, \atmtnState') \in \edges$ if $(\atmtnState, \clr, \atmtnState') \in \atmtnTrans$. Intuitively, $\atmtnArena{\atmtn}$ transforms $\atmtn$ into a \textit{non-blocking} arena thanks to the restriction to essential states.

We may now state formally the link between $\PrefClosure{\lang}$ and the underlying automaton for $\lang$. Our result (and its proof) is similar to~\cite[Lemma 4]{GZ05}.

\begin{lem}
\label{lem:arenaLanguage}
Let $\atmtnfull$ be a (coaccessible) NFA recognizing the regular language $\lang \subseteq \colors^\ast$. Let $\atmtnInitStates' = \atmtnInitStates\cap\atmtnStates_\mathsf{ess}$. The following equality holds:
\[
\PrefClosure{\lang} = \colhat\big( \Plays\left( \atmtnArena{\atmtn}, \atmtnInitStates'\right) \big).
\]
In particular, $\PrefClosure{\lang}$ is non-empty if and only if there exists an essential initial state in $\atmtn$.
\end{lem}

Intuitively, $\PrefClosure{\lang}$ is the language of infinite words that correspond to infinite paths that can always branch and reach a final state, on the automaton $\atmtn$ recognizing $\lang$.

\begin{proof}
If $\atmtnInitStates'$ is empty, the equality trivially holds: $\colhat\big(\Plays\left( \atmtnArena{\atmtn}, \atmtnInitStates'\right)\big)$ and $\PrefClosure{\lang}$ are both empty. Hence, from now on, we assume $\atmtnInitStates' \neq \emptyset$.

We start with the left-to-right inclusion. Let $\word = \clr_1\clr_2\ldots\in\PrefClosure{\lang}$. We first prove that for all $n\ge 1$, it holds that
\[
\clr_1\ldots\clr_n \in \colhat\big(\Paths\left(\atmtnArena{\atmtn}, \atmtnInitStates'\right)\big).
\]
We assume on the contrary that there exists $n\ge 1$ such that
\[
\clr_1\ldots\clr_n \notin \colhat\big(\Paths\left(\atmtnArena{\atmtn}, \atmtnInitStates'\right)\big).
\]
As $\atmtnArena{\atmtn}$ is a restriction of the states of $\atmtn$ to $\atmtnStates_\mathsf{ess}$, this means that no matter how $\clr_1\ldots\clr_n$ is read on $\atmtn$, it goes through a state in $\atmtnStates\setminus\atmtnStates_\mathsf{ess}$. As there is no infinite path from these states, this contradicts that $\word\in\PrefClosure{\lang}$; there cannot be arbitrarily long prefixes starting with $\clr_1\ldots\clr_n$.

We now use the property that we have just proved along with K\"onig's lemma to show that $\word\in\colhat\big( \Plays\left( \atmtnArena{\atmtn}, \atmtnInitStates'\right) \big)$.
We build a forest of trees $\forest$. The vertices of $\forest$ are paths $\pth \in \Paths\left(\atmtnArena{\atmtn}, \atmtnInitStates'\right)$ such that $\colhat(\pth)$ is a prefix of $\word$ and $\edgeIn(\pth)\in\atmtnInitStates'$. For every $\atmtnState\in\atmtnInitStates'$, there is one tree in $\forest$ whose root is the empty path $\emptyPth_\atmtnState$. There is a transition from a vertex $\pth$ to a vertex $\pth'$ if there exists $e\in E$ such that $\pth \cdot \edge = \pth'$.
As there is at least one vertex for each prefix $\clr_1 \ldots \clr_n$, (at least) one of the trees of $\forest$ must be infinite. Moreover, $\forest$ is finitely branching. By K\"onig's lemma, we obtain that there must be an infinite path $\play$ starting from a root $\emptyPth_\atmtnState$ for some $\atmtnState\in\atmtnInitStates'$. By construction, $\colhat(\play) = \word$, so $\word\in\colhat\big( \Plays\left( \atmtnArena{\atmtn}, \atmtnInitStates'\right) \big)$.

We now prove the right-to-left inclusion. Let $\play = \edge_1\edge_2\ldots \in \Plays\left(\atmtnArena{\atmtn}, \atmtnInitStates'\right)$. For $n\ge 1$, the word $\colhat(\edge_1\ldots\edge_n)$ is the color of a path in $\atmtn$, since every edge of $\atmtnArena{\atmtn}$ corresponds to a transition of $\atmtn$. As $\atmtn$ is coaccessible, there is a path in $\atmtn$ from the state corresponding to $\edgeOut(e_n)$ to a final state in $\atmtnFinalStates$. Thus, the word $\colhat(\edge_1\ldots\edge_n)$ is a prefix of an accepted word of $\atmtn$, i.e., a prefix of a word in $\lang$; as this holds for all $n\ge 1$, we obtain that $\colhat(\play)\in \PrefClosure{\lang}$.
\end{proof}

\paragraph{Strategies} A \textit{strategy} $\strat_{i}$ for $\player{i}$, $i\in\{1, 2\}$, on arena $\arenafull$, is a function $\strat_{i}\colon \Paths_{i}(\arena) \to \edges$ such that for all $\pth \in \Paths_i(\arena)$, $\edgeIn(\strat_i(\pth)) = \edgeOut(\pth)$. Let $\strats_{i}(\arena)$ be the set of all strategies of $\player{i}$ on $\arena$.

A \textit{finite-memory strategy} $\strat_{i}$ is a strategy that can be encoded as a \textit{Mealy machine}, i.e., a memory skeleton $\memskelfull$ with transitions over a \textit{finite subset of colors $\colorsSubset \subseteq \colors$}, enriched with a \textit{next-action function} $\memnxt\colon \memstates \times \states_{i} \to \edges$ such that for all $\memstate \in \memstates$, $\state \in \states_i$, $\edgeIn(\memnxt(\memstate, \state)) = \state$. Given a Mealy machine $\mealy_{\strat_i} = (\memskel, \memnxt)$, strategy $\strat_i$ is defined as follows:
\begin{itemize}
\item $\forall\, \state \in \states_i,\, \strat_i(\emptyPth_s) = \memnxt(\meminit, \state)$,
\item $\forall\, \pth\cdot\edge \in \Paths_i(\arena)$, $\edge \in \edges$, $\strat_i(\pth\cdot\edge) = \memnxt\left(\memupdhat\left(\meminit, \colhat\left(\pth\cdot\edge\right)\right), \edgeOut(\edge)\right)$.
\end{itemize}
We denote by $\stratsFM_{i}(\arena)$ the set of all finite-memory strategies of $\player{i}$ on $\arena$. We say that a strategy $\strat_{i} \in \stratsFM_{i}(\arena)$ is \textit{based on memory skeleton} $\memskel$ if it can be encoded as a Mealy machine $\mealy_{\strat_i} = (\memskel, \memnxt)$, as above. We always implicitly assume that strategies of $\stratsFM_{i}(\arena)$ are built by restricting the transitions of their skeleton $\memskel$ to the actual subset of colors appearing in $\arena$. A strategy $\strat_i$ is \textit{memoryless} if it is a function $\strat_i\colon \states_i \to \edges$, or equivalently, if it is based on the trivial memory skeleton $\memskel_{\mathsf{triv}}$. We denote by $\stratsML_{i}(\arena)$ the set of all memoryless strategies of $\player{i}$ on $\arena$.

We denote by $\Plays(\arena, \state, \strat_i)$ the set of plays \textit{consistent} with a strategy $\strat_i$ of $\player{i}$ from an initial state $\state$, i.e., all plays $\play = \edge_1 \edge_2 \ldots \in \Plays(\arena, \state)$ such that for all prefixes $\pth = \edge_1 \ldots \edge_n$, $\edgeOut(\pth) \in \states_i \implies \strat_i(\pth) = \edge_{n+1}$. We write $\Plays(\arena, \state, \strat_1, \strat_2)$ for the singleton set containing the unique play consistent with a couple of strategies for the two players. We use similar notations for histories.

\paragraph{Preference relations}
Let $\pref$ be
a total preorder on $\colors^\omega$, called \emph{preference relation}. We consider \textit{antagonistic} games, where the objective of $\Pone$ is to create the best possible play with regard to $\pref$ whereas the objective of $\Ptwo$ is to obtain the worst possible one. That is, $\Ptwo$ uses the inverse relation $\invpref$. This corresponds to \textit{zero-sum} games when using a quantitative framework.

Given $\word, \word' \in \colors^\omega$, we write $\word \strictpref \word'$ if we have $\neg(\word' \pref \word)$ since the preorder is total. We extend the relation $\pref$ to subsets of $\colors^\omega$ as follows: for $\Words, \Words' \subseteq \colors^\omega$,
\[
\Words \pref \Words' \iff \forall\, \word \in \Words,\, \exists\, \word' \in \Words',\, \word \pref \word'.
\]
We also write
\[
\Words \strictpref \Words' \iff \exists\, \word' \in \Words',\, \forall\, \word \in \Words,\,\word \strictpref \word'.
\]
Note that $\Words \strictpref \Words'$ if and only if $\neg(\Words' \pref \Words)$, and that transitivity is preserved when considering sets.

We sometimes compare words $\word \in \colors^\omega$ with languages $\lang \subseteq \colors^\omega$, by simply identifying word $\word$ to its singleton language $\{\word\}$.

\paragraph{Games}
A (deterministic turn-based two-player) \emph{game} is a tuple $\gamefull$ where $\arena$ is an arena and $\pref$ is
a preference relation. As discussed in Section~\ref{sec:intro}, all the classical objectives from the literature (both qualitative and quantitative) can be expressed in the general framework of preference relations. For $i\in\{1,2\}$, a \emph{$\player{i}$'s one-player game} is a game $\gamefull$ such that $\arena$ is a $\player{i}$'s one-player arena.

\begin{exa}
\label{ex:preferenceRelations}
There are two prominent ways to formalize game objectives in the literature: through \textit{payoff functions} and through \textit{winning conditions}. We take an example of each.

First, consider (lim inf) \textit{mean-payoff} games~\cite{EM79}. In this setting, colors are integers, i.e., $\colors = \mathbb{Z}$, and the goal of $\Pone$ is to create a play $\play$, with $\word = \colhat(\play) = \clr_1\clr_2\ldots{}$, maximizing the following \textit{payoff function}:
\begin{equation*}
\mathsf{MP}(\word) = \liminf_{n \rightarrow \infty}\dfrac{1}{n}\sum_{i = 1}^n \clr_i.
\end{equation*}
Such a payoff function induces a natural preference relation $\pref_{\mathsf{MP}}$ between sequences of colors as follows: for all $\word, \word' \in \colors^\omega$, $\word \pref_{\mathsf{MP}} \word'$ if and only if $\mathsf{MP}(\word) \leq \mathsf{MP}(\word')$. Such quantitative games are \textit{zero-sum}, hence $\Ptwo$ uses the natural inverse relation $\pref_{\mathsf{MP}}^{-1}$: he is a minimizer player in the payoff formulation of these games.

Second, consider \textit{reachability} games (e.g.,~\cite{DBLP:journals/corr/abs-1010-2420}). In this setting, only two colors are needed: one for edges in the \textit{target set}, and one for the other edges. Let us use $\top$ and $\bot$ respectively to color these two sets of edges, i.e., $\colors = \{\top, \bot\}$. Then, the \textit{winning condition} can be simply written as $\wc = \{\word \in \colors^\omega \mid \top \in \word\} \subsetneq \colors^\omega$, i.e., $\wc$ is the set of plays seeing $\top$ at least once. In such games, the goal of $\Pone$ is to create a play $\play$ such that $\colhat(\play) \in \wc$, called \textit{winning} play. Defining a corresponding preference relation $\pref_{\mathsf{reach}}$ is straightforward: for all $\word, \word' \in \colors^\omega$, $\word \strictpref_{\mathsf{reach}} \word'$ if and only if $\word' \in \wc$ and $\word \not\in \wc$. That is, $\pref_{\mathsf{reach}}$ defines two equivalence classes: losing and winning plays. This qualitative setting is \textit{antagonistic}, hence $\Ptwo$ uses the inverse relation $\pref_{\mathsf{reach}}^{-1}$: his winning condition is $\comp{\wc} = \colors^\omega \setminus \wc$ in the classical formulation of these games.

As explained in Section~\ref{sec:intro}, quantitative games are often reduced to qualitative ones by fixing a \textit{threshold} to achieve.\hfill$\lhd$
\end{exa}

\paragraph{Optimal strategies} Let $\gamefull$ be a game on arena $\arenafull$. Given a $\player{i}$-strategy $\strat_i \in \strats_i(\arena)$ and a state $\state \in \states$, we define
\begin{align*}
\UCol_{\pref}(\arena, \state, \strat_i) &= \{\word \in \colors^\omega \mid \exists\, \strat_{3-i} \in \strats_{3-i}(\arena),\, \colhat(\Plays(\arena, \state, \strat_1, \strat_2)) \pref \word\},\\
\DCol_{\pref}(\arena, \state, \strat_i) &= \{\word \in \colors^\omega \mid \exists\, \strat_{3-i} \in \strats_{3-i}(\arena),\, \word \pref \colhat(\Plays(\arena, \state, \strat_1, \strat_2))\}.
\end{align*}
Note that $\DCol_{\pref}(\arena, \state, \strat_i) = \UCol_{\invpref}(\arena, \state, \strat_i)$. Intuitively, $\UCol_{\pref}$ and $\DCol_{\pref}$ represent the \textit{upward} and \textit{downward closures} of sequences of colors (consistent with a strategy) with respect to the preference relation.

Taking the standpoint of $\Pone$, we say that a strategy $\strat_1 \in \strats_1(\arena)$ is \textit{at least as good as} a strategy $\strat'_1 \in \strats_1(\arena)$ from a state $\state \in \states$ if
\[
\UCol_{\pref}(\arena, \state, \strat_1) \subseteq \UCol_{\pref}(\arena, \state, \strat'_1).
\]
Intuitively, $\strat_1$ is at least as good as $\strat'_1$ if the ``worst-case'' plays consistent with $\strat_1$ are at least as good as the ones consistent with $\strat'_1$. The $\UCol$ operator is useful to define this notion properly even in the case where there is no ``worst-case'' play for a strategy (i.e., if the infimum used in the classical quantitative setting is not reached). Similar notions have been used before, e.g., in~\cite{DBLP:journals/corr/abs-1302-3973}.

Symmetrically, for $\Ptwo$, we say that a strategy $\strat_2 \in \strats_2(\arena)$ is at least as good as a strategy $\strat'_2 \in \strats_2(\arena)$ from a state $\state \in \states$ if
\[
\DCol_{\pref}(\arena, \state, \strat_2) \subseteq \DCol_{\pref}(\arena, \state, \strat'_2).
\]

Now, we say that a strategy $\strat_i \in \strats_i(\arena)$ of $\player{i}$ is \textit{optimal} from a state $\state \in \states$, aka $\state$-optimal, if it is at least as good as every other strategy $\strat'_i \in \strats_i(\arena)$ from $\state$. We extend this notation to subsets of states in the natural way, and we say that a strategy $\strat_i$ is \textit{uniformly-optimal} if it is $\states$-optimal.

The goal of our paper is to characterize the preference relations that admit \textit{uniformly-optimal finite-memory (UFM) strategies based on a given skeleton $\memskel$} in all arenas. We also discuss the simpler case of \textit{uniformly-optimal memoryless (UML) strategies}, which corresponds to the subset of preference relations studied by Gimbert and Zielonka~\cite{GZ05}, using the trivial skeleton $\memskel_{\mathsf{triv}}$.

In that respect, the following link is important to observe.

\begin{lem}
\label{lem:UFMiffMLonProduct}
Let $\gamefull$ be a game on arena $\arenafull$. Let $\memskelfull$ be a memory skeleton and let $\strat_i \in \stratsFM_i(\arena)$ be a finite-memory strategy encoded by the Mealy machine $\mealy_{\strat_i} = (\memskel, \memnxt)$.
Then, $\strat_i$ is a UFM strategy in $\game$ if and only if $\memnxt$ corresponds to an $(\states \times \{\meminit\})$-optimal memoryless strategy in $\game' = (\prodAS{\arena}{\memskel}, \pref)$.
\end{lem}

\begin{proof}
We first aim to define a bijection $\pathFun\colon \Paths(\arena)\to \Paths(\prodAS{\arena}{\memskel}, \states\times\{\meminit\})$. Let $\pth = \edge_1\ldots \edge_n\in\Paths(\arena)$, with $\edge_j = (\state_j, \clr_j, \state_{j+1})$. We set $\memstate_1 = \meminit$, and for $2\le j\le n$, $\memstate_j = \memupd(\memstate_{j-1}, \clr_j)$. We define $\edge_j' = ((\state_j, \memstate_j), \clr_j, (\state_{j+1}, \memstate_{j+1}))$, and $\pathFun(\pth) = \edge_1'\ldots \edge_n'$. Notice that $\colhat(\pathFun(\pth)) = \colhat(\pth)$. Furthermore, $\pathFun$ is bijective; as the initial state of the memory $\meminit$ is fixed and the memory skeleton is deterministic, the memory states added to $\pth$ to obtain $\pathFun(\pth)$ are uniquely determined.

We now show that there is a correspondence between strategies of $\strats_i(\arena)$ and strategies of $\strats_i(\prodAS{\arena}{\memskel})$: intuitively, augmenting the arena with the skeleton allows some strategies to be played using less memory, but does not fundamentally change each player's possibilities.
We define a function $\stratFun\colon \strats_i(\arena) \to \strats_i(\prodAS{\arena}{\memskel})$. For $\stratBis_i\in \strats_i(\arena)$ and $\pth'\in\Paths_i(\prodAS{\arena}{\memskel})$ with $\edgeIn(\pth')\in\states\times\{\meminit\}$ and $\edgeOut(\pth') = (\state, \memstate) \in \states_i \times \memstates$, if $\stratBis_i(\inverse{\pathFun}(\pth')) =  (\state, \clr, \state')$, we define $\stratFun(\stratBis_i)(\pth') = ((\state, \memstate), \clr, (\state', \memupd(\memstate, \clr)))$.
The histories induced by strategies $\stratBis_i$ and $\stratFun(\stratBis_i)$ correspond: if $\pth = \inverse{\pathFun}(\pth')$, then we have
\begin{equation} \label{eq:correspondenceStrategies}
\pathFun(\pth\cdot\stratBis_i(\pth)) = \pth'\cdot(\stratFun(\stratBis_i)(\pth')).
\end{equation}
We are only interested in the behavior of strategies of $\strats_i(\prodAS{\arena}{\memskel})$ on histories $\pth'$ with $\edgeIn(\pth')\in \states\times\{\meminit\}$ (in what follows, we will only consider histories and plays starting in such states). If we restrict the image of $\stratFun$ to the set of strategies $\stratBis_i'\colon \Paths_i(\prodAS{\arena}{\memskel}, \states\times\{\meminit\})\to \edges'$, then $\stratFun$ is a bijection.

Consider the next-action function $\memnxt$ of strategy $\strat_i$. Formally, we have to transform $\memnxt$ into a proper memoryless strategy in $\strats^\mathsf{ML}_i(\prodAS{\arena}{\memskel})$. This can be done through the bijection $\stratFun$, yielding the memoryless strategy $\memnxt' = \stratFun(\strat_i)$, which corresponds to $\memnxt$ interpreted over the product arena, and well-defined for all histories starting in $\states\times\{\meminit\}$.

We now show a second fact related to $\stratFun$: we have that for all $\state\in \states$, for all $\stratBis_1\in\strats_1(\arena)$, $\stratBis_2 \in \strats_2(\arena)$,
\begin{equation} \label{eq:colPlaysThroughFun}
	\colhat\big(\Plays(\arena, \state, \stratBis_1, \stratBis_2)\big) = \colhat\big(\Plays(\prodAS{\arena}{\memskel}, (\state, \meminit), \stratFun(\stratBis_1), \stratFun(\stratBis_2))\big).
\end{equation}
This can easily be proved by induction using Equation~\eqref{eq:correspondenceStrategies}.
Indeed, at each step, both strategy $\stratBis_1$ (resp.~$\stratBis_2$) and strategy $\stratFun(\stratBis_1)$ (resp.~$\stratFun(\stratBis_2)$) pick an edge with the same color.

We finish the proof assuming that $\strat_i = \strat_1\in \stratsFM_1(\arena)$; the proof is symmetric for $\Ptwo$. Equation~\eqref{eq:colPlaysThroughFun} implies that for all $\state\in\states$, $\stratBis_1\in\strats_1(\arena)$,
\begin{align} \label{eq:uColThroughFun}
	&\UCol_{\pref}(\arena, \state, \stratBis_1)\notag \\
	&= \{\word \in \colors^\omega \mid \exists\, \stratBis_{2} \in \strats_{2}(\arena),\, \colhat\big(\Plays(\arena, \state, \stratBis_1, \stratBis_2)\big) \pref \word\}\notag \\
	&= \{\word \in \colors^\omega \mid \exists\, \stratBis_{2} \in \strats_{2}(\arena),\, \colhat\big(\Plays(\prodAS{\arena}{\memskel}, (\state, \meminit), \stratFun(\stratBis_1), \stratFun(\stratBis_2))\big) \pref \word\}\notag \\
	&= \{\word \in \colors^\omega \mid \exists\, \stratBis_{2}' \in \strats_{2}(\prodAS{\arena}{\memskel}),\, \colhat\big(\Plays(\prodAS{\arena}{\memskel}, (\state, \meminit), \stratFun(\stratBis_1), \stratBis_2')\big) \pref \word\}\notag \\
	&= \UCol_{\pref}(\prodAS{\arena}{\memskel}, (\state, \meminit), \stratFun(\stratBis_1)).
\end{align}
where the penultimate equality uses that the aforementioned restriction of $\stratFun$ is bijective.

Using Equation~\eqref{eq:uColThroughFun}, we can obtain that a strategy $\stratBis_1$ is uniformly-optimal in $\game = (\arena, \pref)$ if and only if $\stratFun(\stratBis_1)\in\strats_1(\prodAS{\arena}{\memskel})$ is $(\states\times\{\meminit\})$-optimal in $\game' = (\prodAS{\arena}{\memskel}, \pref)$. In particular, $\strat_1$ is uniformly-optimal in $\game$ if and only if $\stratFun(\strat_1) = \memnxt'$ is $(\states\times\{\meminit\})$-optimal in $\game'$.
\end{proof}

\paragraph{Nash equilibria} We use Nash equilibria~\cite{Osborne1994} as tools to establish the existence of optimal strategies in some of our proofs.  Let $\gamefull$ be a game on arena $\arenafull$. Formally, a \textit{Nash equilibrium} (NE) from a state $\state \in \states$ is a couple of strategies $(\strat_1, \strat_2) \in \strats_1(\arena) \times \strats_2(\arena)$ such that, for all $\strat'_1 \in \strats_1(\arena)$, $\strat'_2 \in \strats_2(\arena)$,
\begin{equation}
\label{eq:defNE}
\colhat(\Plays(\arena, \state, \strat'_1, \strat_2)) \pref \colhat(\Plays(\arena, \state, \strat_1, \strat_2)) \pref \colhat(\Plays(\arena, \state, \strat_1, \strat'_2)).
\end{equation}
Similarly to optimal strategies, we call an NE \textit{uniform} if it is an NE from all states $\state \in \states$.

It is worth taking a moment to discuss the link between \textit{optimal strategies} and \textit{Nash equilibria} in our specific context of \textit{antagonistic} games. Both notions seem closely related, and indeed, in~\cite{GZ05}, Gimbert and Zielonka did choose Equation~\eqref{eq:defNE}~--- i.e., the definition of a Nash equilibrium~--- as their definition of \textit{a pair of optimal strategies}. This could lead the reader to believe that both notions coincide. However, they do not in full generality, as we discuss in the following.

Additionally, defining optimality for a \textit{pair} of strategies gives rise to difficulties as one naturally wants to reason about optimal strategies of a player without talking about the (possibly optimal) strategy of its adversary. Our definition of \textit{optimal strategy} has the advantage of giving a clear and precise definition that does not involve the standpoint of the adversary.

As stated before, the ultimate goal of our paper is to characterize preference relations that admit \textit{finite-memory optimal strategies}, but Nash equilibria will serve as tools in our endeavor. Let us establish two interesting properties of Nash equilibria \textit{in antagonistic games}.

First, it is possible to mix different Nash equilibria.

\begin{lem}
\label{lem:mixingNE}
Let $\gamefull$ be a game on arena $\arenafull$, and let $\state \in \states$ be a state. Let $(\strat^a_1, \strat^a_2)$ and $(\strat^b_1, \strat^b_2) \in \strats_1(\arena) \times \strats_2(\arena)$ be two Nash equilibria from $\state$. Then, $(\strat^a_1, \strat^b_2)$ is also a Nash equilibrium from $\state$.
\end{lem}

\begin{proof}
We need to prove that for all $\strat'_1 \in \strats_1(\arena)$, $\strat'_2 \in \strats_2(\arena)$,
\begin{equation}
\label{eq:mixingNEproof}
\colhat(\Plays(\arena, \state, \strat'_1, \strat^b_2)) \pref \colhat(\Plays(\arena, \state, \strat^a_1, \strat^b_2)) \pref \colhat(\Plays(\arena, \state, \strat^a_1, \strat'_2)).
\end{equation}

Since $(\strat^a_1, \strat^a_2)$ is an NE, we know that
\[
\colhat(\Plays(\arena, \state, \strat^b_1, \strat^a_2)) \pref \colhat(\Plays(\arena, \state, \strat^a_1, \strat^a_2)) \pref \colhat(\Plays(\arena, \state, \strat^a_1, \strat^b_2)),
\]
instantiating $\strat'_1$ and $\strat'_2$ to $\strat_1^b$ and $\strat^b_2$ respectively in Equation~\eqref{eq:defNE}.
Similarly, since $(\strat^b_1, \strat^b_2)$ is an NE, we know that
\[
\colhat(\Plays(\arena, \state, \strat^a_1, \strat^b_2)) \pref \colhat(\Plays(\arena, \state, \strat^b_1, \strat^b_2)) \pref \colhat(\Plays(\arena, \state, \strat^b_1, \strat^a_2)),
\]
instantiating $\strat'_1$ and $\strat'_2$ to $\strat_1^a$ and $\strat^a_2$ respectively in Equation~\eqref{eq:defNE}.

Now it is easy to see from the last two lines that all six sequences of colors are equivalent under $\pref$ as the inequalities form a cycle. Hence, since Equation~\eqref{eq:defNE} holds for $(\strat^a_1, \strat^a_2)$ and $(\strat^b_1, \strat^b_2)$, and since $\colhat(\Plays(\arena, \state, \strat^a_1, \strat^b_2))$ is $\pref$-equivalent to both $\colhat(\Plays(\arena, \state, \strat^a_1, \strat^a_2))$ and $\colhat(\Plays(\arena, \state, \strat^b_1, \strat^b_2))$, Equation~\eqref{eq:mixingNEproof} is trivially verified. That concludes our proof.
\end{proof}

\begin{rem}
\label{rmk:antagonistic}
Lemma~\ref{lem:mixingNE} crucially relies on the assumption (transparent in our definition of Nash equilibrium) that we consider \textit{antagonistic} games, that is, $\Ptwo$ uses the inverse preference relation $\invpref$. Actually, our approach almost completely carries over to the general case where $\Pone$ and $\Ptwo$ use two different, unrelated, preference relations: the single breaking point being the use of Lemma~\ref{lem:mixingNE} in Theorem~\ref{thm:conceptsToML} to preserve \textit{memoryless} Nash equilibria in the induction step.\hfill$\lhd$
\end{rem}

We now establish that Nash equilibria induce optimal strategies (again, in our antagonistic context).

\begin{lem}
\label{lem:NEisOpti}
Let $\gamefull$ be a game on arena $\arenafull$, and let $\state \in \states$ be a state. Let $(\strat_1, \strat_2) \in \strats_1(\arena) \times \strats_2(\arena)$ be a Nash equilibrium from $\state$. Then, both $\strat_1$ and $\strat_2$ are $\state$-optimal strategies.
\end{lem}

\begin{proof}
We do the proof for $\Pone$ (it works symmetrically for $\Ptwo$). Let $(\strat_1, \strat_2) \in \strats_1(\arena) \times \strats_2(\arena)$ be a Nash equilibrium from $\state \in \states$. Consider the rightmost inequality of Equation~\eqref{eq:defNE}. From it, we deduce that
\begin{equation}
\label{eq:UColSimplified}
\UCol_{\pref}(\arena, \state, \strat_1) = \{\word \in \colors^\omega \mid \colhat(\Plays(\arena, \state, \strat_1, \strat_2)) \pref \word\}.
\end{equation}
Indeed, $\strat_2$ is the \textit{best response}~\cite{Osborne1994} to $\strat_1$.

We claim that $\strat_1$ is at least as good as every other strategy, hence that $\strat_1$ is optimal.
Let $\strat'_1 \in \strats_1(\arena)$. We need to prove that $\UCol_{\pref}(\arena, \state, \strat_1) \subseteq \UCol_{\pref}(\arena, \state, \strat'_1)$. Let $\word \in \UCol_{\pref}(\arena, \state, \strat_1) $. From Equation~\eqref{eq:UColSimplified} and the leftmost inequality of Equation~\eqref{eq:defNE}, we have $\colhat(\Plays(\arena, \state, \strat'_1, \strat_2)) \pref \word$. Hence, by definition, $\word \in \UCol_{\pref}(\arena, \state, \strat'_1)$, which concludes our proof.
\end{proof}

As noted above, optimal strategies do not always coincide with Nash equilibria. Intuitively, they do coincide in the classical quantitative formulation (using payoff functions) if the \textit{value} of the game exists, that is, if the best payoff that $\Pone$ can guarantee is equal to the worst payoff that $\Ptwo$ can guarantee~\cite{Osborne1994}. In a sense, our concepts of $\UCol_{\pref}$ and $\DCol_{\pref}$ are meant to mimic the classical $\sup\inf$ and $\inf\sup$ formulations in our abstract context where objectives are described as preference relations.

So (quantitative) games do not always have a value, and similarly, in our context, optimal strategies as defined above do not always induce a Nash equilibrium. That being said, they do for arguably all \textit{reasonable} preference relations, as Martin's determinacy result grants the existence of winning strategies~--- in our formalism, Nash equilibria~--- for all Borel winning conditions~\cite{martin_AM75}. That is, for the equivalence to fail, we would need a preference relation capable of inducing non-Borel sets of winning plays (once a threshold is chosen, as explained in Section~\ref{sec:intro}).  Hence, the two notions are virtually equivalent~--- yet, the above clarification is needed to circumvent slight technical issues in the original proofs of Gimbert and Zielonka~\cite{GZ05}.

\begin{rem} \label{rem:UFM1pNE}
In one-player games, the two visions~--- optimal strategies and Nash equilibria~--- coincide.\hfill$\lhd$
\end{rem}

\begin{rem}
\label{rmk:NEonProduct}
Lemma~\ref{lem:UFMiffMLonProduct} can be restated in terms of Nash equilibria, using a similar reasoning.\hfill$\lhd$
\end{rem}

\section{Characterization}
\label{sec:charac}

We are now able to establish our characterization of preference relations admitting finite-memory optimal strategies based on a given memory skeleton $\memskel$. We proceed in three steps. First, in Section~\ref{subsec:concepts}, we present the core concepts of this characterization, i.e., the properties that preference relations must verify to yield UFM strategies. Second, we state our equivalence result in Section~\ref{subsec:results}, alongside a corollary of practical interest that lets one lift results from the one-player case to the two-player one. We defer the formal proofs of both directions of the equivalence to Section~\ref{sec:FMToConcepts} and Section~\ref{sec:conceptsToFM}, and only explain here how to combine them. Finally, we provide an illustrative application of our characterization in Section~\ref{subsec:example}.

\subsection{Concepts}
\label{subsec:concepts}

\subsection*{Generalizing monotony and selectivity} As discussed in Section~\ref{sec:intro}, Gimbert and Zielonka's characterization~\cite{GZ05} relies on notions of \textit{monotony} and \textit{selectivity} of the preference relation.

Intuitively, the main difference between Gimbert and Zielonka's technical approach and ours is the following. In the memoryless setting, all the reasoning can be \textit{abstracted away} from the underlying arena and done at the level of sequences of colors. In the finite-memory one, however, one has to pay attention to how sequences of colors are composed and compared, to maintain consistency with regard to the memory and the underlying game arena. This need to intertwine abstract reasoning on arbitrary sequences of colors with concrete tracking of memory updates is the key obstacle to overcome.

Much of our effort was thus spent on trying to define concepts that would preserve the elegance of monotony and selectivity while allowing us to lift the theory to the finite-memory case. As often the case in these endeavors, the right concepts turned out to be the most natural ones, capturing the intuitive idea that one needs monotony and selectivity \textit{modulo a memory skeleton}.

\begin{defi}[$\memskel$-monotony]
\label{def:monotony}
	Let $\memskelfull$ be a memory skeleton. A preference relation $\pref$ is $\memskel$-\emph{monotone} if for all $\memstate \in \memstates$, for all $\lang_1,
	\lang_2\in \Rec{\colors}$,
	\begin{equation}
	\label{eq:monotony}
	\left( \exists\, \word\in\famelt_{\meminit,\memstate} ,\, \PrefClosure{\word\lang_1}\strictpref\PrefClosure{\word\lang_2}\right)
	\implies \left( \forall\, \word'\in\famelt_{\meminit,\memstate},\, \PrefClosure{\word'\lang_1}\pref\PrefClosure{\word'\lang_2}\right) .
	\end{equation}
\end{defi}

Recall that a memory skeleton $\memskel$ has a fixed initial state $\meminit$. Intuitively, \textit{$\memskel$-monotony} extends Gimbert and Zielonka's monotony by asking one to compare prefixes \textit{belonging to the same language} $\famelt_{\meminit,\memstate}$, that is, prefixes that are deemed equivalent by the memory skeleton. This property roughly captures that $\pref$ is \textit{stable with regard to prefix addition}, for memory-equivalent prefixes.

The original monotony notion is exactly equivalent to our $\memskel$-monotony with $\memskel$ being the trivial skeleton $\memskel_{\mathsf{triv}}$: that is, the memoryless case is naturally a specific subcase of our framework.

\begin{defi}[$\memskel$-selectivity]
\label{def:selectivity}
	Let $\memskelfull$ be a memory skeleton.  A preference relation $\pref$ is $\memskel$-\emph{selective} if for all $\word\in\colors^*$, $\memstate = \memupdhat(\meminit,\word)$,
	for all $\lang_1, \lang_2\in \Rec{\colors}$ such that $\lang_1, \lang_2\subseteq \famelt_{\memstate,\memstate}$, for all $\lang_3\in\Rec{\colors}$,
	\begin{equation}
	\label{eq:selectivity}
	\PrefClosure{\word(\lang_1\cup \lang_2)^*\lang_3} \pref
	\PrefClosure{\word\lang_1^*}\cup\PrefClosure{\word\lang_2^*}\cup\PrefClosure{\word\lang_3}.
	\end{equation}
\end{defi}

Similarly, \textit{$\memskel$-selectivity} extends Gimbert and Zielonka's selectivity by asking one to compare sequences of colors \textit{belonging to the same language} $\famelt_{\memstate,\memstate}$, that is, sequences read as cycles on the memory skeleton. Note also that the memory state $\memstate$ should be consistent with the prefix $\word$ read from the initial memory state $\meminit$. This property roughly captures that $\pref$ is \textit{stable with regard to cycle mixing}, for memory-equivalent cycles.

Again, the original selectivity notion is exactly equivalent to $\memskel_{\mathsf{triv}}$-selectivity.

In a nutshell, $\memskel$-monotony deals with prefixes up to the first cycle (on memory) and $\memskel$-selectivity deals with the cycles thereafter; we will see that memory skeletons can be built in a compositional way based on these two orthogonal yet complementary tasks. We present an example illustrating both concepts and their application in Section~\ref{subsec:example}.

Our notions respect the natural intuition that access to additional memory should always be helpful: if a skeleton $\memskel$ is sufficient to classify sequences of colors in a way that guarantees $\memskel$-monotony and $\memskel$-selectivity, then it should also be the case for ``more powerful'' skeletons.

\begin{lem}
\label{lem:memProdPreserving}
Let $\memskel$ and $\memskel'$ be two memory skeletons. If $\pref$ is $\memskel$-monotone (resp.~$\memskel$-selective) then, it is also $(\memskel \memProduct \memskel')$-monotone (resp.~$(\memskel \memProduct \memskel')$-selective).
\end{lem}

\begin{proof}
We write $\memskelfull$ and $\memskelIndexFull{\prime}$.

Let us assume that $\pref$ is $\memskel$-monotone, that is, for all $\memstate \in \memstates$, for all $\lang_1, \lang_2\in \Rec{\colors}$,
\begin{equation} \label{eq:proofMxM:MMonotone}
\exists\, \word\in\famelt_{\meminit,\memstate} ,\, \PrefClosure{\word\lang_1}\strictpref\PrefClosure{\word\lang_2}
\implies \forall\, \word'\in\famelt_{\meminit,\memstate},\, \PrefClosure{\word'\lang_1}\pref\PrefClosure{\word'\lang_2}.
\end{equation}
We show that $\pref$ is $(\memskel \memProduct \memskel')$-monotone, that is, for all $(\memstate, \memstate')\in\memstates\times\memstates'$, for all $\lang_1, \lang_2\in \Rec{\colors}$,
\begin{equation} \label{eq:proofMxM:MxMMonotone}
	\exists\, \word\in\famelt_{(\meminit, \meminit'),(\memstate, \memstate')} ,\, \PrefClosure{\word\lang_1}\strictpref\PrefClosure{\word\lang_2}
	\implies \forall\, \word'\in\famelt_{(\meminit, \meminit'),(\memstate, \memstate')},\, \PrefClosure{\word'\lang_1}\pref\PrefClosure{\word'\lang_2}.
\end{equation}
To do so, we notice that $\famelt_{(\meminit, \meminit'),(\memstate, \memstate')}\subseteq\famelt_{\meminit,\memstate}$ (the product of memory skeletons simply updates both memories in parallel). Thus, if the premise of Equation~\eqref{eq:proofMxM:MxMMonotone} holds, we obtain by Equation~\eqref{eq:proofMxM:MMonotone} that the conclusion of Equation~\eqref{eq:proofMxM:MxMMonotone} also holds.

A similar argument can be laid out to show that $\memskel$-selectivity implies $(\memskel \memProduct \memskel')$-selectivity. It is enough to notice that for all $(\memstate, \memstate')\in\memstates\times\memstates'$, we have $\famelt_{(\memstate, \memstate'), (\memstate, \memstate')} \subseteq \famelt_{\memstate, \memstate}$: the definition of $\memskel$-selectivity is thus clearly stronger than the definition of $(\memskel \memProduct \memskel')$-selectivity.
\end{proof}

\subsection*{Prefix-covers and cyclic-covers} While the aforementioned concepts of $\memskel$-monotony and $\memskel$-selectivity are the primordial ones for stating the characterization,  we still need two additional notions to prove it.

Let us sketch the issue here already. To prove that monotone and selective preference relations yield UML strategies, Gimbert and Zielonka deploy an \textit{inductive argument} on the number of choices in an arena. Intuitively, we want to use a similar approach for UFM strategies, but because of the unavoidable coupling between the memory skeleton and the arena (e.g., Lemma~\ref{lem:UFMiffMLonProduct}), the induction argument breaks, as adding one choice in the arena results in adding many in the \textit{product arena} (as many as there are memory states), where the reasoning needs to take place. New insight and techniques are thus needed to patch this induction scheme.

To solve this issue, \textit{we decouple the two aspects} (see Section~\ref{sec:conceptsToFM}). Intuitively, we first establish that, on arenas that inherently share the same good properties as product arenas (that is, they already ``classify'' prefixes and cycles as the memory would), we can deploy the induction argument and obtain UML strategies. Then, we obtain the result for UFM strategies on \textit{general} arenas as a corollary. The crux is identifying such ``good'' arenas: this is done through the following notions.

\begin{defi}[Prefix-covers and cyclic-covers]
\label{def:covers}
Let $\memskelfull$ be a memory skeleton and $\arenafull$ be an arena. Let $\covStates \subseteq \states$.

We say that $\memskel$ is a \textit{prefix-cover} of $\covStates$ in $\arena$ if for all $\state \in \states$, there exists $\memstate_\state \in \memstates$ such that, for all $\pth \in \Paths(\arena)$ such that $\edgeIn(\pth) \in \covStates$, $\edgeOut(\pth) = \state$ and such that for all $\pth'$ proper prefix of $\pth$, $\edgeOut(\pth') \neq \state$, we have
$\memupdhat(\meminit, \colhat(\pth)) = \memstate_\state$.

We say that $\memskel$ is a \textit{cyclic-cover} of $\covStates$ in $\arena$ if for all $\pth \in \Paths(\arena)$ such that $\edgeIn(\pth) \in \covStates$, if $\state = \edgeOut(\pth)$ and $\memstate = \memupdhat(\meminit, \colhat(\pth))$, for all $\pth' \in \Paths(\arena)$ such that $\edgeIn(\pth') = \edgeOut(\pth') = \state$, $\memupdhat(\memstate, \colhat(\pth')) = \memstate$.
\end{defi}

Intuitively, $\memskel$ is a prefix-cover for a set of states $\covStates$ if the histories starting in $\covStates$ and visiting a given state $\state \in \states$ for the first time are read up to the same memory state in the memory skeleton. Similarly, $\memskel$ is a cyclic-cover of $\arena$ if the cycles\footnote{Definition~\ref{def:covers} can be equivalently stated by considering simple cycles only.} of $\arena$ are read as cycles in the memory skeleton, once the memory has been initialized properly.

As hinted above, the canonical example of a prefix-covered and cyclic-covered arena is a product arena (but many more arenas can be covered, hence it is beneficial to be general with these concepts).

\begin{lem}
\label{lem:productArenaIsCovered}
Let $\memskelfull$ be a memory skeleton and $\arenafull$ be an arena. Then $\memskel$ is both a prefix-cover and a cyclic-cover for $\covStates = \states \times \{\meminit\}$ in the product arena $\prodAS{\arena}{\memskel}$.
\end{lem}

\begin{proof}
The main argument that we will be using in this proof is that if there is a history $\pth$ with $\edgeIn(\pth) = (\state,\memstate)$ and $\edgeOut(\pth) = (\state', \memstate')$ in the product arena $\prodAS{\arena}{\memskel}$, then reading $\colhat(\pth)$ from $\memstate$ in the memory skeleton $\memskel$ leads to $\memstate'$ (i.e., $\memupdhat(\memstate, \colhat(\pth)) = \memstate'$). This can be easily proved by induction on the length of $\pth$, thanks to how the product arena is built.

We first show that $\memskel$ is a prefix-cover for $\covStates = \states \times \{\meminit\}$ in the product arena $\prodAS{\arena}{\memskel}$. What we have to prove, instantiating the definition of prefix-cover in this case, is that for all $(\state, \memstate)\in \states\times\memstates$, there exists $\memstate_{(\state,\memstate)}\in\memstates$ such that, for all $\pth\in\Paths(\prodAS{\arena}{\memskel})$ such that $\edgeIn(\pth)\in\covStates$, $\edgeOut(\pth) = (\state,\memstate)$ and such that for all $\pth'$ proper prefix of $\pth$, $\edgeOut(\pth') \neq \state$, we have $\memupdhat(\meminit, \colhat(\pth)) = \memstate_{(\state, \memstate)}$.
Let $(\state, \memstate)\in\memstates$; we take $\memstate_{(\state,\memstate)} = \memstate$. Then, if $\pth\in\Paths(\prodAS{\arena}{\memskel})$ is such that $\edgeIn(\pth)\in\covStates$ (that is, is equal to $(\state',\meminit)$ for some $\state'\in\states$), and $\edgeOut(\pth) = (\state, \memstate)$, we have by construction of the product arena that $\memupdhat(\meminit, \colhat(\pth)) = \memstate = \memstate_{(\state,\memstate)}$, as required.

To prove that $\memskel$ is a cyclic-cover for $\covStates$ in $\prodAS{\arena}{\memskel}$, we have to prove that for all $\pth \in \Paths(\prodAS{\arena}{\memskel})$ such that $\edgeIn(\pth) \in \covStates$, if $(\state, \memstate) = \edgeOut(\pth)$ and $\memstate' = \memupdhat(\meminit, \colhat(\pth))$, for all $\pth' \in \Paths(\prodAS{\arena}{\memskel})$ such that $\edgeIn(\pth') = \edgeOut(\pth') = (\state, \memstate)$, $\memupdhat(\memstate', \colhat(\pth')) = \memstate'$. Let $\pth \in \Paths(\prodAS{\arena}{\memskel})$ such that $\edgeIn(\pth) \in \covStates$ (that is, $\edgeIn(\pth) = (\state', \meminit)$ for some $\state'\in\states$). Then, if $(\state, \memstate) = \edgeOut(\pth)$, we have by construction of the product arena that $\memstate' = \memupdhat(\meminit, \colhat(\pth)) = m$. Let $\pth' \in \Paths(\prodAS{\arena}{\memskel})$ such that $\edgeIn(\pth') = \edgeOut(\pth') = (\state, \memstate)$. By construction of the product arena, we therefore have that $\memupdhat(\memstate, \colhat(\pth')) = \memstate$, as required.
\end{proof}

\subsection{Main results}
\label{subsec:results}

\subsection*{Equivalence} We now have the necessary ingredients to state our general equivalence result formally.

\begin{thm}[Equivalence]
\label{thm:equivalence}
Let $\pref$ be a preference relation and let $\memskel$ be a memory skeleton. Then, both players have UFM strategies based on memory skeleton $\memskel$ in all games $\gamefull$ if and only if $\pref$ and $\invpref$ are $\memskel$-monotone and $\memskel$-selective.
\end{thm}

We state this theorem broadly and with a \textit{focus on UFM strategies}. The actual results we have for each direction of the equivalence~--- which we develop in Section~\ref{sec:FMToConcepts} and Section~\ref{sec:conceptsToFM}~--- are a bit stronger, of wider applicability and/or more interesting, but this statement carries the take-home message of our work. It is also meant to mirror the seminal result of Gimbert and Zielonka~\cite[Theorem 2]{GZ05}: their result can be retrieved from Theorem~\ref{thm:equivalence} by taking the trivial memory skeleton $\memskel_{\mathsf{triv}}$. As such, our work brings a \textit{strict generalization of Gimbert and Zielonka's results}~\cite{GZ05} to the finite-memory case.

\begin{rem}
\label{rmk:weakestHypothesesAndCompositionality}
We will refine the statement of our intermediate results  in order to use weaker hypotheses and/or grant stronger conclusions, whenever possible. For example, we only need optimal strategies in the left-to-right direction (Theorem~\ref{thm:FMtoMonotony} and Theorem~\ref{thm:FMtoSelectivity})~--- and not the stronger notion of Nash equilibrium~--- while we do prove the existence of finite-memory Nash equilibria in the other direction (Theorem~\ref{thm:conceptsToML} and Corollary~\ref{cor:UFM}).

Similarly, we study the two implications of the equivalence in a \textit{compositional} way: we split the reasoning for $\memskel$-monotony and $\memskel$-selectivity, using different skeletons for each whenever meaningful, as well as for the players, again when beneficial. Additionally, we distinguish between arenas where the players do not need memory and the ones where they do, the first essentially being arenas that already share the good properties of product arenas (as in ``product with a memory skeleton'').

While such a level of care is not necessary to obtain Theorem~\ref{thm:equivalence}, it has two advantages. First, from a practical standpoint, it permits to obtain \textit{more useful results}\footnote{E.g., of wider applicability, or avoiding the use of memory on covered arenas.} when focusing on a particular direction of the equivalence (as often required in applications). Second, from a theoretical standpoint, it permits to isolate each concept and each element of the reasoning and to \textit{highlight their true roles\footnote{E.g., $\memskel$-monotony deals with prefixes up to the first cycle and $\memskel$-selectivity deals with the cycles thereafter; memory skeletons can be built in a compositional way based on these two orthogonal yet complementary tasks.} in the underlying mechanisms} that lead to the existence of UFM strategies.\hfill$\lhd$
\end{rem}

To prove Theorem~\ref{thm:equivalence}, we invoke the results we will prove in Section~\ref{sec:FMToConcepts} and Section~\ref{sec:conceptsToFM}.

\begin{proof}[Proof of Theorem~\ref{thm:equivalence}]
The left-to-right implication trivially follows from Theorem~\ref{thm:FMtoMonotony} (intuitively, the sufficiency of finite-memory strategies based on $\memskel$ in all one-player arenas implies $\memskel$-monotony) and Theorem~\ref{thm:FMtoSelectivity} (similar, but for $\memskel$-selectivity), applied to each player with respect to his preference relation. The converse implication is established in Corollary~\ref{cor:UFM} (intuitively, $\memskel$-monotony together with $\memskel$-selectivity implies the existence of a uniform finite-memory NE with strategies based on $\memskel$ in all two-player arenas), which can be restated in terms of UFM strategies through Lemma~\ref{lem:NEisOpti}.
\end{proof}

As a by-product of our method, we also obtain a similar equivalence by solely considering one of the two players and the corresponding one-player arenas.

\begin{thm}[One-player equivalence] \label{thm:equivalence1p}
	Let $\pref$ be a preference relation and let $\memskel$ be a memory skeleton. Then, $\Pone$ has UFM strategies based on memory skeleton $\memskel$ in all his one-player games $\gamefull$ if and only if $\pref$ is $\memskel$-monotone and $\memskel$-selective.
\end{thm}

Although this looks like a weak version of Theorem~\ref{thm:equivalence} at first sight, this is actually a distinct result as both sides of the equivalence are weaker: on the left side, it only handles the memory requirements for $\Pone$'s one-player games; on the right side, it does not assume anything about the inverse preference relation $\invpref$.

Albeit close, this is distinct from the \emph{half-positional determinacy} result from~\cite[Theorem~3]{DBLP:journals/amai/BiancoFMM11}, which gives \emph{sufficient} conditions about a winning condition for a player to admit memoryless optimal strategies on every \emph{two-player} arena~--- in Theorem~\ref{thm:equivalence1p}, we give a \emph{necessary and sufficient} condition for a player to admit UFM strategies on his \emph{one-player} arenas only.
The sufficient conditions from~\cite{DBLP:journals/amai/BiancoFMM11} (\emph{strong monotony} and \emph{strong concavity}) imply $\memskelTriv$-monotony and $\memskelTriv$-selectivity, but not the other way around.
Given a preference relation, it is possible for a player to have UFM strategies on his one-player arenas, but not on all two-player arenas: e.g., the example used in~\cite[Lemma~15]{DBLP:journals/amai/BiancoFMM11}. In such an example, Theorem~\ref{thm:equivalence1p} could be applied, but not the result from~\cite{DBLP:journals/amai/BiancoFMM11}.

\begin{proof}[Proof of Theorem~\ref{thm:equivalence1p}]
	The left-to-right implication trivially follows from Theorem~\ref{thm:FMtoMonotony} (for $\memskel$-monotony) and Theorem~\ref{thm:FMtoSelectivity} (for $\memskel$-selectivity). The converse implication is established in Corollary~\ref{cor:UFM1p}.
\end{proof}

\subsection*{Lifting corollary} As discussed in Section~\ref{sec:intro}, the work of Gimbert and Zielonka contains not one, but \textit{two} great results. Alongside the aforementioned equivalence result, Gimbert and Zielonka provide a corollary of high practical interest~\cite[Corollary 7]{GZ05}: they essentially obtain as a by-product of their approach that if memoryless strategies suffice in all one-player games of $\Pone$ and in all one-player games of $\Ptwo$, these strategies also suffice in all two-player games.

This provides an elegant way to prove that a preference relation (or equivalently an objective) admits memoryless optimal strategies \textit{without proving monotony and selectivity at all}: proving it in the two one-player subcases, which is generally much easier\footnote{See examples of one-player vs.~two-player complexity in~\cite{DBLP:journals/acta/BouyerMRLL18,DBLP:conf/fossacs/BouyerHMR017,DBLP:conf/concur/BruyereHRR19}.} as it boils down to graph reasoning, and then lifting the result to the general two-player case through the corollary.

Again, we are able to lift this corollary to the arena-independent finite-memory case, as follows.

\begin{cor}
\label{cor:lifting}
Let $\pref$ be a preference relation and $\memskel_1, \memskel_2$ be two memory skeletons. Assume that
\begin{enumerate}
\item for all one-player arenas $\arena = (\states_1, \states_2 = \emptyset, E)$, $\Pone$ has a UFM strategy $\strat_1 \in \stratsFM_1(\arena)$ based on memory skeleton $\memskel_1$ in $\game = (\arena, \pref)$;
\item for all one-player arenas $\arena = (\states_1 = \emptyset, \states_2, E)$, $\Ptwo$ has a UFM strategy $\strat_2 \in \stratsFM_2(\arena)$ based on memory skeleton $\memskel_2$ in $\game = (\arena, \pref)$.
\end{enumerate}
Then, for all two-player arenas $\arena = (\states_1, \states_2, E)$, both $\Pone$ and $\Ptwo$ have UFM strategies $\strat_i \in \stratsFM_i(\arena)$  based on memory skeleton $\memskel = \memskel_1 \memProduct \memskel_2$ in $\game = (\arena, \pref)$.
\end{cor}

We highlight the two (possibly different) skeletons of the two players to maintain a compositional approach, but if the same skeleton $\memskel$ works in both one-player\footnote{In Corollary~\ref{cor:lifting}~--- and in other places further in this paper~--- we use a slightly more restrictive definition of one-player arenas (no state of the opponent) than in Section~\ref{sec:prelims} (no choice in states of the opponent). Both definitions are morally equivalent, and our use of the more restrictive version here is without loss of generality (as it yields a weaker hypothesis). We use this definition for the sake of readability whenever possible.} versions, it also suffices in the two-player version.

\begin{proof}
By Theorem~\ref{thm:FMtoMonotony} and Theorem~\ref{thm:FMtoSelectivity}~--- which essentially state that the left-to-right implication of Theorem~\ref{thm:equivalence} holds already in one-player games, the hypothesis yields that $\pref$ is $\memskel_1$-monotone and $\memskel_1$-selective, while $\invpref$ is $\memskel_2$-monotone and $\memskel_2$-selective. Now it suffices to apply Corollary~\ref{cor:UFM}~--- essentially the right-to-left implication of Theorem~\ref{thm:equivalence}~--- to get the claim.
\end{proof}

\subsection{Example of application}
\label{subsec:example}

We present an illustrative application of our results, thereby proving the existence of UFM strategies for a specific preference relation: the conjunction of two reachability objectives, a subcase of \textit{generalized reachability games}, studied extensively in~\cite{DBLP:journals/corr/abs-1010-2420}. Let $\colors$ be an arbitrary set of colors, and $\reachT,\reachTT \subseteq \colors$ be two target sets of colors that have to be reached at least once.
Formally, let the winning condition $\wc\subseteq\colors^\omega$ be the set of infinite words $\word = \clr_1\clr_2\ldots$ such that
\[
	\exists\, i,j \in \IN,\, \clr_{i} \in \reachT \land \clr_{j} \in \reachTT.
\]
This winning condition induces a two-level (i.e., win/lose) preference relation $\pref$ as discussed in Example~\ref{ex:preferenceRelations}.

In this example, we will use Theorem~\ref{thm:equivalence} directly in order to provide one thorough illustration of the definitions of $\memskel$-monotony and $\memskel$-selectivity. However, in practice, using Corollary~\ref{cor:lifting} is preferable, as it yields a much shorter proof: by exhibiting the right skeletons for $\Pone$ and $\Ptwo$, we simply have to show that these skeletons are sufficient to play optimally on both players' \textit{one-player arenas}, which amounts to graph reasoning.

We start by showing that this preference relation is not $\memskelTriv$-monotone (that is, is not monotone for~\cite{GZ05}).
Assume $\clr_1 \in \reachT\setminus\reachTT$, $\clr_2\in\reachTT\setminus\reachT$, and $\clr_3\notin\reachT\cup\reachTT$.
Take $\lang_1 = \clr_1^*$, $\lang_2 = \clr_2^*$.
For $\word = \clr_1$, $\word' = \clr_2$, we have $\PrefClosure{\word\lang_1} \strictpref \PrefClosure{\word\lang_2}$, but $\PrefClosure{\word'\lang_2} \strictpref \PrefClosure{\word'\lang_1}$.
This means that the preference relation is not stable with regard to prefix addition (at least, without distinguishing different classes of prefixes).
Similarly, it is not $\memskelTriv$-selective (take $\word$ as the empty word, $\lang_1 = c_1^*$, $\lang_2 = c_2^*$, $\lang_3 = c_3^*$: to win, $\lang_1$ and $\lang_2$ need to be mixed).

We exhibit two memory skeletons $\memskelfullPref$ and $\memskelfullCyc$ such that $\pref$ is $\memskelPref$-monotone and $\memskelCyc$-selective: they are pictured in Figure~\ref{fig:MpMc}. Note that such skeletons are obviously not unique.

Let us prove that $\pref$ is $\memskelPref$-monotone. Let $\memstate\in\memstatesPref$, $\lang_1,\lang_2 \in \Rec{\colors}$; we want to show that Equation~\eqref{eq:monotony} is satisfied.
We assume that there exists $\word \in \famelt_{\meminitPref,\memstate}$ such that $\PrefClosure{\word\lang_1} \strictpref \PrefClosure{\word\lang_2}$: this means that all words of $\PrefClosure{\word\lang_1}$ are losing, and that there exists a winning word in $\PrefClosure{\word\lang_2}$. Let $\word' \in \famelt_{\meminitPref,\memstate}$; we show that we necessarily have that $\PrefClosure{\word'\lang_1} \pref \PrefClosure{\word'\lang_2}$. Note that if $\PrefClosure{\lang_1}$ is empty, this always holds; we now assume that $\PrefClosure{\lang_1}$ is non-empty. We study the two possible values of $\memstate$ separately.
\begin{itemize}
\item If $\memstate = \meminitPref$, then $\word$ and $\word'$ do not reach $\reachT$. If $\word$ does not reach $\reachTT$ either, as there is a winning word in $\PrefClosure{\word\lang_2}$, then there must be a winning word in $\PrefClosure{\lang_2}$. This word is still winning after prepending $\word'$ to it, so there is a winning word in $\PrefClosure{\word'\lang_2}$, and $\PrefClosure{\word'\lang_1} \pref \PrefClosure{\word'\lang_2}$. If $\word$ reaches $\reachTT$, then $\PrefClosure{\lang_1}$ cannot have a word reaching $\reachT$. As $\word'$ does not reach $\reachT$ either, all words of $\PrefClosure{\word'\lang_1}$ are losing, so $\PrefClosure{\word'\lang_1} \pref \PrefClosure{\word'\lang_2}$.
\item If $\memstate = \memstatePref_2$, then $\word$ and $\word'$ reach $\reachT$. Clearly, $\word$ cannot reach $\reachTT$ (as $\PrefClosure{\word\lang_1}$ would be winning). This implies that $\PrefClosure{\lang_2}$ must contain a word reaching $\reachTT$; as $\word'$ reaches $\reachT$, the concatenation of $\word'$ with the word of $\PrefClosure{\lang_2}$ reaching $\reachTT$ means that there is a winning word in $\PrefClosure{\word'\lang_2}$, so $\PrefClosure{\word'\lang_1} \pref \PrefClosure{\word'\lang_2}$.
\end{itemize}

\begin{figure}[tbh]
	\centering
	\begin{minipage}{0.5\columnwidth}
		\centering
		\begin{tikzpicture}[every node/.style={font=\small,inner sep=1pt}]
			\draw (0,0) node[rond,minimum width=23pt] (minit) {$\meminitPref$};
			\draw ($(minit)+(0,-1.8)$) node[rond,minimum width=23pt] (m2) {$\memstatePref_2$};
			\draw (minit) edge[-latex'] node[right,anchor=west,xshift=2pt] {$\reachT$} (m2);
			\draw (minit) edge[-latex',out=150,in=210,looseness=4,distance=0.8cm] node[left,xshift=-2pt] {$\colors\setminus\reachT$} (minit);
			\draw (m2) edge[-latex',out=150,in=210,looseness=4,distance=0.8cm] node[left,anchor=east,xshift=-2pt] {$\colors$} (m2);
			\draw ($(minit)+(1,0)$) edge[-latex'] (minit);
		\end{tikzpicture}
	\end{minipage}%
	\begin{minipage}{0.5\columnwidth}
		\centering
		\begin{tikzpicture}[every node/.style={font=\small,inner sep=1pt}]
			\draw (0,0) node[rond,minimum width=23pt] (minit) {$\meminitCyc$};
			\draw ($(minit)+(0,-1.8)$) node[rond,minimum width=23pt] (m2) {$\memstateCyc_2$};
			\draw (minit) edge[-latex'] node[right,anchor=west,xshift=2pt] {$\reachT\cup\reachTT$} (m2);
			\draw (minit) edge[-latex',out=150,in=210,looseness=4,distance=0.8cm] node[left,xshift=-2pt] {$\colors\setminus(\reachT\cup\reachTT)$} (minit);
			\draw (m2) edge[-latex',out=150,in=210,looseness=4,distance=0.8cm] node[left,anchor=east,xshift=-2pt] {$\colors$} (m2);
			\draw ($(minit)+(1,0)$) edge[-latex'] (minit);
		\end{tikzpicture}
	\end{minipage}%
	\caption{Memory skeletons $\memskelPref$ (left) and $\memskelCyc$ (right) for two-target reachability games.}
	\label{fig:MpMc}
\end{figure}

Let us now prove that $\pref$ is $\memskelCyc$-selective. Let $\word\in\colors^*$, $\memstate = \memupdhatCyc(\meminitCyc,\word)$,
$\lang_1, \lang_2\in \Rec{\colors}$ such that $\lang_1, \lang_2\subseteq \famelt_{\memstate,\memstate}$, and $\lang_3\in\Rec{\colors}$. We show that Equation~\eqref{eq:selectivity} is satisfied, i.e., that
\[
	\PrefClosure{\word(\lang_1\cup \lang_2)^*\lang_3} \pref
	\PrefClosure{\word\lang_1^*}\cup\PrefClosure{\word\lang_2^*}\cup\PrefClosure{\word\lang_3}.
\]
If all words of $\PrefClosure{\word(\lang_1\cup \lang_2)^*\lang_3}$ are losing, this equation trivially holds; we thus assume that this set contains a winning word. We therefore have to show that there is a winning word in $\PrefClosure{\word\lang_1^*}$, $\PrefClosure{\word\lang_2^*}$, or $\PrefClosure{\word\lang_3}$. We study the two possible values of $\memstate$ separately.
\begin{itemize}
	\item If $\memstate = \meminitCyc$, then $\word$ does not reach $\reachT$ nor $\reachTT$, and the same holds for all words of $\lang_1$ and $\lang_2$, as $\lang_1,\lang_2\subseteq \famelt_{\meminitCyc,\meminitCyc}$. Therefore, if a word of $\PrefClosure{\word(\lang_1\cup \lang_2)^*\lang_3}$ is winning, this must be because a word of $\PrefClosure{\word\lang_3}$ is winning.
	\item If $\memstate = \memstateCyc_2$, we distinguish three cases. If $\word$ reaches both $\reachT$ and $\reachTT$, then $\PrefClosure{\word\lang_1^*}\cup\PrefClosure{\word\lang_2^*}\cup\PrefClosure{\word\lang_3}$ trivially contains only winning words. If $\word$ reaches $\reachT$ but not $\reachTT$, then there must be a word reaching $\reachTT$ in $\PrefClosure{(\lang_1\cup \lang_2)^*\lang_3}$. Hence, at least one set among $\PrefClosure{\lang_1^*}$, $\PrefClosure{\lang_2^*}$, and $\PrefClosure{\lang_3}$ must contain a word reaching $\reachTT$, so $\PrefClosure{\word\lang_1^*}$, $\PrefClosure{\word\lang_2^*}$, or $\PrefClosure{\word\lang_3}$ contains a winning word.
	A symmetric argument works if $\word$ reaches $\reachTT$ but not $\reachT$.
\end{itemize}

Similar arguments can be laid out to show that the preference relation $\invpref$ of $\Ptwo$ is $\memskelPref$-monotone and $\memskelTriv$-selective (where $\memskelTriv$ is the trivial memory skeleton defined earlier). Let $\memskel = \memskelPref \memProduct \memskelCyc \memProduct \memskelTriv$ be the product of all the considered skeletons, depicted in Figure~\ref{fig:applicationGame}. Although $\memskel$ formally has four states, only three of them are reachable (no word is read both up to $\memstatePref_2$ in $\memskelPref$ and up to $\meminitCyc$ in $\memskelCyc$).
By Lemma~\ref{lem:memProdPreserving}, we have that both $\pref$ and $\invpref$ are $\memskel$-monotone and $\memskel$-selective. Using Theorem~\ref{thm:equivalence}, we obtain that \textit{both players have UFM strategies based on skeleton $\memskel$ in all games $\gamefull$}. Note that the number of states of memory skeleton $\memskel$ is minimal (no memory skeleton with two states or fewer suffices for $\Pone$ to play optimally in all arenas~\cite{DBLP:journals/corr/abs-1010-2420}).
Notice also that the one-player equivalence (Theorem~\ref{thm:equivalence1p}) gives us a more precise result for one-player games of $\Ptwo$: in these games, $\Ptwo$ can play with memory $\memskelPref \memProduct \memskelTriv$ (which corresponds to $\memskelPref$).

\begin{figure}[tbh]
	\centering
	\begin{minipage}{0.2\columnwidth}
		\centering
		\begin{tikzpicture}[every node/.style={font=\small,inner sep=1pt}]
		\draw (0,0) node[rond] (s1) {$\state_1$};
		\draw ($(s1)+(0,-1.5)$) node[rond] (s2) {$\state_2$};
		\draw ($(s2)+(0,-1.5)$) node[rond] (s3) {$\state_3$};
		\draw (s2) edge[-latex'] node[right,anchor=west,xshift=2pt] {$t_1$} (s3);
		\draw (s1) edge[-latex',out=-120,in=120] (s2);
		\draw (s2) edge[-latex',out=60,in=-60] node[right,anchor=west,xshift=2pt] {$t_2$} (s1);
		\draw (s3) edge[-latex',out=150,in=210,looseness=4,distance=0.8cm] (s3);
		\end{tikzpicture}
	\end{minipage}%
	\begin{minipage}{0.4\columnwidth}
		\centering
		\begin{tikzpicture}[every node/.style={font=\small,inner sep=1pt}]
		\draw (0,0) node[rond] (minit) {$\memstate_1$};
		\draw ($(minit)+(0.9,-1.8)$) node[rond] (m2) {$\memstate_2$};
		\draw ($(minit)+(-0.9,-1.8)$) node[rond] (m3) {$\memstate_3$};
		\draw (minit) edge[-latex'] node[right,anchor=west,xshift=2pt] {$\reachTT\setminus\reachT$} (m2);
		\draw (minit) edge[-latex'] node[left,anchor=east,xshift=-2pt] {$\reachT$} (m3);
		\draw (m2) edge[-latex'] node[above,anchor=south,yshift=2pt] {$\reachT$} (m3);
		\draw (minit) edge[-latex',out=60,in=120,looseness=4,distance=0.8cm] node[above,yshift=2pt] {$\colors\setminus(\reachT\cup\reachTT)$} (minit);
		\draw (m3) edge[-latex',out=150,in=-150,looseness=4,distance=0.8cm] node[left,anchor=east,xshift=-2pt] {$\colors$} (m3);
		\draw (m2) edge[-latex',out=30,in=-30,looseness=4,distance=0.8cm] node[right,anchor=west,xshift=2pt] {$\colors\setminus\reachT$} (m2);
		\draw ($(minit)+(1,0)$) edge[-latex'] (minit);
		\end{tikzpicture}
	\end{minipage}%
	\begin{minipage}{0.4\columnwidth}
		\centering
		\begin{tikzpicture}[every node/.style={font=\small,inner sep=1pt}]
		\draw (0,0)               node[rond] (s1m1) {$\state_1, \memstate_1$};
		\draw ($(s1m1)+(0,-1.5)$) node[rond] (s2m1) {$\state_2, \memstate_1$};
		\draw ($(s2m1)+(0,-1.5)$) node[rond] (s3m1) {$\state_3, \memstate_1$};

		\draw ($(s1m1)+(1.5, 0)$) node[rond] (s1m2) {$\state_1, \memstate_2$};
		\draw ($(s1m2)+(0,-1.5)$) node[rond] (s2m2) {$\state_2, \memstate_2$};
		\draw ($(s2m2)+(1.5,-1.5)$) node[rond] (s3m3) {$\state_3, \memstate_3$};

		\draw (s1m1) edge[-latex',very thick] (s2m1);
		\draw (s2m1) edge[-latex',very thick] node[above left,anchor=south east] {$t_2$} (s1m2);
		\draw (s2m1) edge[-latex',out=-45,in=150] node[below,anchor=north,yshift=-2pt] {$t_1$} (s3m3);
		\draw (s3m1) edge[-latex',out=-30,in=30,looseness=4,distance=0.8cm,very thick] (s3m1);

		\draw (s1m2) edge[-latex',out=-120,in=120,very thick] (s2m2);
		\draw (s2m2) edge[-latex',out=60,in=-60] node[right,anchor=west,xshift=2pt] {$t_2$} (s1m2);
		\draw (s2m2) edge[-latex',very thick] node[above right,anchor=south west, very thick] {$t_1$} (s3m3);

		\draw (s3m3) edge[-latex',out=-30,in=30, looseness=4, distance=0.8cm, very thick] (s3m3);
		\end{tikzpicture}
	\end{minipage}%
	\caption{Arena $\arena$ (left), memory skeleton $\memskel$ (center; with $\meminit = \memstate_1$), and their product arena $\prodAS{\arena}{\memskel}$ (right; only states reachable from $\states\times\{\meminit\}$ are depicted). We assume that $\reachT = \{t_1\}$, $\reachTT = \{t_2\}$. The $(\states \times \{\meminit\})$-optimal memoryless strategy is highlighted with bold arrows.}
	\label{fig:applicationGame}
\end{figure}

We provide an example of a one-player arena $\arena = (\states_1, \states_2 = \emptyset, E)$ in Figure~\ref{fig:applicationGame}, and show that there is a UFM strategy for the preference relation $\pref$ based on skeleton $\memskel$. To do so, we invoke Lemma~\ref{lem:UFMiffMLonProduct}: we show equivalently that the product $\prodAS{\arena}{\memskel}$ admits an $(\states \times \{\meminit\})$-optimal memoryless strategy for $\pref$.
Notice that no memoryless strategy suffices to play optimally in $\gamefull$, as when starting in $\state_2$, $\Pone$ should first visit $\state_1$ before going to $\state_3$.
Also, the $(\states \times \{\meminit\})$-optimal memoryless strategy for the product arena is only optimal if the initial state is in $\states \times \{\meminit\}$; it is for instance not optimal from state $(\state_2,\memstate_2)$.

\subsection{Counterexample to a general lifting corollary}
\label{sec:counterexample}

We discuss in full details the counterexample presented in Section~\ref{sec:intro}. We recall that the goal of this counterexample is to show that a lifting corollary for general finite-memory (instead of \emph{arena-independent} finite-memory) determinacy is not possible.

Let $\colors = \IZ$. We consider the following two winning conditions:
\begin{align*}
	\wc_1 &= \{\clr_1\clr_2\ldots\in\colors^\omega\mid
	\liminf_{n \rightarrow \infty}\sum_{i = 1}^n \clr_i = +\infty\}, \\
	\wc_2 &= \{\clr_1\clr_2\ldots\in\colors^\omega\mid
	\sum_{i = 1}^n \clr_i = 0 \text{ for infinitely many $n$'s}\}.
\end{align*}
If the play obtained by playing a game is $\play$, $\Pone$ wins if and only if $\colhat(\play)$ lies in $\wc = \wc_1 \cup \wc_2$, and $\Ptwo$ wins if and only if $\colhat(\play)$ lies in $\comp{\wc} = \colors^\omega \setminus \wc  =  \comp{\wc_1}\cap\comp{\wc_2}$ (which corresponds to the description given in Section~\ref{sec:intro}).
We prove that $\Pone$ and $\Ptwo$ have finite-memory optimal strategies in their respective one-player games.

Let us fix some terminology beforehand: we say that a cycle in an arena is a \textit{zero} cycle if the sum of its weights is zero, a \textit{positive} cycle if this sum is strictly positive, and a \textit{negative} cycle if this sum is strictly negative.

We first consider $\Pone$'s one-player games. In a one-player arena, $\Pone$ can create a play $\play$ such that $\colhat(\play) \in \wc_1$ if and only if there is a reachable positive cycle. In this case, $\Pone$ can win with a memoryless strategy (simply reaching the cycle and then looping in it). If that is not possible, in order to win, $\Pone$ has to induce a play $\play$ such that $\colhat(\play)\in\wc_2$. We show that if possible, this can be done using finite memory. Let us assume that there exists a play $\play = \edge_1\edge_2\ldots$ such that $\colhat(\play) = \clr_1\clr_2\ldots \in \wc_2$. Let us consider two indices $k, l\in\IN$ such that $k<l$, $e_k = e_l$, $\sum_{i = 1}^k \clr_i = 0$, and $\sum_{i = 1}^l \clr_i = 0$. Such two indices necessarily exist, as there are finitely many edges in the arena, but infinitely many indices for which the running sum of weights is $0$. Notice in particular that $\sum_{i = k+1}^l \clr_i = 0$. Now, consider the play
\[
\play' = \edge_1\ldots\edge_k\edge_{k+1}\ldots\edge_l\edge_{k+1}\ldots\edge_l\edge_{k+1}\ldots,
\]
with the sequence of edges $\edge_{k+1}\ldots\edge_l$ repeating \textit{ad infinitum} ($\play'$ is a ``lasso''). This is a valid play since $\edge_k = \edge_l$. Moreover, we have that $\colhat(\play')\in\wc_2$ as after repeating $m$ times the sequence $\edge_{k+1}\ldots\edge_l$, the sum of the weights equals $\sum_{i=1}^k \clr_i + m \cdot \sum_{i=k+1}^l \clr_i = 0 + m \cdot 0 = 0$.
The play $\play'$ can be implemented with finite memory, as it consists of a finite prefix and a repeated finite sequence, which corresponds to a zero cycle.

We now turn our attention to $\Ptwo$'s one-player games; $\Ptwo$ wins a play $\play$ such that $\colhat(\play) = \clr_1\clr_2\ldots\in\colors^\omega$ if and only if
\[
\liminf_{n \rightarrow \infty}\sum_{i = 1}^n \clr_i < +\infty \land
\sum_{i = 1}^n \clr_i = 0 \text{ for at most finitely many $n$'s}.
\]
In a one-player arena, if there is a reachable negative cycle, $\Ptwo$ can ensure to win by pumping it forever, and can therefore win with a memoryless strategy.
We now consider an arena that has no reachable negative cycle. As we did for $\Pone$, we show that if $\Ptwo$ can win a game in such an arena, then he can do so using finite memory. If $\Ptwo$ can win, let $\play = \edge_1\edge_2\ldots$ be a winning play for $\Ptwo$, i.e., $\colhat(\play) = \clr_1\clr_2\ldots\in\comp{\wc_1}\cap\comp{\wc_2}$. Let $\state$ be a state visited infinitely often when $\play$ is played, and $m\in\IN$ be the first index such that $\edgeOut(\edge_m) = \state$. We can decompose $\play$ into a finite prefix $\edge_1\ldots\edge_m$ followed by an infinite sequence of cycles, all starting in $\state$. Since there is no negative cycle, we cannot have that infinitely many of these cycles are positive, as this would imply that $\colhat(\play) \in \wc_1$. Thus, infinitely many zero cycles are taken from $\state$. As $\colhat(\play) \in \comp{\wc_2}$, there exists such a cycle $\edge_{k}\ldots\edge_l$ (that is, $\edgeIn(\edge_{k}) = \edgeOut(\edge_l) = \state$ and $\sum_{i=k}^l \clr_i = 0$) such that for all $k\le n\le l$, it holds that $\sum_{i = 1}^n \clr_i \neq 0$. This also implies that $\sum_{i = 1}^{k-1} \clr_i \neq 0$, i.e., the history up to this cycle has a non-zero sum.
Now, let us consider the play
\[
\play' = \edge_1\ldots\edge_{k-1}\edge_{k}\ldots\edge_l\edge_{k}\ldots \edge_l\edge_{k}\ldots,
\]
with the sequence of edges $\edge_{k}\ldots\edge_l$ repeating \textit{ad infinitum} ($\play'$ is a ``lasso''). This is a valid play as $\edgeIn(\edge_k) = \edgeOut(\edge_l)$. As $\sum_{i=k}^l \clr_i = 0$, we have that $\colhat(\play') \in\comp{\wc_1}$. Moreover, every time the cycle starts again, the running sum of weights is equal to the same value: $\sum_{i=1}^{k-1} \clr_i \neq 0$. Therefore, as the running sum of weights does not reach zero the first time the cycle is taken, and it also never reaches zero along the cycle, it can never reach zero after index $k-1$. Hence, $\colhat(\play')$ is not in $\wc_2$ either, and $\play'$ is winning for $\Ptwo$.
For the same reason as for $\Pone$, play $\play'$ only requires finite memory to be implemented.

As argued in Section~\ref{sec:intro}, the two-player game from Figure~\ref{fig:counterexample} illustrates that $\Pone$ might need infinite memory to play optimally in the two-player case. This proves that Gimbert and Zielonka's approach cannot work in full generality in the finite-memory case, as we cannot obtain the existence of finite-memory optimal strategies in all two-player games from the existence of finite-memory optimal strategies in all one-player games.

\section{From finite memory based on \texorpdfstring{$\memskel$}{M} to \texorpdfstring{$\memskel$}{M}-monotony and \texorpdfstring{$\memskel$}{M}-selectivity}
\label{sec:FMToConcepts}

\subsection*{Monotony.} We want to keep our approach as compositional as possible, hence we consider the two notions separately. Let us start with $\memskel$-monotony.

\begin{thm}
\label{thm:FMtoMonotony}
Let $\memskelfull$ be a memory skeleton and $\pref$ be a preference relation.
Assume that for all one-player arenas $\arena = (\states_1, \states_2 = \emptyset, E)$, for all $\state, \state' \in \states$, $\Pone$ has an $\state$-optimal and $\state'$-optimal strategy $\strat \in \stratsFM_1(\arena)$, encoded as a Mealy machine $\mealy_{\strat} = (\memskel, \memnxt)$, in $\game = (\arena, \pref)$. Then $\pref$ is $\memskel$-monotone.
\end{thm}

Note that the same holds for $\Ptwo$ and $\invpref$ symmetrically. Also, we do not require full \textit{uniformity} of the strategy, but only uniformity with regard to the fixed pair of states (i.e., strategy $\strat$ does not need to be optimal from other states).

\begin{figure}[tbh]
	\centering
\begin{tikzpicture}[every node/.style={font=\small,inner sep=1pt}]
\draw (0,0) node[oval,minimum width=27mm] (qw) {$\atmtnIndex{\word} \setminus \{\atmtnFinalState^{\word}\}$};
\draw ($(qw)+(3,0)$) node[oval,minimum width=27mm] (qw') {$\atmtnIndex{\word'} \setminus \{\atmtnFinalState^{\word'}\}$};
\draw ($(qw)+(0,-2.5)$) node[oval,minimum width=27mm] (qk1) {$\atmtnIndex{\lang_1} \setminus \{\atmtnInitState^{\lang_1}\}$};
\draw ($(qw')+(0,-2.5)$) node[oval,minimum width=27mm] (qk2) {$\atmtnIndex{\lang_2} \setminus \{\atmtnInitState^{\lang_2}\}$};
\draw ($(qw)!0.5!(qk2)$) node[rond] (t) {$t$};
\draw ($(qw)+(-2,0)$) edge[-latex'] (qw);
\draw ($(qw')+(2,0)$) edge[-latex'] (qw');
\draw (qw) edge[-latex',out=-80,in=135] (t);
\draw (qw') edge[-latex',out=-100,in=45] (t);
\draw (t) edge[-latex',out=-145,in=90] (qk1);
\draw (t) edge[-latex',out=-125,in=70] (qk1);
\draw (t) edge[-latex',out=-35,in=90] (qk2);
\draw (t) edge[-latex',out=-55,in=110] (qk2);
\end{tikzpicture}
	\caption{Automaton $\atmtn$ built to establish $\memskel$-monotony.}
	\label{fig:proofMonotone}
\end{figure}

Our proof can be sketched as follows. We need to establish that Equation~\eqref{eq:monotony} holds. We first instantiate the four languages involved in it: $\{\word\}$, $\{\word'\}$, $\lang_1$ and $\lang_2$. We take NFA recognizing them and build an NFA $\atmtn$ that joins them in such a way that, when $\atmtn$ is considered as a game arena (see Lemma~\ref{lem:arenaLanguage}), its plays correspond exactly to the languages of infinite words considered in Equation~\eqref{eq:monotony}. This arena is essentially composed of two chains emulating the two prefixes $\word$ and $\word'$ and leading to a state $t$ where $\Pone$ has to pick a side corresponding to the two languages $[\lang_1]$ and $[\lang_2]$ (Figure~\ref{fig:proofMonotone}). Now, establishing the $\memskel$-monotony of $\pref$ boils down to invoking an optimal strategy $\strat$ in the corresponding game, the crux being that this strategy always picks the same edge in $t$ (i.e., the same side between subarenas corresponding to $[\lang_1]$ and $[\lang_2]$) as both prefixes $\word$ and $\word'$ are deemed equivalent by the memory skeleton $\memskel$.

\begin{proof}
Let $\memskelfull$ be a memory skeleton and $\pref$ be a preference relation satisfying the hypothesis. Let us prove that $\pref$ is $\memskel$-monotone, i.e.,  that for all $\memstate \in \memstates$, for all $\lang_1,
	\lang_2\in \Rec{\colors}$,
	\begin{equation}
	\label{eq:toProveMonotony}
	\exists\, \word\in\famelt_{\meminit,\memstate} ,\, \PrefClosure{\word\lang_1}\strictpref\PrefClosure{\word\lang_2}
	\implies \forall\, \word'\in\famelt_{\meminit,\memstate},\, \PrefClosure{\word'\lang_1}\pref\PrefClosure{\word'\lang_2}.
	\end{equation}

Let $\memstate \in \memstates$, $\lang_1, \lang_2 \in \Rec{C}$. We assume that $\lang_1, \lang_2 \neq \emptyset$, otherwise Equation~\eqref{eq:toProveMonotony} holds trivially: if $\lang_1$ is empty, the conclusion of the implication is true regardless of $\lang_2$; and if $\lang_2$ is empty, the premise is false. Now, assume there exists $\word \in\famelt_{\meminit,\memstate}$ such that $\PrefClosure{\word\lang_1}\strictpref\PrefClosure{\word\lang_2}$, and let $\word'$ be another prefix in $\famelt_{\meminit,\memstate}$. We will prove that $\PrefClosure{\word'\lang_1}\pref\PrefClosure{\word'\lang_2}$.

Let $\atmtnfullIndex{\word}$, $\atmtnfullIndex{\word'}$, $\atmtnIndex{\lang_{1}} = (\atmtnStates^{\lang_{1}}, \colorsSubset^{\lang_{1}}$, $\atmtnTrans^{\lang_{1}}, \atmtnInitState^{\lang_{1}}, \atmtnFinalStates^{\lang_{1}})$ and $\atmtnfullIndex{\lang_{2}}$ respectively denote NFA recognizing languages $\{\word\}$, $\{\word'\}$, $\lang_1$ and $\lang_2$. They exist since all these languages are regular. 
We assume w.l.o.g.~that automaton $\atmtnIndex{\word}$ (resp.~$\atmtnIndex{\word'}$, $\atmtnIndex{\lang_{1}}$, $\atmtnIndex{\lang_{2}}$) is coaccessible and has only one
initial state $\atmtnInitState^{\word}$ (resp.~$\atmtnInitState^{\word'}$, $\atmtnInitState^{\lang_1}$, $\atmtnInitState^{\lang_2}$) with no ingoing transition. We can do this since $\lang_1$ and $\lang_2$ are non-empty.
We also assume w.l.o.g.~that $\atmtnIndex{\word}$ (resp.~$\atmtnIndex{\word'}$) has only one final state $\atmtnFinalState^{\word}$ (resp.~$\atmtnFinalState^{\word'}$) with no outgoing transition. Actually, $\atmtnIndex{\word}$ and $\atmtnIndex{\word'}$ can be taken as ``chains'' recognizing a unique word, and being coaccessible and deterministic.

We build an automaton $\atmtnfull$ by ``merging'' states $\atmtnInitState^{\lang_1}$, $\atmtnInitState^{\lang_2}$, $\atmtnFinalState^{\word}$, and $\atmtnFinalState^{\word'}$. We call this new merged state $t$. Formally, we built it as follows.
\begin{itemize}
\item $\atmtnStates = (\atmtnStates^\word \cup \atmtnStates^{\word'} \cup \atmtnStates^{\lang_1}  \cup \atmtnStates^{\lang_2} \cup \{t\}) \setminus \{\atmtnInitState^{\lang_1}, \atmtnInitState^{\lang_2},\atmtnFinalState^{\word}, \atmtnFinalState^{\word'}\}$;
\item $\colorsSubset = \colorsSubset^\word \cup \colorsSubset^{\word'} \cup \colorsSubset^{\lang_1}  \cup \colorsSubset^{\lang_2}$;
\item $\atmtnInitStates = \{\atmtnInitState^{\word}, \atmtnInitState^{\word'}\}$ and $\atmtnFinalStates = \atmtnFinalStates^{\lang_1} \cup \atmtnFinalStates^{\lang_2}$;
\item and finally, the transition relation simply takes into account the merging on $t$:
\begin{align*}
\atmtnTrans = &\left\lbrace (q, c, q') \mid (q, c, q') \in (\atmtnTrans^\word \cup \atmtnTrans^{\word'} \cup \atmtnTrans^{\lang_1}  \cup \atmtnTrans^{\lang_2}) \wedge q, q' \not\in  \{\atmtnInitState^{\lang_1}, \atmtnInitState^{\lang_2},\atmtnFinalState^{\word}, \atmtnFinalState^{\word'}\}\right\rbrace\\
\cup &\left\lbrace (q, c, t) \mid (q, c, q') \in (\atmtnTrans^\word \cup \atmtnTrans^{\word'}) \wedge q' \in  \{\atmtnFinalState^{\word}, \atmtnFinalState^{\word'}\}\right\rbrace\\
\cup &\left\lbrace (t, c, q') \mid (q, c, q') \in (\atmtnTrans^{\lang_1}  \cup \atmtnTrans^{\lang_2}) \wedge q \in  \{\atmtnInitState^{\lang_1}, \atmtnInitState^{\lang_2}\}\right\rbrace.
\end{align*}
\end{itemize}
This construction is illustrated in Figure~\ref{fig:proofMonotone}.
The language recognized by $\atmtn$ from $\atmtnInitState^{\word}$ is $\word(\lang_1 \cup \lang_2)$, whereas from $\atmtnInitState^{\word'}$, it is $\word'(\lang_1 \cup \lang_2)$. Observe that $\atmtn$ is coaccessible since both $\atmtnIndex{\lang_1}$ and $\atmtnIndex{\lang_2}$ are coaccessible.

Recall that we assume $\PrefClosure{\word\lang_1}\strictpref\PrefClosure{\word\lang_2}$. By definition, this implies that $\PrefClosure{\word\lang_2} \neq \emptyset$, hence we also have that $\PrefClosure{\lang_2} \neq \emptyset$. From this, we get that $t$ is essential in $\atmtn$ (Lemma~\ref{lem:arenaLanguage}). Thus, it is also the case for $\atmtnInitState^{\word}$ and $\atmtnInitState^{\word'}$.

We will now interpret this NFA as an arena and use the hypothesis. Let $\arena = \atmtnArena{\atmtn}$. By Lemma~\ref{lem:arenaLanguage}, we have that $\colhat(\Plays(\arena, \atmtnInitState^{\word})) = \left[ \word(\lang_1\cup \lang_2)\right] $ and $\colhat(\Plays(\arena, \atmtnInitState^{\word'})) = \left[ \word'(\lang_1\cup \lang_2)\right] $. By hypothesis, $\Pone$ has a $\atmtnInitState^{\word}$-optimal and $\atmtnInitState^{\word'}$-optimal strategy $\strat \in \stratsFM_1(\arena)$, encoded as a Mealy machine $\mealy_{\strat} = (\memskel, \memnxt)$, in $\game = (\arena, \pref)$.

		Let $\play \in \Plays(\arena, \atmtnInitState^{\word}, \strat)$ be the only play consistent with strategy $\strat$ from $\atmtnInitState^{\word}$. By definition of $\arena$, this play $\play$ necessarily contains a history $\pth = \edge_1 \ldots{} \edge_n$ such that $\edgeOut(\edge_n) = t$ and for all $i$, $1 \leq i < n$, $\edgeOut(\edge_i) \neq t$. Observe that $\colhat(\pth) = \word$. Recall that $\memstate = \memupdhat(\meminit, \word)$ is the memory state reached after reading $\word$ since $\word \in\famelt_{\meminit,\memstate}$. Let $\edge = \memnxt(\memstate, t)$ be the edge chosen by $\strat$ in $t$ when $t$ is visited (note that $t$ will be visited only once by construction of $\arena$).

		We will show that $\edge$ belongs to the part generated by $\atmtnIndex{\lang_2}$. By contradiction, assume it belongs to $\atmtnIndex{\lang_1}$. Then, $\play = \pth \cdot \play'$, with $\colhat(\play') \in [\lang_1]$, hence $\colhat(\play) \in [\word \lang_1]$. First, observe that
		\begin{align*}
		 [\word \lang_2] &\pref \colhat(\Plays(\arena, \atmtnInitState^{\word}))
		 \end{align*}
since $[\word \lang_2] \subseteq [\word \lang_1] \cup [\word \lang_2]$, $[\word \lang_1] \cup [\word \lang_2] = \left[ \word(\lang_1\cup \lang_2)\right]$ by Lemma~\ref{lem:languageSplit}, and, as noted above, $\left[ \word(\lang_1\cup \lang_2)\right] = \colhat(\Plays(\arena, \atmtnInitState^{\word}))$. Now since $\strat$ is optimal\footnote{One can easily get from the definition using the $\UCol$-operator that, in this one-player game, $\strat$ is $\atmtnInitState^{\word}$-optimal if and only if for all $\strat' \in \strats(\arena)$, $\colhat(\Plays(\arena, \atmtnInitState^{\word}, \strat')) \pref \colhat(\Plays(\arena, \atmtnInitState^{\word}, \strat))$.} from $\atmtnInitState^{\word}$, we have
		\begin{align*}
		 [\word \lang_2] &\pref \colhat(\play).
		 \end{align*}
Finally, we assumed that $\colhat(\play) \in [\word \lang_1]$, hence we can conclude that
		\begin{align*}
		 [\word \lang_2] &\pref [\word \lang_1],
		 \end{align*}
	 which contradicts the hypothesis that $[\word \lang_1] \strictpref [\word \lang_2]$. Hence, we have established that $\edge$ belongs to $\atmtnIndex{\lang_2}$.

	 Now let us consider $\play'' \in \Plays(\arena, \atmtnInitState^{\word'}, \strat)$, the only play consistent with strategy $\strat$ from $\atmtnInitState^{\word'}$. Again, by definition of $\arena$, this play $\play''$ necessarily contains a history $\pth'' = \edge_1 \ldots{} \edge_n$ such that $\edgeOut(\edge_n) = t$ and for all $i$, $1 \leq i < n$, $\edgeOut(\edge_i) \neq t$. Observe that $\colhat(\pth'') = \word'$. Since $\word' \in \famelt_{\meminit,\memstate}$, we also have that $\memupdhat(\meminit, \word') = \memstate$, i.e., the memory state reached after reading $\word'$ is the same as the one reached after reading $\word$. Recall that $\memnxt$ is deterministic by definition: i.e., for a given memory state and state of the arena, it always prescribes the same edge. Hence, we have that $\edge = \memnxt(\memstate, t)$ is exactly the same as before, and therefore belongs to $\atmtnIndex{\lang_2}$. Thus, $\colhat(\play'') \in [\word'\lang_2]$.

	 Finally, since $\strat$ is also $\atmtnInitState^{\word'}$-optimal and applying the same reasoning as above, we have that
		\begin{align*}
		 [\word' \lang_1] \pref [\word' \lang_1] \cup [\word' \lang_2] = \left[ \word'(\lang_1\cup \lang_2)\right] &= \colhat(\Plays(\arena, \atmtnInitState^{\word'}))\\
		 &\pref \colhat(\play'')\\
		 &\pref [\word' \lang_2],
		 \end{align*}
		 which proves Equation~\eqref{eq:toProveMonotony} and concludes our proof.
\end{proof}

\subsection*{Selectivity.} We now turn to selectivity, which focuses on stability with regard to cycle mixing.

\begin{thm}
\label{thm:FMtoSelectivity}
Let $\memskelfull$ be a memory skeleton and $\pref$ be a preference relation. Assume that for all one-player arenas $\arena = (\states_1, \states_2 = \emptyset, E)$, for all $\state \in \states$, $\Pone$ has an $\state$-optimal strategy $\strat \in \stratsFM_1(\arena)$, encoded as a Mealy machine $\mealy_{\strat} = (\memskel, \memnxt)$, in $\game = (\arena, \pref)$. Then $\pref$ is $\memskel$-selective.
\end{thm}

Note that the same holds for $\Ptwo$ and $\invpref$ symmetrically. Again, observe that our hypothesis is as weak as possible as no uniformity is required.

Our proof bears similarities with the case of monotony. We need to establish that Equation~\eqref{eq:selectivity} holds. We first instantiate the four languages involved in it: $\{\word\}$, $\lang_1$, $\lang_2$ and $\lang_3$. We take NFA recognizing them and build an NFA $\atmtn$ that joins them in such a way that, when $\atmtn$ is considered as a game arena (see Lemma~\ref{lem:arenaLanguage}), its plays correspond exactly to the languages of infinite words considered in Equation~\eqref{eq:selectivity}. This arena is essentially composed of a chain emulating the prefix $\word$ and leading to a state $t$ where $\Pone$ can visit sides that generate cycles from $\lang_1$ and $\lang_2$~--- forever or for a finite time~--- or branch to a side corresponding to $\lang_3$ (Figure~\ref{fig:proofSelective}). Now, establishing the $\memskel$-selectivity of $\pref$ boils down to invoking an optimal strategy $\strat$ in the corresponding game, the crux being that this strategy always picks the same edge in $t$ (i.e., the same side between subarenas corresponding to $[\lang^\ast_1]$, $[\lang_2^*]$ and $[\lang_3]$) as all cycles on $t$ are deemed equivalent by the memory skeleton $\memskel$. The main difference with the previous construction is clear in the last sentence: it is now possible to come back to $t$, possibly infinitely often, and our proof takes that into account (as illustrated in Figure~\ref{fig:proofSelective}).

\begin{proof}
Let $\memskelfull$ be a memory skeleton and $\pref$ a preference relation satisfying the hypothesis. Let us prove that $\pref$ is $\memskel$-selective, i.e., that for all $\word\in\colors^*$, $\memstate = \memupdhat(\meminit,\word)$,
for all $\lang_1, \lang_2\in \Rec{C}$ such that $\lang_1, \lang_2\subseteq \famelt_{\memstate,\memstate}$, for all $\lang_3\in\Rec{\colors}$,
\begin{equation}
	\label{eq:toProveSelectivity}
	\PrefClosure{\word(\lang_1\cup \lang_2)^*\lang_3} \pref
	\PrefClosure{\word\lang_1^*}\cup\PrefClosure{\word\lang_2^*}\cup\PrefClosure{\word\lang_3}.
\end{equation}

Let $\word \in \colors^\ast$ and $\memstate = \memupdhat(\meminit,\word)$. Let $\lang_1, \lang_2, \lang_3\in \Rec{C}$, with $\lang_1, \lang_2\subseteq \famelt_{\memstate,\memstate}$. In the following, we assume all three languages $\lang_1$, $\lang_2$ and $\lang_3$ to be non-empty. Indeed, if $\lang_3$ is empty, so is the left-hand side of Equation~\eqref{eq:toProveSelectivity}, hence it trivially holds. If both $\lang_1$ and $\lang_2$ are empty,  Equation~\eqref{eq:toProveSelectivity} compares $\PrefClosure{\word\lang_3}$ to itself, hence it trivially holds again. Finally, if $\lang_1$ is the only empty language among the three, then Equation~\eqref{eq:toProveSelectivity} can be restated as follows:
	\begin{equation*}
	\PrefClosure{\word(\lang_1\cup \lang_2)^*\lang_3} = \PrefClosure{\word(\lang_2\cup \lang_2)^*\lang_3}
	\pref \PrefClosure{\word\lang_2^*}\cup\PrefClosure{\word\lang_2^*}\cup\PrefClosure{\word\lang_3} = \PrefClosure{\word\lang_1^*}\cup\PrefClosure{\word\lang_2^*}\cup\PrefClosure{\word\lang_3},
	\end{equation*}
where the middle inequality~--- the one to prove~--- involves three non-empty sets. A symmetric argument holds if $\lang_2$ is the only empty language. We also assume that $\lang_1$ and $\lang_2$ do not contain the empty word for technical convenience: this is w.l.o.g.~thanks to the Kleene stars used in the regular expressions to consider.

As for monotony, we start by considering NFA for all these languages: let $\atmtnfullIndex{\word}$, $\atmtnfullIndex{\lang_{1}}$, $\atmtnIndex{\lang_{2}} = (\atmtnStates^{\lang_{2}}, \colorsSubset^{\lang_{2}}$, $\atmtnTrans^{\lang_{2}}, \atmtnInitState^{\lang_{2}}$, $\atmtnFinalStates^{\lang_{2}})$ and $\atmtnfullIndex{\lang_{3}}$ respectively denote NFA recognizing languages $\{\word\}$, $\lang_1$, $\lang_2$ and $\lang_3$. 
They exist since all these languages are regular. We assume w.l.o.g.~that automaton $\atmtnIndex{\word}$ (resp.~$\atmtnIndex{\lang_{1}}$, $\atmtnIndex{\lang_{2}}$, $\atmtnIndex{\lang_{3}}$) is coaccessible and has only one initial state $\atmtnInitState^{\word}$ (resp.~$\atmtnInitState^{\lang_1}$, $\atmtnInitState^{\lang_2}$, $\atmtnInitState^{\lang_3}$) with no ingoing transition. We can do this since $\lang_1$, $\lang_2$ and $\lang_3$ are non-empty.
We also assume w.l.o.g.~that $\atmtnIndex{\word}$ (resp.~$\atmtnIndex{\lang_{1}}$, $\atmtnIndex{\lang_{2}}$) has only one final state $\atmtnFinalState^{\word}$ (resp.~$\atmtnFinalState^{\lang_1}$,
$\atmtnFinalState^{\lang_2}$) with no outgoing transition. Again $\atmtnIndex{\word}$ can simply be a ``chain'' recognizing a unique word, being both coaccessible and deterministic.

Similarly to Theorem~\ref{thm:FMtoMonotony}, we build an automaton $\atmtnfull$ by ``merging'' states $\atmtnInitState^{\lang_1}$, $\atmtnInitState^{\lang_2}$, $\atmtnInitState^{\lang_3}$, $\atmtnFinalState^{\word}$, $\atmtnFinalState^{\lang_1}$, and $\atmtnFinalState^{\lang_2}$. We call this new merged state $t$. Formally, we build it as follows.
\begin{itemize}
\item $\atmtnStates = (\atmtnStates^\word \cup \atmtnStates^{\lang_1} \cup \atmtnStates^{\lang_2}  \cup \atmtnStates^{\lang_3} \cup \{t\}) \setminus \{\atmtnInitState^{\lang_1}, \atmtnInitState^{\lang_2}, \atmtnInitState^{\lang_3},\atmtnFinalState^{\word}, \atmtnFinalState^{\lang_1}, \atmtnFinalState^{\lang_2}\}$;
\item $\colorsSubset = \colorsSubset^\word \cup \colorsSubset^{\lang_1} \cup \colorsSubset^{\lang_2}  \cup \colorsSubset^{\lang_3}$;
\item $\atmtnInitStates = \atmtnInitStates^{\word} = \{\atmtnInitState^{\word}\}$ and $\atmtnFinalStates = \atmtnFinalStates^{\lang_3}$;
\item and finally, the transition relation simply takes into account the merging on $t$:
\begin{align*}
\atmtnTrans = &\left\lbrace (q, c, q') \mid (q, c, q') \in (\atmtnTrans^\word \cup \atmtnTrans^{\lang_1}  \cup \atmtnTrans^{\lang_2} \cup \atmtnTrans^{\lang_3}) \wedge q, q' \not\in  \{\atmtnInitState^{\lang_1}, \atmtnInitState^{\lang_2}, \atmtnInitState^{\lang_3},\atmtnFinalState^{\word}, \atmtnFinalState^{\lang_1}, \atmtnFinalState^{\lang_2}\}\right\rbrace\\
\cup &\left\lbrace (q, c, t) \mid (q, c, q') \in (\atmtnTrans^\word \cup \atmtnTrans^{\lang_1}  \cup \atmtnTrans^{\lang_2}) \wedge q' \in  \{\atmtnFinalState^{\word}, \atmtnFinalState^{\lang_1}, \atmtnFinalState^{\lang_2}\}\right\rbrace\\
\cup &\left\lbrace (t, c, q') \mid (q, c, q') \in (\atmtnTrans^{\lang_1}  \cup \atmtnTrans^{\lang_2} \cup \atmtnTrans^{\lang_3}) \wedge q \in  \{\atmtnInitState^{\lang_1}, \atmtnInitState^{\lang_2}, \atmtnInitState^{\lang_3}\}\right\rbrace.
\end{align*}
\end{itemize}
This construction is illustrated in Figure~\ref{fig:proofSelective}. The language recognized by $\atmtn$ is $\word(\lang_1\cup \lang_2)^\ast\lang_3$. Observe that $\atmtn$ is coaccessible since $\atmtnIndex{\lang_1}$, $\atmtnIndex{\lang_2}$ and $\atmtnIndex{\lang_3}$ are coaccessible. Also observe that $t$ is essential by construction: by merging the initial and final states of $\lang_1$ (resp.~$\lang_2$), we created cycles on $t$. Thus, $\atmtnInitState^{\word}$ is also essential.

\begin{figure}[tbh]
	\centering
\begin{tikzpicture}[every node/.style={font=\small,inner sep=1pt}]
\draw (0,0) node[oval,minimum width=27mm] (qw) {$\atmtnIndex{\word} \setminus \{\atmtnFinalState^{\word}\}$};
\draw ($(qw)+(0,-1.2)$) node[rond] (t) {$t$};
\draw ($(t)+(-2.5,-1.2)$) node[oval,minimum width=27mm] (qk1) {$\atmtnIndex{\lang_1} \setminus \{\atmtnInitState^{\lang_1}, \atmtnFinalState^{\lang_1}\}$};
\draw ($(t)+(2.5,-1.2)$) node[oval,minimum width=27mm] (qk2) {$\atmtnIndex{\lang_2} \setminus \{\atmtnInitState^{\lang_2}, \atmtnFinalState^{\lang_2}\}$};
\draw ($(t)+(0,-2.2)$) node[oval,minimum width=27mm] (qk3) {$\atmtnIndex{\lang_3} \setminus \{\atmtnInitState^{\lang_3}\}$};
\draw ($(qw)+(0,0.8)$) edge[-latex'] (qw);
\draw (qw) edge[-latex'] (t);
\draw (t) edge[-latex',out=-140,in=40] (qk1);
\draw (t) edge[-latex',out=-150,in=50] (qk1);
\draw (qk1) edge[-latex',out=80,in=-180] (t);
\draw (qk1) edge[-latex',out=90,in=-190] (t);
\draw (t) edge[-latex',out=-40,in=140] (qk2);
\draw (t) edge[-latex',out=-30,in=130] (qk2);
\draw (qk2) edge[-latex',out=100,in=0] (t);
\draw (qk2) edge[-latex',out=90,in=10] (t);
\draw (t) edge[-latex'] (qk3);
\end{tikzpicture}
	\caption{Automaton $\atmtn$ built to establish $\memskel$-selectivity.}
	\label{fig:proofSelective}
\end{figure}

We will now interpret this NFA as an arena and use the hypothesis. Let $\arena = \atmtnArena{\atmtn}$. By Lemma~\ref{lem:arenaLanguage}, we have that $\colhat(\Plays(\arena, \atmtnInitState^{\word})) = \left[ \word(\lang_1\cup \lang_2)^\ast\lang_3\right]$. By hypothesis, $\Pone$ has a $\atmtnInitState^{\word}$-optimal strategy $\strat \in \stratsFM_1(\arena)$, encoded as a Mealy machine $\mealy_{\strat} = (\memskel, \memnxt)$, in $\game = (\arena, \pref)$.

Let $\play \in \Plays(\arena, \atmtnInitState^{\word}, \strat)$ be the only play consistent with $\strat$ from $\atmtnInitState^{\word}$. By $\atmtnInitState^{\word}$-optimality, we have that
\begin{equation}
\label{eq:optimalPlay}
	\left[ \word(\lang_1\cup \lang_2)^\ast\lang_3\right] \pref \colhat(\play).
\end{equation}

By definition of $\arena$, this play $\play$ necessarily contains a history $\pth = \edge_1 \ldots{} \edge_n$ such that $\edgeOut(\edge_n) = t$ and for all $i$, $1 \leq i < n$, $\edgeOut(\edge_i) \neq t$. Observe that $\colhat(\pth) = \word$. Recall that $\memstate = \memupdhat(\meminit, \word)$ is the memory state reached after reading $\word$ since $\word \in\famelt_{\meminit,\memstate}$. Let $\edge = \memnxt(\memstate, t)$ be the edge chosen by $\strat$ in $t$ when $t$ is first visited. Note that in contrast to the construction in Theorem~\ref{thm:FMtoMonotony}, $t$ could be visited many times here, and even infinitely often (using cycles from $\lang_1$ and $\lang_2$). We consider two cases in the following.

First, assume that $\edge$ belongs to the part of the arena generated by $\atmtnIndex{\lang_3}$. Since $t$ (originally $\atmtnInitState^{\lang_3}$) has no incoming transition in $\atmtnIndex{\lang_3}$, we conclude that $\play$ never visits $t$ again, and that $\colhat(\play) \in [\word \lang_3]$. By Equation~\eqref{eq:optimalPlay}, we verify Equation~\eqref{eq:toProveSelectivity}.

Now, assume that $\edge$ belongs to the part of the arena generated by $\atmtnIndex{\lang_1}$ (the same reasoning will apply symmetrically for $\atmtnIndex{\lang_2}$). We want to show that $\colhat(\play) \in [\word \lang^\ast_1]$, i.e., that $\strat$ never switches to another part of the arena. Two cases are possible: either $(a)$ $\play$ visits $t$ only once, or $(b)$ $\play$ visits $t$ at least twice.

Case $(a)$. Since $\play$ visits $t$ only once and $t$ is the only state where the play could switch to a different automaton, we have that $\play = \pth\cdot \play'$ for a suffix $\play'$ starting in $t$ and entirely contained in $\atmtnIndex{\lang_1}$. Hence, we have $\colhat(\play) = \word\cdot \colhat(\play')$ with $\colhat(\play') \in [\lang_1]$. Thus, $\colhat(\play) \in [\word \lang_1] \subseteq [\word \lang^\ast_1]$.

Case $(b)$. Let $\play = \pth \cdot \pth' \cdot \play'$, such that $\pth'$ ends with the second visit of $t$. Recall that $\word = \colhat(\pth)$, $\memstate = \memupdhat(\meminit, \word)$, and $\edge = \memnxt(\memstate, t)$. Now, by definition of $\lang_1$, we have that $\colhat(\pth') \in \famelt_{\memstate,\memstate}$. Hence, $\memupdhat(\memstate, \colhat(\pth'))  =\memstate$. Intuitively, the memory skeleton is back to the same memory state after reading the cycle $\pth'$. As argued in Theorem~\ref{thm:FMtoMonotony}, $\memnxt$ is deterministic, and both the state of the arena and the memory state are identical after $\pth$ and after $\pth\cdot\pth'$. Therefore $\strat(\pth\cdot \pth') = \strat(\pth) = \edge$. Iterating this reasoning (as all cycles on $t$ in $\atmtnIndex{\lang_1}$ are read as cycles on $m$ in the memory), we conclude that $\play = \pth \cdot (\pth')^\omega$. This implies that $\colhat(\play) \in [\word \lang_1^\ast]$.

Hence, in both cases, we have that $\colhat(\play) \in [\word \lang_1^\ast]$. Now, by Equation~\eqref{eq:optimalPlay}, we verify Equation~\eqref{eq:toProveSelectivity}.

Wrapping everything up, we have that whatever the part of the arena to which $e$ belongs, Equation~\eqref{eq:toProveSelectivity} is verified. Therefore, we have shown that $\pref$ is indeed $\memskel$-selective.
\end{proof}

\subsection*{Wrap-up.} We have established that the existence of finite-memory optimal strategies based on a skeleton $\memskel$ in one-player games implies both $\memskel$-monotony and $\memskel$-selectivity of the preference relation, under mild uniformity assumptions. It is interesting to observe that this holds already for \textit{one-player} games (a fortiori, for \textit{two-player} games too). Next, we consider the converse: we will prove that $\memskel$-monotony and $\memskel$-selectivity implies the existence of UFM strategies, not only in one-player games, but even in \textit{two-player} ones, when satisfied by the preference relation \textit{and its inverse}.

\section{From \texorpdfstring{$\memskel$}{M}-monotony and \texorpdfstring{$\memskel$}{M}-selectivity to finite memory based on \texorpdfstring{$\memskel$}{M}}
\label{sec:conceptsToFM}

\subsection*{Induction step.} To prove the sought implication (Theorem~\ref{thm:conceptsToML}), we first focus on \textit{memoryless} strategies in ``covered'' arenas, as discussed in Section~\ref{subsec:concepts}. Intuitively, a ``covered'' arena resembles a product arena (with a memory skeleton): hence studying memoryless strategies on such arenas is very close to studying finite-memory strategies on general arenas.

We will proceed by induction on the number of choices in an arena, as sketched in Section~\ref{subsec:concepts}. This induction will require us to mix different Nash equilibria (one for each player) in a proper way, to maintain the desired property. For the sake of readability, we thus start by proving the induction step for one player.

For an arena $\arenafull$, we write $n_\arena = \vert \edges \vert - \vert \states \vert$ for its number of choices. We also define the notion of \emph{subarena}:
we say that an arena $\arena' = (\states_1', \states_2', \edges')$ is a \textit{subarena} of an arena $\arenafull$ if $\states_1 = \states_1'$, $\states_2 = \states_2'$, and $\edges' \subseteq \edges$.
That is, arena $\arena'$ is a subarena of $\arena$ if it can be obtained from $\arena$ by removing some edges of $\arena$ (while keeping it non-blocking). We say that a set of arenas $\arenaClass$ is closed under the subarena operation if for all $\arena\in\arenaClass$, for all subarenas $\arena'$ of $\arena$, $\arena'\in\arenaClass$.

\begin{lem}
\label{lem:inductionStep}
Let $\pref$ be a preference relation, $\memskelPref$ and $\memskelCyc$ be two memory skeletons, and $\arenaClass$ be a set of arenas closed under the subarena operation. Assume that $\pref$ is $\memskelPref$-monotone and $\memskelCyc$-selective, and that for all $\Ptwo$'s one-player arenas $\arenafull \in \arenaClass$, for all subsets of states $\covStates \subseteq \states$ for which $\memskelPref$ is a prefix-cover and $\memskelCyc$ is a cyclic-cover, $\Ptwo$ has an optimal strategy from $\covStates$.

Let $n \in \mathbb{N}$. Assume that for all arenas $\arena' = (\states'_1, \states'_2, \edges') \in \arenaClass$ such that $n_{\arena'} < n$, for all subsets of states $\covStates' \subseteq \states'$ for which $\memskelPref$ is a prefix-cover and $\memskelCyc$ is a cyclic-cover, there exists a memoryless Nash equilibrium $(\strat'_1, \strat'_2) \in \stratsML_1(\arena') \times \stratsML_2(\arena')$ from $\covStates'$ in $\game' = (\arena', \pref)$.

Then, for all arenas $\arenafull \in \arenaClass$ such that $n_{\arena} = n$, for all subsets of states $\covStates \subseteq \states$ for which $\memskelPref$ is a prefix-cover and $\memskelCyc$ is a cyclic-cover, there exists a Nash equilibrium $(\strat_1, \strat_2) \in \stratsML_1(\arena) \times \strats_2(\arena)$ from $\covStates$ in $\gamefull$ such that $\strat_1$ is memoryless.
\end{lem}

Note that the same holds for $\Ptwo$ and $\invpref$ symmetrically.

Intuitively, Lemma~\ref{lem:inductionStep} states that under the hypotheses of $\memskelPref$-monotony and $\memskelCyc$-selectivity, if both players can play optimally with memoryless strategies in ``small'' and ``covered'' arenas, the same property holds for at least $\Pone$ in ``covered'' arenas where an additional choice exists.

This lemma has to be commented. First, observe that the property is about Nash equilibria. Indeed, as explained in Section~\ref{sec:prelims}, the result we prove is actually slightly stronger than the existence of optimal strategies, as it can be stated for Nash equilibria.

Second, this lemma is focused on proving the existence of an NE in which $\Pone$'s strategy is memoryless: proving that this holds for both players will be done in Theorem~\ref{thm:conceptsToML}.

Third, as motivated in Section~\ref{subsec:results}, we state our result as the existence of memoryless optimal strategies in ``covered'' arenas: the existence of UFM strategies in general arenas will follow (Corollary~\ref{cor:UFM}), but taking this road allows us to keep optimal strategies memoryless for many arenas (which already share the ``classifying'' properties that a product with a memory skeleton would grant).

Fourth, we use two different skeletons, one for monotony (i.e., dealing with prefixes) and one for selectivity (i.e., dealing with cycles). Obviously, one can use a single combined skeleton using Lemma~\ref{lem:memProdPreserving} and Lemma~\ref{lem:productArenaIsCovered}, but our approach has the advantage of being \textit{compositional} and highlighting how each skeleton / property impacts the reasoning in the proof: we will see that they have different uses.

Lastly, the notions of prefix-covers and cyclic-covers are defined with regard to a covered set of states $\covStates$ in order to keep the need for uniformity minimal, in the same spirit as what we did in Section~\ref{sec:FMToConcepts}.

As mentioned above, our proof is essentially an induction step. Starting from an arena $\arena$ with $n_\arena = n$ choices, we identify a state $t$ in which $\Pone$ has at least two
outgoing edges (the proof is symmetric for $\Ptwo$). By splitting the edges in $t$ into two sets, we obtain two corresponding subarenas $\arena_a$ and $\arena_b$ such that $n_{\arena_a}, n_{\arena_b} < n$, along with the corresponding subgames. The induction hypothesis gives us two memoryless Nash equilibria (from $\covStates$) in these subgames: $(\strat^a_1, \strat^a_2)$ and $(\strat^b_1, \strat^b_2)$. The arguments can then be unfolded intuitively as follows. First, using $\memskelPref$-monotony and $\memskelPref$ being a prefix-cover, we identify one subarena (say $\arena_a$) which is clearly at least as good as the other for $\Pone$. Second, we build a strategy profile $(\strat^{\#}_1, \strat^{\#}_2)$, that we claim to be an NE in $\game$, in the following way: $\Pone$ uses strategy $\strat_1^a$ (the one from the best subarena) and $\Ptwo$ reacts to $\Pone$'s actions by playing the corresponding best-response strategy. I.e., if $\Pone$ plays in $\arena_a$, $\Ptwo$ plays according to $\strat_2^a$, and otherwise he plays according to $\strat_2^b$. Third, it remains to prove the two inequalities of Equation~\eqref{eq:defNE}. The rightmost one is easy, as well as the leftmost one in the subcase where the unique play $\pi \in \Plays(\arena, \state, \strat^{\#}_1, \strat^{\#}_2)$ does not visit state $t$: they can both be proved essentially thanks to the induction hypothesis and easy construction arguments. The crux of the proof is thus in the last step: proving that the leftmost inequality holds when the play visits $t$. This can be achieved thanks to $\memskelCyc$-selectivity and $\memskelCyc$ being a cyclic-cover, Lemma~\ref{lem:languageSplit}, inherent properties of the preference relation, $\arena_a$ being the best subarena thanks to $\memskelPref$-monotony, and the induction hypothesis, in that order.

Obviously, $\memskel$-monotony (Definition~\ref{def:monotony}), $\memskel$-selectivity (Definition~\ref{def:selectivity}), prefix-covers and cyclic-covers (Definition~\ref{def:covers}) were defined to be sufficient to provide Lemma~\ref{lem:inductionStep}: one of the main challenges was to have them not too powerful as to keep them also necessary, as proved in Section~\ref{sec:FMToConcepts}.

\begin{proof}
Let $\memskelPref = (\memstates^\mathsf{p}, \meminit^\mathsf{p}, \memupd^\mathsf{p})$ and $\memskelCyc = (\memstates^\mathsf{c}, \meminit^\mathsf{c}, \memupd^\mathsf{c})$. Let $\pref$ be a preference relation that is $\memskelPref$-monotone and $\memskelCyc$-selective. Let $n \in \mathbb{N}$ and assume that for all arenas $\arena' = (\states'_1, \states'_2, \edges') \in \arenaClass$ such that $n_{\arena'} < n$, for all subsets of states $\covStates' \subseteq \states'$ for which $\memskelPref$ is a prefix-cover and $\memskelCyc$ is a cyclic-cover, there exists a memoryless Nash equilibrium $(\strat'_1, \strat'_2) \in \stratsML_1(\arena') \times \stratsML_2(\arena')$ in $\game' = (\arena', \pref)$.

Now, let $\arenafull \in \arenaClass$ be an arena such that $n_\arena = n$, and let $\covStates \subseteq \states$ be a subset of states for which $\memskelPref$ is a prefix-cover and $\memskelCyc$ is a cyclic-cover. Our goal is to prove that there exists an NE $(\strat_1, \strat_2) \in \stratsML_1(\arena) \times \strats_2(\arena)$ from $\covStates$ in $\gamefull$ such that $\strat_1$ is memoryless. The same proof can be done with respect to $\player{2}$ symmetrically, using its preference relation $\pref^{-1}$ and the appropriate skeletons.

If $\arena$ is such that $\Pone$ has no choice (i.e., all his states have only one outgoing edge), then $\arena$ is a $\Ptwo$'s one-player arena. There is only one (memoryless) strategy in $\strats_1(\arena)$, and by hypothesis, $\Ptwo$ has a strategy to play in $\arena$ that is optimal from states prefix-covered by $\memskelPref$ and cyclic-covered by $\memskelCyc$. Thus there is indeed a Nash equilibrium in $\stratsML_1(\arena) \times \strats_2(\arena)$ from $\covStates$.
Let us now assume that $\Pone$ has at least one choice and let $t \in \states_1$ be a state with at least two outgoing edges, i.e.,  $\vert\{\edge \in \edges \mid \edgeIn(\edge) = t\}\vert > 1$. We partition $\{\edge \in \edges \mid \edgeIn(\edge) = t\}$ into two (non-empty) sets $\edges_a$ and $\edges_b$, and we define two corresponding subarenas, $\arena_a = (\states_1, \states_2, \edges \setminus \edges_b)$, and $\arena_b = (\states_1, \states_2, \edges \setminus \edges_a)$, which are in $\arenaClass$ as this set is closed under the subarena operation.
Observe that it remains true that $\memskelPref$ is a prefix-cover and $\memskelCyc$ is a cyclic-cover of $\covStates$, both in $\arena_a$ and $\arena_b$, by definition of prefix- and cyclic-covers (intuitively, we quantify universally over fewer histories than in $\arena$).

Thus, by induction hypothesis (since $n_{\arena_a}, n_{\arena_b} < n_\arena = n$), we have memoryless NE from $\covStates$ in the subgames $\game_a = (\arena_a, \pref)$ and $\game_b = (\arena_b, \pref)$. Let us denote them by $(\strat_1^j, \strat^j_2)$ for game $\game_j$, $j \in \{a, b\}$.

Since $\memskelPref$ is a prefix-cover of $\covStates$ in $\arena$, there exists $\memstate^\mathsf{p}_t \in \memstates^\mathsf{p}$ such that, for all $\pth \in \Paths(\arena)$ such that $\edgeIn(\pth) \in \covStates$, $\edgeOut(\pth) = t$ and such that for all $\pth'$ proper prefix of $\pth$, $\edgeOut(\pth') \neq t$, we have $\memupdhat(\meminit^\mathsf{p}, \colhat(\pth)) = \memstate^\mathsf{p}_t$. Now, let $\lang^\mathsf{p}_j = \colhat\big( \Paths(\arena_j, t, \strat^j_2)\big)$, for $j \in \{a, b\}$, that is, $\lang^\mathsf{p}_j$ contains all (projections to colors of) histories consistent with $\strat_2^j$ and starting in $t$ in subarena $\arena_j$.

By $\memskelPref$-monotony, we can easily deduce that we have
\begin{align}
&\forall\, \word \in \famelt^\mathsf{p}_{\meminit^\mathsf{p}, \memstate_t^\mathsf{p}},\; [\word \lang^\mathsf{p}_a] \pref [\word \lang^\mathsf{p}_b],\nonumber\\
\text{or } &\forall\, \word \in \famelt^\mathsf{p}_{\meminit^\mathsf{p}, \memstate_t^\mathsf{p}},\; [\word \lang^\mathsf{p}_b] \pref [\word \lang^\mathsf{p}_a]\label{eq:KaIsBetter}
\end{align}
where $\famelt^\mathsf{p}_{\meminit^\mathsf{p}, \memstate_t^\mathsf{p}}$ stands for the usual language of sequences of colors read from $\meminit^\mathsf{p}$ to $\memstate_t^\mathsf{p}$, the additional superscript being used to highlight that we are considering skeleton $\memskelPref$ here. From now on, we assume w.l.o.g.~that \eqref{eq:KaIsBetter} holds, i.e., that $\forall\, \word \in \famelt^\mathsf{p}_{\meminit^\mathsf{p}, \memstate_t^\mathsf{p}},\; [\word \lang^\mathsf{p}_b] \pref [\word \lang^\mathsf{p}_a]$. Intuitively, this means that, for $\Pone$, committing to the subarena $\arena_a$ is always at least as good as committing to the subarena $\arena_b$. Note that this does not imply anything with regard to alternating between the two subarenas, which could a priori be beneficial: we will deal with that soon thanks to $\memskelCyc$-selectivity.

Let us define the strategy $\strat^{\#}_1 \in \stratsML_1(\arena)$ of $\player{1}$ in the game $\game = (\arena, \pref)$ as $\strat^{\#}_1 = \strat^{a}_1$, the strategy used in the NE from $\covStates$ in the subgame $\game_a$ (we chose this one because of assumption \eqref{eq:KaIsBetter}: $\game_a$ is the better subgame of the two for $\Pone$). Strategy $\strat^{\#}_1$ is thus memoryless by definition. Note that $\strat^{\#}_1$ is well-defined on $\arena$ even though the original strategy was on $\arena_a$, since it is memoryless (i.e., whether the prefix did visit $\arena_b$ or not does not matter). Now we define a corresponding strategy $\strat^{\#}_2 \in \strats_2(\arena)$ for $\player{2}$ in $\game$ that uses a small amount of memory, as follows:
\begin{equation*}
\forall\, \pth \in \Paths_2(\arena),\; \strat^{\#}_2(\pth) = \begin{cases}
\strat^a_2(\pth) \text{ if } \pth \text{ never visited } t,\\
\strat^a_2(\pth) \text{ if the last visit of } t \text{ was followed by an edge in } \arena_a,\\
\strat^b_2(\pth) \text{ otherwise}.
\end{cases}
\end{equation*}
Again this strategy is well-defined on $\arena$. Our goal is to show that the strategy profile $(\strat^{\#}_1, \strat^{\#}_2)$ is an NE from all states in $\covStates$ in the larger game $\game$. In particular, Lemma~\ref{lem:NEisOpti} then implies that this profile is a couple of $\covStates$-optimal strategies.

Formally, we will establish that for all $\state \in \covStates$, for all $\strat_1 \in \strats_1(\arena)$, for all $\strat_2 \in \strats_2(\arena)$, we have
\begin{equation}
\label{eq:NE}
\colhat(\Plays(\arena, \state, \strat_1, \strat^{\#}_2)) \pref \colhat(\Plays(\arena, \state, \strat^{\#}_1, \strat^{\#}_2)) \pref \colhat(\Plays(\arena, \state, \strat^{\#}_1, \strat_2)).
\end{equation}

We begin with the rightmost inequality of Equation~\eqref{eq:NE}. Let $\state \in \covStates$ and let $\strat_2 \in \strats_2(\arena)$ be an arbitrary strategy for $\player{2}$ in $\game$. We denote by $\strat_2[\arena_a]$ its restriction to (histories of) $\arena_a$: note that this strategy is well-defined as only edges belonging to $\player{1}$ have been removed in $\arena_a$.

We have
\begin{align*}
&\colhat(\Plays(\arena, \state, \strat^{\#}_1, \strat^{\#}_2)) \\
&= \colhat(\Plays(\arena_a, \state, \strat^{a}_1, \strat^{a}_2)) &\text{because these strategies stay in } \arena_a,\\
&\pref \colhat(\Plays(\arena_a, \state, \strat^{a}_1, \strat_2[\arena_a])) &\text{because } (\strat_1^{a}, \strat^{a}_2) \text{ is an NE from } s \text{ in } \arena_a,\\
&= \colhat(\Plays(\arena, \state, \strat^{\#}_1, \strat_2)) &\text{because these strategies stay in } \arena_a,
\end{align*}
hence the rightmost inequality is verified.

Now, consider the leftmost inequality of Equation~\eqref{eq:NE}. Let $\state \in \covStates$ and let $\strat_1 \in \strats_1(\arena)$ be an arbitrary strategy for $\player{1}$ in $\game$. Let $\play \in \Plays(\arena, \state, \strat_1, \strat^{\#}_2)$ be the only play consistent with $\strat_1$ and $\strat^{\#}_2$ from $\state$. We first consider the case where $\play$ never visits $t$. If this is the case, then $\play$ is also a play in $\arena_a$. Let $\strat'_1 \in \strats_1(\arena_a)$ be a strategy of $\player{1}$ that mimics $\strat_1$ on all histories that belong to $\arena_a$, except the ones ending in $t$, where it plays an arbitrary edge in $\edges_a$. We have
\begin{align*}
\colhat(\play) &= \colhat(\Plays(\arena, \state, \strat_1, \strat^{\#}_2))\\
&= \colhat(\Plays(\arena_a, \state, \strat'_1, \strat^{a}_2)) &\text{because } \play \text{ stays in } \arena_a \text{ and never visits } t,\\
&\pref \colhat(\Plays(\arena_a, \state, \strat^{a}_1, \strat^a_2)) &\text{because } (\strat_1^{a}, \strat^{a}_2) \text{ is an NE from } s \text{ in } \arena_a,\\
&= \colhat(\Plays(\arena, \state, \strat^{\#}_1, \strat^{\#}_2)) &\text{because these strategies stay in } \arena_a,
\end{align*}
hence the leftmost inequality is verified in the case where $\play$ never visits $t$.

It remains to consider the case where $\play$ does visit $t$. Observe that for the moment, we have not used $\memskelCyc$-selectivity and the fact that $\state \in \covStates$: they will be crucial to solve this (more complex) case.

We define $\lang^\mathsf{c}_j = \colhat\big(\{\pth \in \Paths(\arena_j, t, \strat_2^j) \mid \edgeOut(\pth) = t\}\big)$, for $j \in \{a, b\}$, that is, $\lang^\mathsf{c}_j$ contains all (projections to colors of) cycles on $t$ consistent with $\strat_2^j$ (i.e., the strategy from the subgame NE) in subarena $\arena_j$. Since the unique play $\play \in \Plays(\arena, \state, \strat_1, \strat^{\#}_2)$ visits $t$ at least once, we write $\play = \pth \cdot \play'$ for $\pth$, the prefix ending with the first visit of $t$. Let $\word = \colhat(\pth)$. Observe that
\[
\colhat(\play) \in [\word(\lang^\mathsf{c}_a \cup \lang^\mathsf{c}_b)^\ast (\lang^\mathsf{p}_a \cup \lang^\mathsf{p}_b)]
\]
since $\strat_2^{\#}$ alternates between $\strat_2^a$ and $\strat_2^b$ depending on what $\Pone$ plays in $t$.
Intuitively, either $\play$ cycles infinitely often on $t$ using cycles of $(\lang^\mathsf{c}_a \cup \lang^\mathsf{c}_b)$, or it does it for a while, then switches to $(\lang^\mathsf{p}_a \cup \lang^\mathsf{p}_b)$, which induces that $\play$ commits to a subarena.
Thus, we trivially have $\colhat(\play) \pref [\word(\lang^\mathsf{c}_a \cup \lang^\mathsf{c}_b)^\ast (\lang^\mathsf{p}_a \cup \lang^\mathsf{p}_b)]$.

Since $\memskelCyc$ is a cyclic-cover of $\covStates$ in $\arena$, and $\pth$ starts in $\covStates$, we know that for $\memstate^\mathsf{c} = \widehat{\memupd^\mathsf{c}}(\meminit^\mathsf{c}, \colhat(\pth))$, and for all $\pth' \in \Paths(\arena)$ such that $\edgeIn(\pth') = \edgeOut(\pth') = t$, we have $\widehat{\memupd^\mathsf{c}}(\memstate^\mathsf{c}, \colhat(\pth')) = \memstate^\mathsf{c}$. That is, all cycles on $t$ are read as cycles on $\memstate^\mathsf{c}$ in $\memskelCyc$. This implies that $\lang^\mathsf{c}_a, \lang^\mathsf{c}_b \subseteq  \famelt_{\memstate^\mathsf{c},\memstate^\mathsf{c}}$. Knowing that, we can invoke the $\memskelCyc$-selectivity (Equation~\eqref{eq:selectivity}) of the preference relation to obtain
\begin{align*}
\colhat(\play) &\pref [\word(\lang^\mathsf{c}_a)^\ast] \cup [\word(\lang^\mathsf{c}_b)^\ast] \cup [\word (\lang^\mathsf{p}_a \cup \lang^\mathsf{p}_b)].
\end{align*}
Using Lemma~\ref{lem:languageSplit}, we have
\begin{align*}
\colhat(\play) \pref [\word(\lang^\mathsf{c}_a)^\ast] \cup [\word(\lang^\mathsf{c}_b)^\ast] \cup [\word \lang^\mathsf{p}_a] \cup [\word \lang^\mathsf{p}_b].
\end{align*}
Observe that $[\word(\lang^\mathsf{c}_j)^\ast] \subseteq [\word \lang^\mathsf{p}_j]$, for $j \in \{a, b\}$. Hence, we have
\begin{align*}
\colhat(\play) &\pref [\word \lang^\mathsf{p}_a] \cup [\word \lang^\mathsf{p}_b].
\end{align*}
Now, recall that using the $\memskelPref$-monotony of $\pref$, we assumed that $\forall\, \word \in \famelt^\mathsf{p}_{\meminit^\mathsf{p}, \memstate_t^\mathsf{p}},\; [\word \lang^\mathsf{p}_b] \pref [\word \lang^\mathsf{p}_a]$. Since $\pth$ starts in $\covStates$, ends in its first visit to $t$ and $\memskelPref$ is a prefix-cover of $\covStates$, this inequality is in particular true for $\word = \colhat(\pth)$. Hence, we have
\begin{align*}
\colhat(\play) &\pref [\word \lang^\mathsf{p}_a].
\end{align*}
Now, recall that $(\strat_1^{a}, \strat_2^a)$ is an NE from $\covStates$ in $\game_a$. Recall also that $\word$ represents the history up to the first visit of $t$ consistent with $(\strat_1, \strat_2^{\#})$; it is also consistent with $(\strat_1, \strat_2^{a})$ since $\strat_2^{\#}$ follows $\strat_2^{a}$ up to the first visit of $t$. Hence, we also have
\[
[\word \lang^\mathsf{p}_a] \subseteq \colhat(\Plays(\arena_a, s, \strat_2^{a})) \pref \colhat(\Plays(\arena_a, \state, \strat_1^{a}, \strat_2^{a})).
\]
Therefore,
\begin{align*}
\colhat(\play) &\pref \colhat(\Plays(\arena_a, \state, \strat_1^{a}, \strat_2^{a}))\\
&= \colhat(\Plays(\arena, \state, \strat_1^{\#}, \strat_2^{\#})) &\text{ because these strategies stay in } \arena_a.
\end{align*}
Recalling that $\play$ is the only play in $\Plays(\arena, \state, \strat_1, \strat^{\#}_2)$, we are done with proving the leftmost inequality of Equation~\eqref{eq:NE}.

Summing up our arguments, we have established that the couple of strategies $(\strat^{\#}_1, \strat^{\#}_2)$ is indeed a Nash equilibrium in $\game$ from $\covStates$. Note that this in particular implies, via Lemma~\ref{lem:NEisOpti}, that $\strat^{\#}_1$ is an $\covStates$-optimal memoryless strategy in $\game$.
\end{proof}

\subsection*{Memoryless Nash equilibria.} We are now armed to establish the implication sketched earlier. As motivated before, we first state the result in the context of memoryless NE on ``covered'' arenas, the finite-memory case on general arenas will follow almost trivially. We first show the result for one-player arenas, and use it to obtain the two-player case.

\begin{lem}
\label{lem:conceptsToML1p}
Let $\pref$ be a preference relation and $\memskelPref$, $\memskelCyc$ be two memory skeletons. Assume that $\pref$ is $\memskelPref$-monotone and $\memskelCyc$-selective. Then, for all $\Pone$'s one-player arenas $\arenafull$, for all subsets of states $\covStates \subseteq \states$ for which $\memskelPref$ is a prefix-cover and $\memskelCyc$ is a cyclic-cover, there exists a memoryless optimal strategy $\strat_1 \in \stratsML_1(\arena)$ from $\covStates$ in $\game = (\arena, \pref)$.
\end{lem}

\begin{proof}
Let $\arenaClass$ be the set of all $\Pone$'s one-player arenas, which is closed under the subarena operation. By hypothesis, $\pref$ is $\memskelPref$-monotone and $\memskelCyc$-selective, and clearly $\Ptwo$ has optimal strategies on all his one-player arenas in $\arenaClass$, as $\Ptwo$ has no choice in any of these arenas.

We proceed by induction on the number of choices in the arena. The base case, $n_\arena = 0$, is trivial. Now let $n \in \IN \setminus \{0\}$ and assume the result holds for $n_\arena < n$. Let $\arenafull\in\arenaClass$ be a $\Pone$'s one-player arena such that $n_\arena = n$, and let $\covStates \subseteq \states$ be a subset of states for which $\memskelPref$ is a prefix-cover and $\memskelCyc$ is a cyclic-cover. We can invoke Lemma~\ref{lem:inductionStep} (note that as we only consider $\Pone$'s one-player arenas, the existence of a Nash equilibrium coincides with the existence of an optimal strategy for $\Pone$~--- see Remark~\ref{rem:UFM1pNE}), and obtain an optimal strategy for $\Pone$ from $\covStates$.
\end{proof}

We are now ready for the two-player case, which requires monotony and selectivity assumptions for both relation $\pref$ and relation $\invpref$.

\begin{thm}
\label{thm:conceptsToML}
Let $\pref$ be a preference relation and $\memskelPref_1$,  $\memskelPref_2$, $\memskelCyc_1$ and $\memskelCyc_2$ be four memory skeletons. Assume that $\pref$ is $\memskelPref_1$-monotone and $\memskelCyc_1$-selective, and that $\invpref$ is $\memskelPref_2$-monotone and $\memskelCyc_2$-selective. Then, for all arenas $\arenafull$, for all subsets of states $\covStates \subseteq \states$ for which $\memskelPref_1$ and $\memskelPref_2$ are prefix-covers, and $\memskelCyc_1$ and $\memskelCyc_2$ are cyclic-covers, there exists a memoryless Nash equilibrium $(\strat_1, \strat_2) \in \stratsML_1(\arena) \times \stratsML_2(\arena)$ from $\covStates$ in $\game = (\arena, \pref)$.
\end{thm}

As always, we want to keep our results as general and compositional as possible, hence we consider different skeletons for the two players. As argued before, one can always take a single skeleton for the two players, as well as for the two notions, by taking their product and using Lemma~\ref{lem:memProdPreserving} and Lemma~\ref{lem:productArenaIsCovered}.

As discussed previously, this theorem in particular implies the existence of memoryless $\covStates$-optimal strategies for both players (via Lemma~\ref{lem:NEisOpti}).

It is fairly straightforward to prove Theorem~\ref{thm:conceptsToML} once Lemma~\ref{lem:inductionStep} is established: the main idea is to invoke Lemma~\ref{lem:inductionStep} for both players while doing the induction, and obtain two Nash equilibria, both of which being memoryless for only one player. Then, to conclude, we resort to Lemma~\ref{lem:mixingNE} which gives us the possibility to mix these two NE into one that is now memoryless for \textit{both} players.

\begin{rem}
\label{rmk:generalNE}
Recall that a crucial hypothesis for Lemma~\ref{lem:mixingNE} to hold is that our games are \textit{antagonistic}, i.e., that we consider $\pref$ and its inverse relation $\invpref$. It is quite interesting to observe that our use of Lemma~\ref{lem:mixingNE} is the only circumstance in which this hypothesis matters (and it is indeed essential) in all our reasoning.\footnote{To be more precise: we wrote everything in the antagonistic setting, but Equation~\eqref{eq:defNE} can be written as two inequalities in the general setting~--- $\colhat(\Plays(\arena, \state, \strat'_1, \strat_2)) \pref_1 \colhat(\Plays(\arena, \state, \strat_1, \strat_2))$ and $\colhat(\Plays(\arena, \state, \strat_1, \strat'_2)) \pref_2 \colhat(\Plays(\arena, \state, \strat_1, \strat_2))$~--- and all our previous reasoning can be rewritten accordingly.} In other words, most of our arguments would hold for two different preference relations, $\pref_1$ and $\pref_2$, without the hypothesis that $\pref_2$ equals $(\pref_1)^{-1}$. The problem would be that we cannot mix the two equilibria in a single equilibrium with both strategies being memoryless~--- while we do need it in the hypothesis of the induction step Lemma~\ref{lem:inductionStep}.

Whether the same reasoning can be extended to (general) Nash equilibria by adapting Lemma~\ref{lem:inductionStep} to take into account the unavoidable blow-up of memory is a question we leave open for future work. Note that the memory bounds would be awful in any case: as the induction would unroll, the memory needed in the equilibria would build up (essentially one bit of memory is added at each call of the induction step in our easier setting, which is then discarded thanks to Lemma~\ref{lem:inductionStep}).\hfill$\lhd$
\end{rem}

\begin{proof}
Let $\pref$ be a preference relation and $\memskelPref_1$,  $\memskelPref_2$, $\memskelCyc_1$ and $\memskelCyc_2$ be four memory skeletons such that $\pref$ is $\memskelPref_1$-monotone and $\memskelCyc_1$-selective, and $\invpref$ is $\memskelPref_2$-monotone and $\memskelCyc_2$-selective.
We consider the set of all arenas, $\arenaClass$, which is closed under the subarena operation. By Lemma~\ref{lem:conceptsToML1p}, we immediately obtain that for all $\Pone$'s (resp.~$\Ptwo$'s) one-player arenas $\arenafull$, for all subsets of states $\covStates \subseteq \states$ for which $\memskelPref_1$ and $\memskelPref_2$ are prefix-covers, and $\memskelCyc_1$ and $\memskelCyc_2$ are cyclic-covers, $\Pone$ (resp.~$\Ptwo$) has an optimal strategy from $\covStates$.

We will proceed by induction on the number of choices in the arena, as described before. The base case, $n_\arena = 0$, is trivial. Now let $n \in \mathbb{N} \setminus \{0\}$ and assume the result holds for $n_\arena < n$. Let $\arenafull \in\arenaClass$ be an arena such that $n_\arena = n$, and let $\covStates \subseteq \states$ be a subset of states for which $\memskelPref_1$ and $\memskelPref_2$ are prefix-covers, and $\memskelCyc_1$ and $\memskelCyc_2$ are cyclic-covers.

Focusing on $\Pone$ and $\pref$, we invoke Lemma~\ref{lem:inductionStep} (using $\memskelPref_1\memProduct\memskelPref_2$ and $\memskelCyc_1\memProduct\memskelCyc_2$, and the induction hypothesis) and obtain an NE $(\strat^\spadesuit_1, \strat^\spadesuit_2) \in \stratsML_1(\arena) \times \strats_2(\arena)$ from $\covStates$ in $\game = (\arena, \pref)$. Note that this NE is only memoryless for $\Pone$! Symmetrically, focusing on $\Ptwo$ and $\invpref$, we invoke Lemma~\ref{lem:inductionStep} (using $\memskelPref_1\memProduct\memskelPref_2$ and $\memskelCyc_1\memProduct\memskelCyc_2$, and the induction hypothesis) and obtain an NE $(\strat^\clubsuit_1, \strat^\clubsuit_2) \in \strats_1(\arena) \times \stratsML_2(\arena)$ from $\covStates$ in $\game = (\arena, \pref)$. Again, note that this NE is only memoryless for $\Ptwo$.

To conclude, it suffices to use Lemma~\ref{lem:mixingNE}: $(\strat^\spadesuit_1, \strat^\spadesuit_2)$ can be mixed with $(\strat^\clubsuit_1, \strat^\clubsuit_2)$ into an equivalent NE $(\strat^\spadesuit_1, \strat^\clubsuit_2) \in \stratsML_1(\arena) \times \stratsML_2(\arena)$, which is now memoryless for \textit{both} players. This concludes our induction step and our proof.
\end{proof}

\subsection*{Finite-memory Nash equilibria and UFM strategies.} Finally, we conclude this section by establishing our result as a corollary.

\begin{cor}
\label{cor:UFM}
Let $\pref$ be a preference relation and $\memskelPref_1$,  $\memskelPref_2$, $\memskelCyc_1$ and $\memskelCyc_2$ be four memory skeletons. Assume that $\pref$ is $\memskelPref_1$-monotone and $\memskelCyc_1$-selective, and that $\invpref$ is $\memskelPref_2$-monotone and $\memskelCyc_2$-selective. Then, for all arenas $\arenafull$, there exists a uniform finite-memory Nash equilibrium $(\strat_1, \strat_2) \in \stratsFM_1(\arena) \times \stratsFM_2(\arena)$ in $\game = (\arena, \pref)$, such that strategies $\strat_i$ are encoded as Mealy machines $\mealy_{\strat_i} = (\memskel, \memnxt^i)$ based on the joint memory skeleton $\memskel = \memskelPref_1 \memProduct \memskelPref_2 \memProduct \memskelCyc_1 \memProduct \memskelCyc_2$.
\end{cor}

As usual in this section, we state our result for the slightly stronger notion of Nash equilibria: it involves in particular the \textit{existence of UFM strategies for both players}. As for Theorem~\ref{thm:conceptsToML}, we use four memory skeletons to keep the approach compositional and player-based, and we provide strategies based on their product memory. However, if there exists a skeleton $\memskel$ that is already such that both $\pref$ and $\invpref$ are $\memskel$-monotone and $\memskel$-selective, this skeleton suffices to build both strategies (this is clear in the following proof).

This corollary is fairly easy to obtain. We build the joint memory skeleton $\memskel$ as defined above. By Lemma~\ref{lem:memProdPreserving} and Lemma~\ref{lem:productArenaIsCovered}, we can invoke Theorem~\ref{thm:conceptsToML} on the product arena $\prodAS{\arena}{\memskel}$ and obtain a memoryless NE on it, or equivalently, a finite-memory one on the original arena, through Lemma~\ref{lem:UFMiffMLonProduct}.

\begin{proof}
Let $\pref$ be a preference relation and let $\memskelPref_1$,  $\memskelPref_2$, $\memskelCyc_1$ and $\memskelCyc_2$ be four memory skeletons such that $\pref$ is $\memskelPref_1$-monotone and $\memskelCyc_1$-selective, and that $\invpref$ is $\memskelPref_2$-monotone and $\memskelCyc_2$-selective. Let $\arenafull$ be an arena.

We define $\memskelfull = \memskelPref_1 \memProduct \memskelPref_2 \memProduct \memskelCyc_1 \memProduct \memskelCyc_2$, the joint memory skeleton.  By Lemma~\ref{lem:memProdPreserving}, $\pref$ and $\invpref$ are both $\memskel$-monotone and $\memskel$-selective.

Consider the product arena $\arena' =\prodAS{\arena}{\memskel}$, as defined in Section~\ref{sec:prelims}. Recall that $\states' =  \states\times\memstates$. By Lemma~\ref{lem:productArenaIsCovered}, the set of states $\covStates' = \states \times \{ \meminit\} \subseteq \states'$ is both prefix-covered and cyclic-covered by $\memskel$.

Putting the last two arguments together, we may invoke Theorem~\ref{thm:conceptsToML} on $\arena'$ and obtain a memoryless Nash equilibrium $(\strat'_1, \strat'_2) \in \stratsML_1(\arena') \times \stratsML_2(\arena')$ from $\covStates'$ in $\game' = (\arena', \pref)$.

To conclude, it suffices to use Lemma~\ref{lem:UFMiffMLonProduct} (stated using NE, as discussed in Remark~\ref{rmk:NEonProduct}): the memoryless equilibrium $(\strat'_1, \strat'_2)$ in the product game $\game'$ can be seen as a finite-memory equilibrium $(\strat_1, \strat_2) \in \stratsFM_1(\arena) \times \stratsFM_2(\arena)$ in $\game = (\arena, \pref)$, where strategies $\strat_i$ are encoded as Mealy machines $\mealy_{\strat_i} = (\memskel, \memnxt^i)$.
\end{proof}

We can also formulate a version of this last result focusing on one-player arenas.

\begin{cor}
	\label{cor:UFM1p}
	Let $\pref$ be a preference relation and $\memskelPref$,  $\memskelCyc$ be two memory skeletons. Assume that $\pref$ is $\memskelPref$-monotone and $\memskelCyc$-selective. Then, for all $\Pone$'s one-player arenas $\arenafull$, there exists a UFM strategy $\strat_1 \in \stratsFM_1(\arena)$ in $\game = (\arena, \pref)$, such that strategy $\strat_1$ is encoded as a Mealy machine $\mealy_{\strat_1} = (\memskel, \memnxt)$ based on the joint memory skeleton $\memskel = \memskelPref \memProduct \memskelCyc$.
\end{cor}

\begin{proof}[Proof sketch]
	The proof is very similar to (and easier than) the proof of Corollary~\ref{cor:UFM}, but uses the one-player implication of Lemma~\ref{lem:conceptsToML1p} instead of Theorem~\ref{thm:conceptsToML}.
\end{proof}

\begin{rem}
Our whole induction scheme is bottom-up, and presented through the prism of covered arenas. It is possible to obtain a similar proof scheme by going top-down and starting from product arenas~--- which are particular cases of covered arenas (Lemma~\ref{lem:productArenaIsCovered}). Our approach pursues two objectives. First, extracting the main technical elements needed and describing them through the concepts of prefix- and cyclic-covers to give a better grasp of how things work and where. Second, providing memoryless optimal strategies in \textit{all} covered arenas (Theorem~\ref{thm:conceptsToML}).\hfill$\lhd$
\end{rem}

\section{Discussion}
\label{sec:discussion}

We close our paper with a discussion of the assets and limits of our approach, its applicability with regard to the current research landscape, and the directions we aim to follow in future work.

\paragraph{Technical features of our approach} As observed through Remark~\ref{rmk:weakestHypothesesAndCompositionality}, our results are established using fine-grained assumptions and conclusions, in an effort to push the approach to its limits. They also preserve \textit{compositionality}, splitting the reasoning for $\memskel$-monotony and $\memskel$-selectivity, and for the two players.

Alongside $\memskel$-monotony and $\memskel$-selectivity, we define two other key concepts to solve the technical issues related to the induction on product arenas: \textit{prefix-covers} and \textit{cyclic-covers}. These notions are crucial tools to prove the results in Section~\ref{sec:conceptsToFM}.

\paragraph{Some advantages} The aforementioned concepts of prefix-covers and cyclic-covers also have benefits from a practical point of view: given a preference relation $\pref$ and the corresponding memory skeleton $\memskel$, they let us \textit{identify game arenas where memoryless strategies suffice} whereas finite memory (based on $\memskel$) might be necessary in general. Such arenas are the ones covered by $\memskel$.\footnote{The follow-up paper~\cite{DBLP:conf/concur/BouyerORV21} further discusses how to know if an arena is covered.} Hence in practice, this approach permits to obtain UML strategies for many arenas where a coarser approach would only provide UFM ones.

Our approach yields \textit{two methods} to establish that a preference relation (or equivalently a payoff function or a winning condition) admits UFM strategies. The first one, exhibiting appropriate memory skeletons and proving $\memskel$-monotony and $\memskel$-selectivity, is based on Theorem~\ref{thm:equivalence} and can be used \textit{compositionally} through Corollary~\ref{cor:UFM}. The second one follows the \textit{lifting corollary}, Corollary~\ref{cor:lifting}: one only has to study the one-player subcases then invoke this result to lift the existence of UFM strategies to the two-player case, without checking for $\memskel$-monotony and $\memskel$-selectivity at all. Hence this second method is often painless in practice.

Two interesting facts can be seen through Corollary~\ref{cor:lifting}. First, there is \textit{no blow-up in the memory} required when going from one-player games to two-player games: the overall memory simply combines the memory skeletons of the two players. Second, assuming that one has an algorithm to solve\footnote{I.e., decide who has a winning strategy from a given state.} one-player games~--- say for $\Pone$~--- for a winning condition satisfying our hypotheses, this lifting corollary also induces a \textit{naive algorithm for the two-player case for free}: thanks to the bounds on memory, one may enumerate the strategies of the adversary, $\Ptwo$~--- or guess one if one aims for a non-deterministic algorithm~--- and solve the corresponding $\Pone$'s game(s) where the strategy of $\Ptwo$ is fixed. Note that while such a simple algorithm might not be optimal, it does correspond to the approach giving the best complexity class known for the renowned family of games in $\mathsf{NP} \cap \mathsf{coNP}$, such as, e.g., parity or mean-payoff games (e.g.,~\cite{DBLP:journals/ipl/Jurdzinski98}). These last two cases could already be dealt with thanks to Gimbert and Zielonka's result since they involve memoryless strategies, but now a similar road can be taken for any objective that admits arena-independent finite-memory optimal strategies, such as, e.g., generalized parity games.

\paragraph{Applicability} Let us give a quick tour of some classical (combinations of) objectives~--- expressed through winning conditions, payoffs or preference relations~--- and assess whether our approach permits to establish the existence of UFM strategies in the corresponding games.

Note that when considering multiple (quantitative) objectives, optimal strategies usually do not exist, and one has to settle for \textit{Pareto-optimal} ones (e.g.,~\cite{DKQR20}). However, in many cases, the (decision) problem under study is as follows: given a threshold (vector), define the winning condition as all the plays achieving at least this threshold, and check for a winning strategy. Hence multi-objective quantitative games are often de facto reduced to qualitative win-lose games for this so-called \textit{threshold problem}. Observe that, given a multi-objective setting, if UFM strategies exist for all threshold problems, then finite-memory strategies suffice to realize the Pareto front (as each point of this front can be considered as a threshold). Therefore, \textit{our approach also enables reasoning about the existence of finite-memory Pareto-optimal strategies in multi-objective games}.

We start our overview with some game settings that fall under the scope of our approach. Obviously, \textit{all memoryless-determined objectives} are among them, since we generalize Gimbert and Zielonka's work~\cite{GZ05}: this includes, e.g., mean-payoff~\cite{EM79}, parity~\cite{DBLP:conf/focs/EmersonJ88,DBLP:journals/tcs/Zielonka98}, energy~\cite{DBLP:conf/emsoft/ChakrabartiAHS03} or average-energy games~\cite{DBLP:journals/acta/BouyerMRLL18}. As established in Section~\ref{sec:intro}, our results encompass all cases where \textit{arena-independent} memory suffices. Hence they permit to rediscover the existence of UFM strategies for games such as, e.g., generalized reachability~\cite{DBLP:journals/corr/abs-1010-2420}, generalized parity~\cite{DBLP:conf/fossacs/ChatterjeeHP07}, Muller~\cite{DBLP:conf/lics/DziembowskiJW97,DBLP:journals/corr/abs-2105-12009}, window parity~\cite{DBLP:journals/corr/BruyereHR16}, some variants of window mean-payoff~\cite{DBLP:journals/iandc/Chatterjee0RR15}, or lower- and upper-bounded (multi-dimension) energy games~\cite{DBLP:conf/formats/BouyerFLMS08,DBLP:journals/acta/BouyerMRLL18,DBLP:conf/fossacs/BouyerHMR017}. Our approach can also be useful to extend these known results to more general combinations, either via appropriate memory skeletons or through the lifting corollary (see an application in Section~\ref{subsec:example}).

There are many games that do not fit our approach for \textit{good reasons}, as they do not admit UFM strategies in general: e.g., multi-dimension mean-payoff~\cite{DBLP:journals/iandc/VelnerC0HRR15}, mean-payoff parity~\cite{DBLP:conf/lics/ChatterjeeHJ05}, finitary parity and Streett~\cite{DBLP:journals/tocl/ChatterjeeHH09}, or energy mean-payoff games~\cite{DBLP:conf/concur/BruyereHRR19}. More interesting are games for which finite-memory strategies exist, but the memory is \textit{arena-dependent}. These notably include games with multi-dimension lower-bounded energy objectives and no upper bound~\cite{DBLP:journals/acta/ChatterjeeRR14,DBLP:conf/icalp/JurdzinskiLS15}, or other variants of window mean-payoff games~\cite{DBLP:journals/iandc/Chatterjee0RR15}. In such games, the players usually have to keep track of information such as, e.g., the sum of weights along an acyclic path, which is bounded for any given arena, but by a value that grows when the arena grows. Hence the need for memory that grows with the arena parameters. Our results cannot be applied directly to such cases in order to obtain the existence of finite-memory strategies for all games. An adaptation of our approach could potentially be used for subclasses of arenas where the parameters are bounded (in order to regain a skeleton working on all arenas of the class).

\paragraph{Comparison with related work} We already discussed extensively the most important related articles~\cite{DBLP:conf/mfcs/GimbertZ04,GZ05,DBLP:conf/icalp/Kopczynski06,DBLP:journals/amai/BiancoFMM11,DBLP:journals/iandc/AminofR17,martin_AM75,DBLP:conf/fsttcs/0001PR18} in Section~\ref{sec:intro}, alongside a technical comparison between our work and Gimbert and Zielonka's seminal result~\cite{GZ05}. Here, we simply want to highlight interesting directions of research inspired by some of these papers.

First, Aminof and Rubin provide a simpler (but incomplete) approach to memoryless determinacy through the prism of first-cycle games in~\cite{DBLP:journals/iandc/AminofR17}: a similar take on finite-memory determinacy could be appealing~--- it could provide sufficient conditions easier to test than $\memskel$-monotony and $\memskel$-selectivity.

Second, Bianco et al.~establish sufficient (and relaxed) conditions to ensure the existence of UML strategies \textit{for one player, in two-player games,} in~\cite{DBLP:journals/amai/BiancoFMM11}: it would be interesting to study the corresponding problem in the finite-memory case. Indeed, in many games where infinite memory is needed, it is only the case for one of the players (e.g.,~\cite{DBLP:journals/iandc/VelnerC0HRR15,DBLP:conf/lics/ChatterjeeHJ05,DBLP:conf/concur/BruyereHRR19}) and such conditions could thus prove useful. Note that this is different from Theorem~\ref{thm:equivalence1p}, which gives a sufficient \textit{and necessary} condition but for one-player games only.

Finally, recall that Le Roux et al.~give a rather tight characterization of combinations of objectives preserving the sufficiency of finite-memory strategies in~\cite{DBLP:conf/fsttcs/0001PR18}. Their techniques, as well as the scope of their results, are somewhat orthogonal to ours. Whether both approaches can be intertwined to obtain results on more general settings remains an open question.

\paragraph{Limits and future work} To close this paper, we recall three limits of our approach, and the corresponding open problems.

First, as explained throughout the paper, our results cover all cases where \textit{arena-independent} memory suffices, and are \textit{limited to these cases}. We have argued that the approach cannot be fully lifted to the general case, for good reasons, as the lifting corollary breaks in some situations (see Sections~\ref{sec:intro} and~\ref{sec:counterexample}). Still, we have hope to generalize our approach to some extent to the \textit{arena-dependent} case, through some \textit{function} associating memory skeletons to arenas, as discussed in Section~\ref{sec:intro}. Obtaining a lifting corollary~--- under well-chosen conditions~--- in the arena-dependent case would be of tremendous help in practice: see for example~\cite{DBLP:journals/acta/BouyerMRLL18,DBLP:conf/fossacs/BouyerHMR017,DBLP:conf/concur/BruyereHRR19}. Hence this is clearly the next step in our quest.

Second, our result is a characterization \textit{instantiated by a memory skeleton $\memskel$}. While the lifting corollary is helpful in applications, it would be fantastic to be able to find an appropriate skeleton automatically, and to be able to determine if a given skeleton is minimal (with regard to a preference relation). This paper is a first step toward these long-term objectives.

Lastly, as explained in Remark~\ref{rmk:antagonistic} and Remark~\ref{rmk:generalNE}, most of our arguments carry over to the case of \textit{general Nash equilibria}. That is, when considering not necessarily antagonistic games where the two players use different, not necessarily inverse, preference relations. Whether our approach can be adapted in this case, at the price of an unavoidable blow-up of memory, is an open question worth considering. In particular, we want to study the links between our results (including the lifting from one-player to two-player games) and recent results lifting finite-memory determinacy in two-player games to the existence of finite-memory Nash equilibria in multi-player games~\cite{DBLP:journals/iandc/RouxP18}.

\bibliographystyle{alphaurl}
\bibliography{fm}

\newcommand{\etalchar}[1]{$^{#1}$}
\begin{thebibliography}{CdAHS03}

\bibitem[AR17]{DBLP:journals/iandc/AminofR17}
Benjamin Aminof and Sasha Rubin.
\newblock First-cycle games.
\newblock {\em Inf. Comput.}, 254:195--216, 2017.
\newblock \href {https://doi.org/10.1016/j.ic.2016.10.008}
  {\path{doi:10.1016/j.ic.2016.10.008}}.

\bibitem[BCH{\etalchar{+}}16]{DBLP:conf/lata/BrenguierCHPRRS16}
Romain Brenguier, Lorenzo Clemente, Paul Hunter, Guillermo~A. P{\'{e}}rez,
  Mickael Randour, Jean{-}Fran{\c{c}}ois Raskin, Ocan Sankur, and Mathieu
  Sassolas.
\newblock Non-zero sum games for reactive synthesis.
\newblock In Adrian{-}Horia Dediu, Jan Janousek, Carlos Mart{\'{\i}}n{-}Vide,
  and Bianca Truthe, editors, {\em Language and Automata Theory and
  Applications - 10th International Conference, {LATA} 2016, Prague, Czech
  Republic, March 14-18, 2016, Proceedings}, volume 9618 of {\em Lecture Notes
  in Computer Science}, pages 3--23. Springer, 2016.
\newblock \href {https://doi.org/10.1007/978-3-319-30000-9\_1}
  {\path{doi:10.1007/978-3-319-30000-9\_1}}.

\bibitem[BCJ18]{DBLP:reference/mc/BloemCJ18}
Roderick Bloem, Krishnendu Chatterjee, and Barbara Jobstmann.
\newblock Graph games and reactive synthesis.
\newblock In Edmund~M. Clarke, Thomas~A. Henzinger, Helmut Veith, and Roderick
  Bloem, editors, {\em Handbook of Model Checking}, pages 921--962. Springer,
  2018.
\newblock \href {https://doi.org/10.1007/978-3-319-10575-8\_27}
  {\path{doi:10.1007/978-3-319-10575-8\_27}}.

\bibitem[BDOR19]{DBLP:conf/concur/BrihayeDOR19}
Thomas Brihaye, Florent Delgrange, Youssouf Oualhadj, and Mickael Randour.
\newblock Life is random, time is not: {M}arkov decision processes with window
  objectives.
\newblock In Fokkink and van Glabbeek \cite{DBLP:conf/concur/2019}, pages
  8:1--8:18.
\newblock \href {https://doi.org/10.4230/LIPIcs.CONCUR.2019.8}
  {\path{doi:10.4230/LIPIcs.CONCUR.2019.8}}.

\bibitem[BFL{\etalchar{+}}08]{DBLP:conf/formats/BouyerFLMS08}
Patricia Bouyer, Ulrich Fahrenberg, Kim~Guldstrand Larsen, Nicolas Markey, and
  Jir{\'{\i}} Srba.
\newblock Infinite runs in weighted timed automata with energy constraints.
\newblock In Franck Cassez and Claude Jard, editors, {\em Formal Modeling and
  Analysis of Timed Systems, 6th International Conference, {FORMATS} 2008,
  Saint Malo, France, September 15-17, 2008. Proceedings}, volume 5215 of {\em
  Lecture Notes in Computer Science}, pages 33--47. Springer, 2008.
\newblock \href {https://doi.org/10.1007/978-3-540-85778-5\_4}
  {\path{doi:10.1007/978-3-540-85778-5\_4}}.

\bibitem[BFMM11]{DBLP:journals/amai/BiancoFMM11}
Alessandro Bianco, Marco Faella, Fabio Mogavero, and Aniello Murano.
\newblock Exploring the boundary of half-positionality.
\newblock {\em Ann. Math. Artif. Intell.}, 62(1-2):55--77, 2011.
\newblock \href {https://doi.org/10.1007/s10472-011-9250-1}
  {\path{doi:10.1007/s10472-011-9250-1}}.

\bibitem[BHM{\etalchar{+}}17]{DBLP:conf/fossacs/BouyerHMR017}
Patricia Bouyer, Piotr Hofman, Nicolas Markey, Mickael Randour, and Martin
  Zimmermann.
\newblock Bounding average-energy games.
\newblock In Javier Esparza and Andrzej~S. Murawski, editors, {\em Foundations
  of Software Science and Computation Structures - 20th International
  Conference, {FOSSACS} 2017, Held as Part of the European Joint Conferences on
  Theory and Practice of Software, {ETAPS} 2017, Uppsala, Sweden, April 22-29,
  2017, Proceedings}, volume 10203 of {\em Lecture Notes in Computer Science},
  pages 179--195, 2017.
\newblock \href {https://doi.org/10.1007/978-3-662-54458-7\_11}
  {\path{doi:10.1007/978-3-662-54458-7\_11}}.

\bibitem[BHR16]{DBLP:journals/corr/BruyereHR16}
V{\'{e}}ronique Bruy{\`{e}}re, Quentin Hautem, and Mickael Randour.
\newblock Window parity games: an alternative approach toward parity games with
  time bounds.
\newblock In Domenico Cantone and Giorgio Delzanno, editors, {\em Proceedings
  of the Seventh International Symposium on Games, Automata, Logics and Formal
  Verification, GandALF 2016, Catania, Italy, 14-16 September 2016}, volume 226
  of {\em {EPTCS}}, pages 135--148, 2016.
\newblock \href {https://doi.org/10.4204/EPTCS.226.10}
  {\path{doi:10.4204/EPTCS.226.10}}.

\bibitem[BHRR19]{DBLP:conf/concur/BruyereHRR19}
V{\'{e}}ronique Bruy{\`{e}}re, Quentin Hautem, Mickael Randour, and
  Jean{-}Fran{\c{c}}ois Raskin.
\newblock Energy mean-payoff games.
\newblock In Fokkink and van Glabbeek \cite{DBLP:conf/concur/2019}, pages
  21:1--21:17.
\newblock \href {https://doi.org/10.4230/LIPIcs.CONCUR.2019.21}
  {\path{doi:10.4230/LIPIcs.CONCUR.2019.21}}.

\bibitem[BLO{\etalchar{+}}20]{fmCONCUR}
Patricia Bouyer, St{\'{e}}phane {Le~Roux}, Youssouf Oualhadj, Mickael Randour,
  and Pierre Vandenhove.
\newblock Games where you can play optimally with arena-independent finite
  memory.
\newblock In Igor Konnov and Laura Kov{\'{a}}cs, editors, {\em 31th
  International Conference on Concurrency Theory, {CONCUR} 2020, September 1-4,
  2020, Vienna, Austria}, volume 171 of {\em LIPIcs}. Schloss Dagstuhl -
  Leibniz-Zentrum f{\"{u}}r Informatik, 2020.

\bibitem[BMR{\etalchar{+}}18]{DBLP:journals/acta/BouyerMRLL18}
Patricia Bouyer, Nicolas Markey, Mickael Randour, Kim~G. Larsen, and Simon
  Laursen.
\newblock Average-energy games.
\newblock {\em Acta Inf.}, 55(2):91--127, 2018.
\newblock \href {https://doi.org/10.1007/s00236-016-0274-1}
  {\path{doi:10.1007/s00236-016-0274-1}}.

\bibitem[BORV21]{DBLP:conf/concur/BouyerORV21}
Patricia Bouyer, Youssouf Oualhadj, Mickael Randour, and Pierre Vandenhove.
\newblock Arena-independent finite-memory determinacy in stochastic games.
\newblock In Serge Haddad and Daniele Varacca, editors, {\em 32nd International
  Conference on Concurrency Theory, {CONCUR} 2021, August 24-27, 2021, Virtual
  Conference}, volume 203 of {\em LIPIcs}, pages 26:1--26:18. Schloss Dagstuhl
  - Leibniz-Zentrum f{\"{u}}r Informatik, 2021.
\newblock \href {https://doi.org/10.4230/LIPIcs.CONCUR.2021.26}
  {\path{doi:10.4230/LIPIcs.CONCUR.2021.26}}.

\bibitem[BRV21]{DBLP:journals/corr/abs-2110-01276}
Patricia Bouyer, Mickael Randour, and Pierre Vandenhove.
\newblock Characterizing omega-regularity through finite-memory determinacy of
  games on infinite graphs.
\newblock {\em CoRR}, abs/2110.01276, 2021.
\newblock \href {http://arxiv.org/abs/2110.01276} {\path{arXiv:2110.01276}}.

\bibitem[Cas21]{DBLP:journals/corr/abs-2105-12009}
Antonio Casares.
\newblock On the minimisation of transition-based {R}abin automata and the
  chromatic memory requirements of {M}uller conditions.
\newblock {\em CoRR}, abs/2105.12009, 2021.
\newblock URL: \url{https://arxiv.org/abs/2105.12009}, \href
  {http://arxiv.org/abs/2105.12009} {\path{arXiv:2105.12009}}.

\bibitem[CD12]{DBLP:journals/tcs/ChatterjeeD12}
Krishnendu Chatterjee and Laurent Doyen.
\newblock Energy parity games.
\newblock {\em Theor. Comput. Sci.}, 458:49--60, 2012.
\newblock \href {https://doi.org/10.1016/j.tcs.2012.07.038}
  {\path{doi:10.1016/j.tcs.2012.07.038}}.

\bibitem[CdAHS03]{DBLP:conf/emsoft/ChakrabartiAHS03}
Arindam Chakrabarti, Luca de~Alfaro, Thomas~A. Henzinger, and Mari{\"e}lle
  Stoelinga.
\newblock Resource interfaces.
\newblock In Rajeev Alur and Insup Lee, editors, {\em EMSOFT}, volume 2855 of
  {\em Lecture Notes in Computer Science}, pages 117--133. Springer, 2003.

\bibitem[CDRR15]{DBLP:journals/iandc/Chatterjee0RR15}
Krishnendu Chatterjee, Laurent Doyen, Mickael Randour, and
  Jean{-}Fran{\c{c}}ois Raskin.
\newblock Looking at mean-payoff and total-payoff through windows.
\newblock {\em Inf. Comput.}, 242:25--52, 2015.
\newblock \href {https://doi.org/10.1016/j.ic.2015.03.010}
  {\path{doi:10.1016/j.ic.2015.03.010}}.

\bibitem[CHH09]{DBLP:journals/tocl/ChatterjeeHH09}
Krishnendu Chatterjee, Thomas~A. Henzinger, and Florian Horn.
\newblock Finitary winning in omega-regular games.
\newblock {\em {ACM} Trans. Comput. Log.}, 11(1):1:1--1:27, 2009.
\newblock \href {https://doi.org/10.1145/1614431.1614432}
  {\path{doi:10.1145/1614431.1614432}}.

\bibitem[CHJ05]{DBLP:conf/lics/ChatterjeeHJ05}
Krishnendu Chatterjee, Thomas~A. Henzinger, and Marcin Jurdzinski.
\newblock Mean-payoff parity games.
\newblock In {\em 20th {IEEE} Symposium on Logic in Computer Science {(LICS}
  2005), 26-29 June 2005, Chicago, IL, USA, Proceedings}, pages 178--187.
  {IEEE} Computer Society, 2005.
\newblock \href {https://doi.org/10.1109/LICS.2005.26}
  {\path{doi:10.1109/LICS.2005.26}}.

\bibitem[CHP07]{DBLP:conf/fossacs/ChatterjeeHP07}
Krishnendu Chatterjee, Thomas~A. Henzinger, and Nir Piterman.
\newblock Generalized parity games.
\newblock In Helmut Seidl, editor, {\em Foundations of Software Science and
  Computational Structures, 10th International Conference, {FOSSACS} 2007, Held
  as Part of the Joint European Conferences on Theory and Practice of Software,
  {ETAPS} 2007, Braga, Portugal, March 24-April 1, 2007, Proceedings}, volume
  4423 of {\em Lecture Notes in Computer Science}, pages 153--167. Springer,
  2007.
\newblock \href {https://doi.org/10.1007/978-3-540-71389-0\_12}
  {\path{doi:10.1007/978-3-540-71389-0\_12}}.

\bibitem[CRR14]{DBLP:journals/acta/ChatterjeeRR14}
Krishnendu Chatterjee, Mickael Randour, and Jean{-}Fran{\c{c}}ois Raskin.
\newblock Strategy synthesis for multi-dimensional quantitative objectives.
\newblock {\em Acta Inf.}, 51(3-4):129--163, 2014.
\newblock \href {https://doi.org/10.1007/s00236-013-0182-6}
  {\path{doi:10.1007/s00236-013-0182-6}}.

\bibitem[DJW97]{DBLP:conf/lics/DziembowskiJW97}
Stefan Dziembowski, Marcin Jurdzinski, and Igor Walukiewicz.
\newblock How much memory is needed to win infinite games?
\newblock In {\em Proceedings, 12th Annual {IEEE} Symposium on Logic in
  Computer Science, Warsaw, Poland, June 29 - July 2, 1997}, pages 99--110.
  {IEEE} Computer Society, 1997.
\newblock \href {https://doi.org/10.1109/LICS.1997.614939}
  {\path{doi:10.1109/LICS.1997.614939}}.

\bibitem[DKQR20]{DKQR20}
Florent Delgrange, Joost{-}Pieter Katoen, Tim Quatmann, and Mickael Randour.
\newblock Simple strategies in multi-objective {MDP}s.
\newblock In Armin Biere and David Parker, editors, {\em Tools and Algorithms
  for the Construction and Analysis of Systems - 26th International Conference,
  {TACAS} 2020, Held as Part of the European Joint Conferences on Theory and
  Practice of Software, {ETAPS} 2020, Dublin, Ireland, April 25-30, 2020,
  Proceedings, Part {I}}, volume 12078 of {\em Lecture Notes in Computer
  Science}, pages 346--364. Springer, 2020.
\newblock \href {https://doi.org/10.1007/978-3-030-45190-5\_19}
  {\path{doi:10.1007/978-3-030-45190-5\_19}}.

\bibitem[EJ88]{DBLP:conf/focs/EmersonJ88}
E.~Allen Emerson and Charanjit~S. Jutla.
\newblock The complexity of tree automata and logics of programs.
\newblock In {\em FOCS}, pages 328--337. IEEE Computer Society, 1988.

\bibitem[EM79]{EM79}
Andrzej Ehrenfeucht and Jan Mycielski.
\newblock Positional strategies for mean payoff games.
\newblock {\em Int. Journal of Game Theory}, 8(2):109--113, 1979.

\bibitem[FH10]{DBLP:journals/corr/abs-1010-2420}
Nathana{\"{e}}l Fijalkow and Florian Horn.
\newblock The surprizing complexity of reachability games.
\newblock {\em CoRR}, abs/1010.2420, 2010.
\newblock \href {http://arxiv.org/abs/1010.2420} {\path{arXiv:1010.2420}}.

\bibitem[FHKM15]{DBLP:conf/icalp/FijalkowHKS15}
Nathana{\"{e}}l Fijalkow, Florian Horn, Denis Kuperberg, and {Micha{l}
  Skrzypczak}.
\newblock Trading bounds for memory in games with counters.
\newblock In Halld{\'{o}}rsson et~al. \cite{DBLP:conf/icalp/2015-2}, pages
  197--208.
\newblock \href {https://doi.org/10.1007/978-3-662-47666-6\_16}
  {\path{doi:10.1007/978-3-662-47666-6\_16}}.

\bibitem[FvG19]{DBLP:conf/concur/2019}
Wan Fokkink and Rob van Glabbeek, editors.
\newblock {\em 30th International Conference on Concurrency Theory, {CONCUR}
  2019, August 27-30, 2019, Amsterdam, the Netherlands}, volume 140 of {\em
  LIPIcs}. Schloss Dagstuhl - Leibniz-Zentrum f{\"{u}}r Informatik, 2019.
\newblock URL: \url{http://www.dagstuhl.de/dagpub/978-3-95977-121-4}.

\bibitem[Gim07]{DBLP:conf/stacs/Gimbert07}
Hugo Gimbert.
\newblock Pure stationary optimal strategies in {M}arkov decision processes.
\newblock In Wolfgang Thomas and Pascal Weil, editors, {\em {STACS} 2007, 24th
  Annual Symposium on Theoretical Aspects of Computer Science, Aachen, Germany,
  February 22-24, 2007, Proceedings}, volume 4393 of {\em Lecture Notes in
  Computer Science}, pages 200--211. Springer, 2007.
\newblock \href {https://doi.org/10.1007/978-3-540-70918-3\_18}
  {\path{doi:10.1007/978-3-540-70918-3\_18}}.

\bibitem[GK14]{gimbert2014twoplayer}
Hugo Gimbert and Edon Kelmendi.
\newblock Two-player perfect-information shift-invariant submixing stochastic
  games are half-positional.
\newblock Unpublished, 2014.

\bibitem[GTW02]{DBLP:conf/dagstuhl/2001automata}
Erich Gr{\"{a}}del, Wolfgang Thomas, and Thomas Wilke, editors.
\newblock {\em Automata, Logics, and Infinite Games: {A} Guide to Current
  Research [outcome of a Dagstuhl seminar, February 2001]}, volume 2500 of {\em
  Lecture Notes in Computer Science}. Springer, 2002.

\bibitem[GZ04]{DBLP:conf/mfcs/GimbertZ04}
Hugo Gimbert and Wieslaw Zielonka.
\newblock When can you play positionally?
\newblock In Jir{\'{\i}} Fiala, V{\'{a}}clav Koubek, and Jan Kratochv{\'{\i}}l,
  editors, {\em Mathematical Foundations of Computer Science 2004, 29th
  International Symposium, {MFCS} 2004, Prague, Czech Republic, August 22-27,
  2004, Proceedings}, volume 3153 of {\em Lecture Notes in Computer Science},
  pages 686--697. Springer, 2004.
\newblock \href {https://doi.org/10.1007/978-3-540-28629-5\_53}
  {\path{doi:10.1007/978-3-540-28629-5\_53}}.

\bibitem[GZ05]{GZ05}
Hugo Gimbert and Wieslaw Zielonka.
\newblock Games where you can play optimally without any memory.
\newblock In Mart{\'{\i}}n Abadi and Luca de~Alfaro, editors, {\em {CONCUR}
  2005 - Concurrency Theory, 16th International Conference, {CONCUR} 2005, San
  Francisco, CA, USA, August 23-26, 2005, Proceedings}, volume 3653 of {\em
  Lecture Notes in Computer Science}, pages 428--442. Springer, 2005.
\newblock \href {https://doi.org/10.1007/11539452\_33}
  {\path{doi:10.1007/11539452\_33}}.

\bibitem[HIKS15]{DBLP:conf/icalp/2015-2}
Magn{\'{u}}s~M. Halld{\'{o}}rsson, Kazuo Iwama, Naoki Kobayashi, and Bettina
  Speckmann, editors.
\newblock {\em Automata, Languages, and Programming - 42nd International
  Colloquium, {ICALP} 2015, Kyoto, Japan, July 6-10, 2015, Proceedings, Part
  {II}}, volume 9135 of {\em Lecture Notes in Computer Science}. Springer,
  2015.
\newblock \href {https://doi.org/10.1007/978-3-662-47666-6}
  {\path{doi:10.1007/978-3-662-47666-6}}.

\bibitem[JLS15]{DBLP:conf/icalp/JurdzinskiLS15}
Marcin Jurdzinski, Ranko Lazic, and Sylvain Schmitz.
\newblock Fixed-dimensional energy games are in pseudo-polynomial time.
\newblock In Halld{\'{o}}rsson et~al. \cite{DBLP:conf/icalp/2015-2}, pages
  260--272.
\newblock \href {https://doi.org/10.1007/978-3-662-47666-6\_21}
  {\path{doi:10.1007/978-3-662-47666-6\_21}}.

\bibitem[Jur98]{DBLP:journals/ipl/Jurdzinski98}
Marcin Jurdzinski.
\newblock Deciding the winner in parity games is in {UP} $\cap$ co-{UP}.
\newblock {\em Inf. Process. Lett.}, 68(3):119--124, 1998.
\newblock \href {https://doi.org/10.1016/S0020-0190(98)00150-1}
  {\path{doi:10.1016/S0020-0190(98)00150-1}}.

\bibitem[Kop06]{DBLP:conf/icalp/Kopczynski06}
Eryk Kopczy\'nski.
\newblock Half-positional determinacy of infinite games.
\newblock In Michele Bugliesi, Bart Preneel, Vladimiro Sassone, and Ingo
  Wegener, editors, {\em Automata, Languages and Programming, 33rd
  International Colloquium, {ICALP} 2006, Venice, Italy, July 10-14, 2006,
  Proceedings, Part {II}}, volume 4052 of {\em Lecture Notes in Computer
  Science}, pages 336--347. Springer, 2006.
\newblock \href {https://doi.org/10.1007/11787006\_29}
  {\path{doi:10.1007/11787006\_29}}.

\bibitem[Kop08]{KopThesis}
Eryk Kopczy\'nski.
\newblock {\em Half-positional Determinacy of Infinite Games}.
\newblock PhD thesis, Warsaw University, 2008.

\bibitem[{Le~}13]{DBLP:journals/corr/abs-1302-3973}
St{\'{e}}phane {Le~Roux}.
\newblock Infinite sequential {N}ash equilibrium.
\newblock {\em Logical Methods in Computer Science}, 9(2), 2013.
\newblock \href {https://doi.org/10.2168/LMCS-9(2:3)2013}
  {\path{doi:10.2168/LMCS-9(2:3)2013}}.

\bibitem[{Le~}18]{DBLP:conf/mfcs/Roux18}
St{\'{e}}phane {Le~Roux}.
\newblock Concurrent games and semi-random determinacy.
\newblock In Igor Potapov, Paul~G. Spirakis, and James Worrell, editors, {\em
  43rd International Symposium on Mathematical Foundations of Computer Science,
  {MFCS} 2018, August 27-31, 2018, Liverpool, {UK}}, volume 117 of {\em
  LIPIcs}, pages 40:1--40:15. Schloss Dagstuhl - Leibniz-Zentrum fuer
  Informatik, 2018.
\newblock \href {https://doi.org/10.4230/LIPIcs.MFCS.2018.40}
  {\path{doi:10.4230/LIPIcs.MFCS.2018.40}}.

\bibitem[LP18]{DBLP:journals/iandc/RouxP18}
St{\'{e}}phane {Le~Roux} and Arno Pauly.
\newblock Extending finite-memory determinacy to multi-player games.
\newblock {\em Inf. Comput.}, 261(Part):676--694, 2018.
\newblock \href {https://doi.org/10.1016/j.ic.2018.02.024}
  {\path{doi:10.1016/j.ic.2018.02.024}}.

\bibitem[LPR18]{DBLP:conf/fsttcs/0001PR18}
St{\'{e}}phane {Le~Roux}, Arno Pauly, and Mickael Randour.
\newblock Extending finite-memory determinacy by {B}oolean combination of
  winning conditions.
\newblock In Sumit Ganguly and Paritosh~K. Pandya, editors, {\em 38th {IARCS}
  Annual Conference on Foundations of Software Technology and Theoretical
  Computer Science, {FSTTCS} 2018, December 11-13, 2018, Ahmedabad, India},
  volume 122 of {\em LIPIcs}, pages 38:1--38:20. Schloss Dagstuhl -
  Leibniz-Zentrum fuer Informatik, 2018.
\newblock \href {https://doi.org/10.4230/LIPIcs.FSTTCS.2018.38}
  {\path{doi:10.4230/LIPIcs.FSTTCS.2018.38}}.

\bibitem[Mar75]{martin_AM75}
Donald~A. Martin.
\newblock Borel determinacy.
\newblock {\em Annals of Mathematics}, 102(2):363--371, 1975.

\bibitem[OR94]{Osborne1994}
Martin~J. Osborne and Ariel Rubinstein.
\newblock {\em A course in game theory}.
\newblock The MIT Press, Cambridge, USA, 1994.

\bibitem[Ran13]{rECCS}
Mickael Randour.
\newblock Automated synthesis of reliable and efficient systems through game
  theory: A case study.
\newblock In {\em Proc. of ECCS 2012}, Springer Proceedings in Complexity XVII,
  pages 731--738. Springer, 2013.
\newblock \href {https://doi.org/10.1007/978-3-319-00395-5\_90}
  {\path{doi:10.1007/978-3-319-00395-5\_90}}.

\bibitem[VCD{\etalchar{+}}15]{DBLP:journals/iandc/VelnerC0HRR15}
Yaron Velner, Krishnendu Chatterjee, Laurent Doyen, Thomas~A. Henzinger,
  Alexander~Moshe Rabinovich, and Jean{-}Fran{\c{c}}ois Raskin.
\newblock The complexity of multi-mean-payoff and multi-energy games.
\newblock {\em Inf. Comput.}, 241:177--196, 2015.
\newblock \href {https://doi.org/10.1016/j.ic.2015.03.001}
  {\path{doi:10.1016/j.ic.2015.03.001}}.

\bibitem[Zie98]{DBLP:journals/tcs/Zielonka98}
Wieslaw Zielonka.
\newblock Infinite games on finitely coloured graphs with applications to
  automata on infinite trees.
\newblock {\em Theor. Comput. Sci.}, 200(1-2):135--183, 1998.

\end{thebibliography}

\end{document}